\documentclass[12pt,a4paper]{article}
\usepackage{jheplikemod,ifpdf,tocloft,rotating,amsmath,textcomp,mathrsfs,mathtools,xcolor}
\usepackage{tikz}
\usetikzlibrary{calc,decorations.markings,shapes.geometric,decorations.pathreplacing}
\usepackage[backgroundcolor=green!5, linecolor=blue!60]{todonotes}
\def\boundingDraw{red}
\def\boundingDraw{none}

\definecolor{mhvblue}{rgb}{0.6,0.6,0.7765}
\definecolor{nmhvred}{rgb}{0.6765,0.15,0.3}
\definecolor{ampgrey}{rgb}{0.9,0.9,0.9}
\definecolor{unord}{rgb}{0,0,0}
\definecolor{ord}{rgb}{0,0,0.575}
\definecolor{anchorLeg}{rgb}{0.575,0.0,0.225}
\definecolor{optLegColour2}{rgb}{0.4,0.4,0.4}
\definecolor{optLegColour}{rgb}{0.8,0.8,0.8}
\definecolor{optLegColour3}{rgb}{0.8,0.8,0.8}
\definecolor{ord}{rgb}{0,0,0.575}
\definecolor{cut1}{rgb}{0.109,0.459,0.737}
\definecolor{cut2}{rgb}{0.757,0.153,0.176}
\definecolor{totalCount}{rgb}{0,0,0.575}
\definecolor{topCount}{rgb}{0.575,0.0,0.225}
\definecolor{dim}{rgb}{0.55,0.55,0.55}
\definecolor{ndotColor}{rgb}{0.65,0.25,0.25}

\def\figScale{1}
\def\legSpread{4}

\def\edgeLen{1*\figScale}

\pgfmathsetmacro{\pLen}{\edgeLen/(2*sin(72/2))}
\def\legLen{\edgeLen*0.45}

\def\labelDist{\legLen*1.2}
\def\lineThickness{(1pt)}
\def\dotSize{(\figScale*12pt)}
\def\ampSize{(1*\figScale*12pt)}

\def\boxRad{0.707\figScale}
\def\legSpread{30}
\def\legSpreadB{40}
\def\markStroke{0.65}

\tikzset{fullamp/.style={coordinate,minimum size=0.8*\ampSize,ball color=black!50,circle,text=white,inner sep=0}}
\tikzset{fullmhv/.style={coordinate,minimum size=0.8*\ampSize,ball color=mhvblue,circle,text=white,inner sep=0}}
\tikzset{fullnmhv/.style={coordinate,minimum size=0.8*\ampSize,ball color=nmhvred,circle,text=white,inner sep=0}}
\tikzset{fullmhvBar/.style={coordinate,minimum size=0.8*\ampSize,ball color=white,circle,text=white,inner sep=0}}
\tikzset{ordAmp/.style={fill=ampgrey,circle,draw=black,line width=\lineThickness,minimum size=0.6*\ampSize,text=white,inner sep=0}}
\tikzset{mhv/.style={fill=mhvblue,circle,draw=black,line width=\lineThickness,minimum size=0.8*\ampSize,text=white,inner sep=0}}
\tikzset{mhvBar/.style={fill=white,circle,draw=black,line width=\lineThickness,minimum size=0.8*\ampSize,text=white,inner sep=0}}
\tikzset{fgraphEdge/.style={anchorLeg,line width=\lineThickness,line cap=round}}
\tikzset{fgraphExt/.style={ord,line width=\lineThickness,line cap=round}}
\tikzset{fgraphOpt/.style={ord,dotted,line width=\lineThickness,line cap=round}}
\tikzset{fdot/.style={fill=anchorLeg,circle,minimum size=0.35*\ampSize,inner sep=0}}
\tikzset{bdot/.style={fill=black,circle,minimum size=0.35*\ampSize,inner sep=0}}
\tikzset{ext/.style={black,line width=\lineThickness,line cap=round}}
\tikzset{under/.style={white,line width=4*\lineThickness,line cap=round}}
\tikzset{optExt/.style={black,dotted,line width=\lineThickness,line cap=round,rounded corners=10pt}}
\tikzset{optExtSc/.style={black,line width=\lineThickness,line cap=round,rounded corners=10pt}}
\tikzset{dashed/.style={black!70,dotted,line width=\lineThickness,line cap=round,rounded corners=10pt}}
\tikzset{dashed2/.style={black,dotted,line width=\lineThickness,line cap=round,rounded corners=10pt}}
\tikzset{ddot/.style={fill=black,circle,minimum size=0.35*\dotSize,inner sep=0}}
\tikzset{int/.style={black,line width=\lineThickness,line cap=round,rounded corners=1.5pt}}

\tikzset{intInfR/.style={nmhvred,line width=\lineThickness,line cap=round,rounded corners=1.5pt}}
\tikzset{blueDot/.style={fill=mhvblue,circle,draw=black,line width=\lineThickness,minimum size=0.5*\ampSize,text=white,inner sep=0}}
\tikzset{whiteDot/.style={fill=white,circle,draw=black,line width=\lineThickness,minimum size=0.5*\ampSize,text=white,inner sep=0}}
\tikzset{blackDot/.style={fill=black,circle,minimum size=0.5*\ampSize,inner sep=0}}
\tikzset{compositeDot/.style={fill=none,draw=black,line width=\lineThickness,circle,minimum size=0.75*\ampSize,inner sep=0}}
\tikzset{ndot/.style={transform shape,scale=0.35*\figScale,aspect=0.65,draw=ndotColor,line width=\markStroke*\figScale,shape=circle,fill=none}}
\tikzset{markedEdgeR/.style={draw=none,decoration={markings,mark connection node=connode,mark=at position 0.5 with {\node[ndot] (connode) {};}},postaction={decorate}}}

\tikzset{directedEdge/.style={draw=none,decoration={markings,mark connection node=connode,mark=at position 0.5 with {\node[transform shape, scale=0.205*\figScale,shape=dart,aspect=0.5,fill=black,draw] (connode) {};}},postaction={decorate}}}

\tikzset{directedEdgeBend/.style={rounded corners=10pt,draw=none,decoration={markings,mark connection node=connode,mark=at position 0.5 with {\node[transform shape, scale=0.205,shape=dart,aspect=0.5,fill=black,draw] (connode) {};}},postaction={decorate}}}

\newcommand{\lsVerts}[4]{
\ifthenelse{#1=1}{\node at (v1) [fullmhvBar] {};}{\ifthenelse{#1=2}{\node at (v1) [fullmhv] {};}{\ifthenelse{#1=3}{\node at (v1) [fullnmhv] {};}{\ifthenelse{#1=0}{\node at (v1) [bdot] {};}{\ifthenelse{#1=4}{\node at (v1) [fullamp] {};}{}}}}}
\ifthenelse{#2=1}{\node at (v2) [fullmhvBar] {};}{\ifthenelse{#2=2}{\node at (v2) [fullmhv] {};}{\ifthenelse{#2=3}{\node at (v2) [fullnmhv] {};}{\ifthenelse{#2=0}{\node at (v2) [bdot] {};}{\ifthenelse{#2=4}{\node at (v2) [fullamp] {};}{}}}}}
\ifthenelse{#3=1}{\node at (v3) [fullmhvBar] {};}{\ifthenelse{#3=2}{\node at (v3) [fullmhv] {};}{\ifthenelse{#3=3}{\node at (v3) [fullnmhv] {};}{\ifthenelse{#3=0}{\node at (v3) [bdot] {};}{\ifthenelse{#3=4}{\node at (v3) [fullamp] {};}{}}}}}
\ifthenelse{#4=1}{\node at (v4) [fullmhvBar] {};}{\ifthenelse{#4=2}{\node at (v4) [fullmhv] {};}{\ifthenelse{#4=3}{\node at (v4) [fullamp] {};}{\ifthenelse{#4=0}{\node at (v4) [bdot] {};}{}}}}}

\newcommand{\tikzBox}[1]{\begin{tikzpicture}[scale=1,baseline=-3.05,rotate=0]\useasboundingbox ($\figScale*(-1.3,-1.3)$) rectangle ($\figScale*(1.3,1.3)$);\draw[int, line width=0.1,draw=\boundingDraw] ($\figScale*(-1.3,-1.3)$) rectangle ($\figScale*(1.3,1.3)$);
#1\end{tikzpicture}}

\newcommand{\tikzBoxBub}[1]{\begin{tikzpicture}[scale=1,baseline=-3.05,rotate=0]\useasboundingbox ($\figScale*(-1.3,-0.6)$) rectangle ($\figScale*(1.3,0.6)$);\draw[int, line width=0.1,draw=\boundingDraw] ($\figScale*(-1.3,-0.6)$) rectangle ($\figScale*(1.3,0.6)$);
#1\end{tikzpicture}}

\newcommand{\boxVerts}{\coordinate (v0) at (0,0);
\coordinate (v1) at ($(-135:\boxRad)$);\coordinate (v2) at ($(135:\boxRad)$);\coordinate (v3) at ($(45:\boxRad)$);\coordinate (v4) at ($(-45:\boxRad)$);
\coordinate (a0) at ($(v1)+(-135+\legSpread:0.95*\labelDist)$);\coordinate (a1) at ($(v1)+(-135:0.9*\labelDist)$);\coordinate (a2) at ($(v1)+(-135-\legSpread:0.95*\labelDist)$);
\coordinate (b0) at ($(v2)+(135+\legSpread:\labelDist)$);\coordinate (b1) at ($(v2)+(135:\labelDist)$);\coordinate (b2) at ($(v2)+(135-\legSpread:\labelDist)$);
\coordinate (c0) at ($(v3)+(45+\legSpread:\labelDist)$);\coordinate (c1) at ($(v3)+(45:\labelDist)$);\coordinate (c2) at ($(v3)+(45-\legSpread:\labelDist)$);
\coordinate (d0) at ($(v4)+(-45+\legSpread:\labelDist)$);\coordinate (d1) at ($(v4)+(-45:\labelDist)$);\coordinate (d2) at ($(v4)+(-45-\legSpread:\labelDist)$);
}

\newcommand{\triVerts}{\coordinate (v0) at (180:0.25);
\coordinate (v1) at ($(-120:0.8*\boxRad)-(0:0.25)$);\coordinate (v2) at ($(120:0.8*\boxRad)-(0:0.25)$);\coordinate (v3) at ($(0:0.8*\boxRad)-(0:0.25)$);
\coordinate (a0) at ($(v1)+(-120+\legSpread:0.95*\labelDist)$);\coordinate (a1) at ($(v1)+(-120:0.9*\labelDist)$);\coordinate (a2) at ($(v1)+(-120-\legSpread:0.95*\labelDist)$);
\coordinate (b0) at ($(v2)+(120+\legSpread:\labelDist)$);\coordinate (b1) at ($(v2)+(120:\labelDist)$);\coordinate (b2) at ($(v2)+(120-\legSpread:\labelDist)$);
\coordinate (c0) at ($(v3)+(0+\legSpread:\labelDist)$);\coordinate (c1) at ($(v3)+(0:\labelDist)$);\coordinate (c2) at ($(v3)+(0-\legSpread:\labelDist)$);
}

\newcommand{\boxEdges}{\draw[int](v1)--(v2);\draw[int](v2)--(v3);\draw[int](v3)--(v4);\draw[int](v4)--(v1);}
\newcommand{\triEdges}{\draw[int](v1)--(v2);\draw[int](v2)--(v3);\draw[int](v3)--(v1);}
\newcommand{\singleLeg}[3]{\draw[ext] #1--($#1+(#2:\legLen)$);\node at ($#1+(#2:1.2*\labelDist)$)[]{{\footnotesize $#3$}};}
\newcommand{\singleLegIn}[3]{\draw[ext] #1--($#1+(#2:1.1*\legLen)$);\draw[directedEdge] ($#1+(#2:1.35*\legLen)$)--#1;\node at ($#1+(#2:1.2*\labelDist)$)[]{{\footnotesize $#3$}};}
\newcommand{\singleLegOut}[3]{\draw[ext] #1--($#1+(#2:1.1*\legLen)$);\draw[directedEdge] #1--($#1+(#2:1.35*\legLen)$);\node at ($#1+(#2:1.2*\labelDist)$)[]{{\footnotesize $#3$}};}

\newcommand{\massiveLeg}[4]{\coordinate (aa0) at #1;
\coordinate (aa5) at ($#1+(#2+\legSpread:1.04*\legLen)$);
\coordinate (bb5) at ($#1+(#2-\legSpread:1.04*\legLen)$);
\fill[optLegColour] (aa0)--(aa5)--(bb5)--(aa0);
\draw[ext] #1--($#1+(#2+\legSpread:\legLen)$);\node at ($#1+(#2+\legSpread:1.15*\labelDist)$)[]{{\footnotesize $#3$}};
\draw[ext] #1--($#1+(#2-\legSpread:\legLen)$);\node at ($#1+(#2-\legSpread:1.15*\labelDist)$)[]{{\footnotesize $#4$}};}

\newcommand{\optLeg}[4]{\coordinate (aa0) at #1;
\coordinate (aa5) at ($#1+(#2+\legSpread:1.04*\legLen)$);
\coordinate (bb5) at ($#1+(#2-\legSpread:1.04*\legLen)$);
\coordinate (aa1) at ($(aa0)!0.4!(aa5)$);\coordinate (aa2) at ($(aa0)!0.55!(aa5)$);\coordinate (aa3) at ($(aa0)!0.7!(aa5)$);\coordinate (aa4) at ($(aa0)!0.85!(aa5)$);
\coordinate (bb1) at ($(aa0)!0.4!(bb5)$);\coordinate (bb2) at ($(aa0)!0.55!(bb5)$);\coordinate (bb3) at ($(aa0)!0.7!(bb5)$);\coordinate (bb4) at ($(aa0)!0.85!(bb5)$);
\fill[optLegColour] (aa0)--(aa1)--(bb1)--(aa0);
\fill[optLegColour] (aa2)--(aa3)--(bb3)--(bb2)--(aa2);
\fill[optLegColour] (aa4)--(aa5)--(bb5)--(bb4)--(aa4);
\draw[ext] #1--($#1+(#2+\legSpread:\legLen)$);\node at ($#1+(#2+\legSpread:1.15*\labelDist)$)[]{{\footnotesize $#3$}};
\draw[dashed2] #1--($#1+(#2-\legSpread:\legLen)$);
\node at ($#1+(#2-\legSpread:1.15*\labelDist)$)[]{{\footnotesize $#4$}};
}
\newcommand{\optLegs}[4]{\coordinate (aa0) at #1;
\coordinate (aa5) at ($#1+(#2+\legSpreadB:1.04*\legLen)$);
\coordinate (bb5) at ($#1+(#2-\legSpreadB:1.04*\legLen)$);
\coordinate (aa1) at ($(aa0)!0.4!(aa5)$);\coordinate (aa2) at ($(aa0)!0.55!(aa5)$);\coordinate (aa3) at ($(aa0)!0.7!(aa5)$);\coordinate (aa4) at ($(aa0)!0.85!(aa5)$);
\coordinate (bb1) at ($(aa0)!0.4!(bb5)$);\coordinate (bb2) at ($(aa0)!0.55!(bb5)$);\coordinate (bb3) at ($(aa0)!0.7!(bb5)$);\coordinate (bb4) at ($(aa0)!0.85!(bb5)$);
\fill[optLegColour3] (aa0)--(aa1)--(bb1)--(aa0);
\fill[optLegColour3] (aa2)--(aa3)--(bb3)--(bb2)--(aa2);
\fill[optLegColour3] (aa4)--(aa5)--(bb5)--(bb4)--(aa4);
\draw[dashed2] #1--($#1+(#2+\legSpreadB:\legLen)$);\node at ($#1+(#2+\legSpreadB:1.15*\labelDist)$)[]{{\footnotesize $#3$}};
\draw[dashed2] #1--($#1+(#2-\legSpreadB:\legLen)$);
\node at ($#1+(#2-\legSpreadB:1.15*\labelDist)$)[]{{\footnotesize $#4$}};
}
\newcommand{\optLegsM}[4]{\coordinate (aa0) at #1;
\coordinate (aa5) at ($#1+(#2+\legSpreadB:1.04*\legLen)$);
\coordinate (bb5) at ($#1+(#2-\legSpreadB:1.04*\legLen)$);
\coordinate (aa1) at ($(aa0)!0.4!(aa5)$);\coordinate (aa2) at ($(aa0)!0.55!(aa5)$);\coordinate (aa3) at ($(aa0)!0.7!(aa5)$);\coordinate (aa4) at ($(aa0)!0.85!(aa5)$);
\coordinate (bb1) at ($(aa0)!0.4!(bb5)$);\coordinate (bb2) at ($(aa0)!0.55!(bb5)$);\coordinate (bb3) at ($(aa0)!0.7!(bb5)$);\coordinate (bb4) at ($(aa0)!0.85!(bb5)$);
\fill[optLegColour3] (aa0)--(aa1)--(bb1)--(aa0);
\fill[optLegColour3] (aa2)--(aa3)--(bb3)--(bb2)--(aa2);
\fill[optLegColour3] (aa4)--(aa5)--(bb5)--(bb4)--(aa4);
\draw[ext] #1--($#1+(#2+\legSpreadB:\legLen)$);\node at ($#1+(#2+\legSpreadB:1.15*\labelDist)$)[]{{\footnotesize $#3$}};
\draw[dashed2] #1--($#1+(#2-\legSpreadB:\legLen)$);
\node at ($#1+(#2-\legSpreadB:1.15*\labelDist)$)[]{{\footnotesize $#4$}};
}

\newcommand{\bubVerts}{\coordinate (v0) at (0:0);
\draw[draw=none] ($(v0)+(0:14.15pt)$) arc (0:180:14.15pt and 10pt)coordinate[pos=0](v2)coordinate[pos=0.5](e2)coordinate[pos=1](v1);
\draw[draw=none] (v1) arc (180:360:14.15pt and 10pt)coordinate[pos=0.5](e1);
\coordinate (e2a) at ($(e2)+(180:0.5)$);\coordinate (e2b) at ($(e2)+(0:0.5)$);\coordinate (e1a) at ($(e1)+(180:0.5)$);\coordinate (e1b) at ($(e1)+(0:0.5)$);
\coordinate (a0) at ($(v1)+(180+\legSpread:0.95*\labelDist)$);\coordinate (a1) at ($(v1)+(180:0.95*\labelDist)$);\coordinate (a2) at ($(v1)+(180-\legSpread:0.95*\labelDist)$);
\coordinate (b0) at ($(v2)+(0+\legSpread:\labelDist)$);\coordinate (b1) at ($(v2)+(0:\labelDist)$);\coordinate (b2) at ($(v2)+(0-\legSpread:\labelDist)$);
}

\newcommand{\bubEdges}{\bubVerts\draw[int] ($(v0)+(0:14.15pt)$) arc (0:180:14.15pt and 10pt);
\draw[int] (v1) arc (180:360:14.15pt and 10pt);}

\newcommand{\triNonSinglet}{\begin{scope}\path[clip](v1)--(v2)--(v3)--(v1);
\node at (v1) [fill=nmhvred,circle,minimum size=1.2*\ampSize,inner sep=0] {};
\node at (v2) [fill=nmhvred,circle,minimum size=1.2*\ampSize,inner sep=0] {};
\node at (v3) [fill=nmhvred,circle,minimum size=1.2*\ampSize,inner sep=0] {};
\draw[nmhvred,line width=6*\lineThickness,line cap=round,rounded corners=1.5pt](v3)--(v1)--(v2)--(v3);
\draw[white,line width=4*\lineThickness,line cap=round,rounded corners=1.5pt](v3)--(v1)--(v2)--(v3);
\node at (v1) [fill=white,circle,minimum size=1*\ampSize,inner sep=0] {};
\node at (v2) [fill=white,circle,minimum size=1*\ampSize,inner sep=0] {};
\node at (v3) [fill=white,circle,minimum size=1*\ampSize,inner sep=0] {};
\node at (v0) [] {{\scriptsize${\color{nmhvred}\pm}$}};
\end{scope}}

\newcommand{\boxNonSinglet}{\begin{scope}\path[clip](v1)--(v2)--(v3)--(v4)--(v1);
\node at (v1) [fill=nmhvred,circle,minimum size=1.2*\ampSize,inner sep=0] {};
\node at (v2) [fill=nmhvred,circle,minimum size=1.2*\ampSize,inner sep=0] {};
\node at (v3) [fill=nmhvred,circle,minimum size=1.2*\ampSize,inner sep=0] {};
\node at (v4) [fill=nmhvred,circle,minimum size=1.2*\ampSize,inner sep=0] {};
\draw[nmhvred,line width=6*\lineThickness,line cap=round,rounded corners=1.5pt](v4)--(v1)--(v2)--(v3)--(v4);
\draw[white,line width=4*\lineThickness,line cap=round,rounded corners=1.5pt](v4)--(v1)--(v2)--(v3)--(v4);
\node at (v1) [fill=white,circle,minimum size=1*\ampSize,inner sep=0] {};
\node at (v2) [fill=white,circle,minimum size=1*\ampSize,inner sep=0] {};
\node at (v3) [fill=white,circle,minimum size=1*\ampSize,inner sep=0] {};
\node at (v4) [fill=white,circle,minimum size=1*\ampSize,inner sep=0] {};
\node at (v0) [] {{\scriptsize${\color{nmhvred}\pm}$}};
\end{scope}}
\newcommand{\bubNonSinglet}{\begin{scope}\path[clip] ($(v0)+(0:14.15pt)$) arc (0:360:14.15pt and 10pt);
\node at (v1) [fill=nmhvred,circle,minimum size=1.2*\ampSize,inner sep=0] {};
\node at (v2) [fill=nmhvred,circle,minimum size=1.2*\ampSize,inner sep=0] {};
\draw[nmhvred,line width=6*\lineThickness,line cap=round,rounded corners=1.5pt]($(v0)+(0:14.15pt)$) arc (0:360:14.15pt and 10pt);
\draw[white,line width=4*\lineThickness,line cap=round,rounded corners=1.5pt]($(v0)+(0:14.15pt)$) arc (0:360:14.15pt and 10pt);
\node at (v1) [fill=white,circle,minimum size=1*\ampSize,inner sep=0] {};
\node at (v2) [fill=white,circle,minimum size=1*\ampSize,inner sep=0] {};
\node at (v0) [] {{\scriptsize${\color{nmhvred}\pm}$}};
\end{scope}}

\newcommand{\boxIntegrandSchematic}{\tikzBox{\boxVerts\boxEdges\optLeg{(v1)}{-135}{}{}\optLeg{(v2)}{135}{}{}\optLeg{(v3)}{45}{}{}\optLeg{(v4)}{-45}{}{}
\draw[decoration={brace},decorate,line width=0.7pt,rotate=135] (a0)--(a2);\node at ($(a1)+(-135:0.3)$) [] {$A\!$};
\draw[decoration={brace},decorate,line width=0.7pt,rotate=-135] (b0)--(b2);\node at ($(b1)+(135:0.3)$) [] {$B\!$};
\draw[decoration={brace},decorate,line width=0.7pt,rotate=45] (c0)--(c2);\node at ($(c1)+(45:0.3)$) [] {$C\!$};
\draw[decoration={brace},decorate,line width=0.7pt,rotate=-45] (d0)--(d2);\node at ($(d1)+(-45:0.3)$) [] {$D\!$};
\lsVerts{0}{0}{0}{0}
\draw[directedEdge](v1)--(v2);\draw[directedEdge](v2)--(v3);\draw[directedEdge](v3)--(v4);\draw[directedEdge](v4)--(v1);
\node at ($(v1)!0.5!(v2)+(180:0.2)$) [] {{\footnotesize$\ell_b$}};
\node at ($(v2)!0.5!(v3)+(90:0.25)$) [] {{\footnotesize$\ell_c$}};
\node at ($(v3)!0.5!(v4)+(0:0.25)$) [] {{\footnotesize$\ell_d$}};
\node at ($(v4)!0.5!(v1)+(-90:0.25)$) [] {{\footnotesize$\ell_a$}};
\node at (v0) [] {$i$};
}}

\newcommand{\boxIntegrandsSchematic}{\tikzBox{\boxVerts\boxEdges\optLeg{(v1)}{-135}{}{}\optLeg{(v2)}{135}{}{}\optLeg{(v3)}{45}{}{}\optLeg{(v4)}{-45}{}{};
\draw[decoration={brace},decorate,line width=0.7pt,rotate=135] (a0)--(a2);\node at ($(a1)+(-135:0.3)$) [] {$A\!$};
\draw[decoration={brace},decorate,line width=0.7pt,rotate=-135] (b0)--(b2);\node at ($(b1)+(135:0.3)$) [] {$B\!$};
\draw[decoration={brace},decorate,line width=0.7pt,rotate=45] (c0)--(c2);\node at ($(c1)+(45:0.3)$) [] {$C\!$};
\draw[decoration={brace},decorate,line width=0.7pt,rotate=-45] (d0)--(d2);\node at ($(d1)+(-45:0.3)$) [] {$D\!$};
\lsVerts{0}{0}{0}{0}
\draw[markedEdgeR](v1)--(v2);\draw[markedEdgeR](v2)--(v3);\draw[directedEdge](v4)--(v1);
\node at ($(v4)!0.5!(v1)+(-90:0.25)$) [] {{\footnotesize$\ell$}};
}}

\newcommand{\triangleIntegrandSchematic}{\tikzBox{\triVerts\triEdges\optLeg{(v1)}{-120}{}{}\optLeg{(v2)}{120}{}{}\optLeg{(v3)}{0}{}{}
\draw[decoration={brace},decorate,line width=0.7pt,rotate=120] (a0)--(a2);\node at ($(a1)+(-120:0.3)$) [] {$A\!$};
\draw[decoration={brace},decorate,line width=0.7pt,rotate=-120] (b0)--(b2);\node at ($(b1)+(120:0.25)$) [] {$B\!$};
\draw[decoration={brace},decorate,line width=0.7pt,rotate=0] (c0)--(c2);\node at ($(c1)+(0:0.2)$) [] {$C\!$};
\lsVerts{0}{0}{0}{5}
\draw[directedEdge](v1)--(v2);\draw[directedEdge](v2)--(v3);\draw[directedEdge](v3)--(v1);
\node at ($(v1)!0.5!(v2)+(180:0.2)$) [] {{\footnotesize$\ell_b$}};
\node at ($(v2)!0.5!(v3)+(60:0.25)$) [] {{\footnotesize$\ell_c$}};
\node at ($(v3)!0.5!(v1)+(-60:0.25)$) [] {{\footnotesize$\ell_a$}};
\node at (v0) [] {$I$};
}}

\newcommand{\triangleIntegrandsSchematic}{\tikzBox{\triVerts\triEdges\optLeg{(v1)}{-120}{}{}\optLeg{(v2)}{120}{}{}\optLeg{(v3)}{0}{}{}
\draw[decoration={brace},decorate,line width=0.7pt,rotate=120] (a0)--(a2);\node at ($(a1)+(-120:0.3)$) [] {$A\!$};
\draw[decoration={brace},decorate,line width=0.7pt,rotate=-120] (b0)--(b2);\node at ($(b1)+(120:0.25)$) [] {$B\!$};
\draw[decoration={brace},decorate,line width=0.7pt,rotate=0] (c0)--(c2);\node at ($(c1)+(0:0.2)$) [] {$C\!$};
\lsVerts{0}{0}{0}{5}
\draw[markedEdgeR](v1)--(v2);\draw[directedEdge](v3)--(v1);
\node at ($(v3)!0.5!(v1)+(-60:0.25)$) [] {{\footnotesize$\ell$}};
}}

\newcommand{\bubbleIntegrandSchematic}{\tikzBoxBub{\bubVerts\bubEdges\optLeg{(v1)}{180}{}{}\optLeg{(v2)}{0}{}{}
\draw[decoration={brace},decorate,line width=0.7pt,rotate=180] (a0)--(a2);\node at ($(a1)+(-180:0.25)$) [] {$A\!$};
\draw[decoration={brace},decorate,line width=0.7pt,rotate=0] (b0)--(b2);\node at ($(b1)+(0:0.15)$) [] {$B\!$};
\draw[directedEdge] (e1b)--(e1a);\draw[directedEdge] (e2a)--(e2b);
\node at ($(e1a)!0.5!(e1b)+(-90:0.2)$) [] {{\footnotesize$\ell_a\,\,$}};
\node at ($(e2a)!0.5!(e2b)+(90:0.2)$) [] {{\footnotesize$\,\,\ell_b$}};
\lsVerts{0}{0}{5}{5}}}

\newcommand{\bubbleIntegrandsSchematic}{\tikzBoxBub{\bubVerts\bubEdges\optLeg{(v1)}{180}{}{}\optLeg{(v2)}{0}{}{}
\draw[decoration={brace},decorate,line width=0.7pt,rotate=180] (a0)--(a2);\node at ($(a1)+(-180:0.25)$) [] {$A\!$};
\draw[decoration={brace},decorate,line width=0.7pt,rotate=0] (b0)--(b2);\node at ($(b1)+(0:0.15)$) [] {$B\!$};
\draw[directedEdge] (e1b)--(e1a);
\node at ($(e1a)!0.5!(e1b)+(-90:0.2)$) [] {{\footnotesize$\ell\,\,$}};
\lsVerts{0}{0}{5}{5}}}

\newcommand{\fourMassInts}{\tikzBox{\boxVerts\boxEdges\massiveLeg{(v1)}{-135}{}{}\massiveLeg{(v2)}{135}{}{}\massiveLeg{(v3)}{45}{}{}\massiveLeg{(v4)}{-45}{}{}
\node at ($(a1)+(-135:0.2)$) [] {$A$};
\node at ($(b1)+(135:0.2)$) [] {$B$};
\node at ($(c1)+(45:0.2)$) [] {$C$};
\node at ($(d1)+(-45:0.2)$) [] {$D$};\draw[directedEdge](v1)--(v2);\draw[directedEdge](v2)--(v3);\draw[directedEdge](v3)--(v4);\draw[directedEdge](v4)--(v1);
\node at ($(v1)!0.5!(v2)+(180:0.2)$) [] {{\footnotesize$\ell_b$}};
\node at ($(v2)!0.5!(v3)+(90:0.25)$) [] {{\footnotesize$\ell_c$}};
\node at ($(v3)!0.5!(v4)+(0:0.25)$) [] {{\footnotesize$\ell_d$}};
\node at ($(v4)!0.5!(v1)+(-90:0.25)$) [] {{\footnotesize$\ell_a$}};
\lsVerts{0}{0}{0}{0}
\node at (v0) [] {$i$};
}}

\newcommand{\threeMassInts}{\tikzBox{\boxVerts\boxEdges\singleLeg{(v1)}{-135}{a}\massiveLeg{(v2)}{135}{}{}\massiveLeg{(v3)}{45}{}{}\massiveLeg{(v4)}{-45}{}{}
\node at ($(b1)+(135:0.2)$) [] {$B$};
\node at ($(c1)+(45:0.2)$) [] {$C$};
\node at ($(d1)+(-45:0.2)$) [] {$D$};
\draw[directedEdge](v1)--(v2);\draw[directedEdge](v2)--(v3);\draw[directedEdge](v3)--(v4);\draw[directedEdge](v4)--(v1);
\node at ($(v1)!0.5!(v2)+(180:0.2)$) [] {{\footnotesize$\ell_b$}};
\node at ($(v2)!0.5!(v3)+(90:0.25)$) [] {{\footnotesize$\ell_c$}};
\node at ($(v3)!0.5!(v4)+(0:0.25)$) [] {{\footnotesize$\ell_d$}};
\node at ($(v4)!0.5!(v1)+(-90:0.25)$) [] {{\footnotesize$\ell_a$}};
\lsVerts{0}{0}{0}{0}
\node at (v0) [] {$i$};
}}

\newcommand{\twoMassHardInts}{\tikzBox{\boxVerts\boxEdges\singleLeg{(v1)}{-135}{a}\singleLeg{(v2)}{135}{b}\massiveLeg{(v3)}{45}{}{}\massiveLeg{(v4)}{-45}{}{}
\node at ($(c1)+(45:0.2)$) [] {$C$};
\node at ($(d1)+(-45:0.2)$) [] {$D$};
\draw[directedEdge](v1)--(v2);\draw[directedEdge](v2)--(v3);\draw[directedEdge](v3)--(v4);\draw[directedEdge](v4)--(v1);
\node at ($(v1)!0.5!(v2)+(180:0.2)$) [] {{\footnotesize$\ell_b$}};
\node at ($(v2)!0.5!(v3)+(90:0.25)$) [] {{\footnotesize$\ell_c$}};
\node at ($(v3)!0.5!(v4)+(0:0.25)$) [] {{\footnotesize$\ell_d$}};
\node at ($(v4)!0.5!(v1)+(-90:0.25)$) [] {{\footnotesize$\ell_a$}};
\lsVerts{0}{0}{0}{0}
\node at (v0) [] {$i$};
}}

\newcommand{\twoMassEasyInts}{\tikzBox{\boxVerts\boxEdges\singleLeg{(v1)}{-135}{a}\optLeg{(v2)}{135}{}{}\singleLeg{(v3)}{45}{c}{}\optLeg{(v4)}{-45}{}{}
\node at ($(b1)+(135:0.2)$) [] {$B$};
\node at ($(d1)+(-45:0.2)$) [] {$D$};
\draw[directedEdge](v1)--(v2);\draw[directedEdge](v2)--(v3);\draw[directedEdge](v3)--(v4);\draw[directedEdge](v4)--(v1);
\node at ($(v1)!0.5!(v2)+(180:0.2)$) [] {{\footnotesize$\ell_b$}};
\node at ($(v2)!0.5!(v3)+(90:0.25)$) [] {{\footnotesize$\ell_c$}};
\node at ($(v3)!0.5!(v4)+(0:0.25)$) [] {{\footnotesize$\ell_d$}};
\node at ($(v4)!0.5!(v1)+(-90:0.25)$) [] {{\footnotesize$\ell_a$}};

\lsVerts{0}{0}{0}{0}
\node at (v0) [] {$i$};
}}

\newcommand{\threeMassTriInts}{\tikzBox{\triVerts\triEdges\massiveLeg{(v1)}{-120}{}{}\massiveLeg{(v2)}{120}{}{}\massiveLeg{(v3)}{0}{}{}
\node at ($(a1)+(-120:0.145)$) [] {$A\!$};
\node at ($(b1)+(120:0.125)$) [] {$B\!$};
\node at ($(c1)+(0:0.125)$) [] {$C\!$};
\lsVerts{0}{0}{0}{5}\draw[directedEdge](v1)--(v2);\draw[directedEdge](v2)--(v3);\draw[directedEdge](v3)--(v1);
\node at ($(v1)!0.5!(v2)+(180:0.2)$) [] {{\footnotesize$\ell_b$}};
\node at ($(v2)!0.5!(v3)+(60:0.25)$) [] {{\footnotesize$\ell_c$}};
\node at ($(v3)!0.5!(v1)+(-60:0.25)$) [] {{\footnotesize$\ell_a$}};
\node at (v0) [] {$I$};
}}

\newcommand{\twoMassTriInts}{\tikzBox{\triVerts\triEdges\singleLeg{(v1)}{-120}{a}\massiveLeg{(v2)}{120}{}{}\massiveLeg{(v3)}{0}{}{}
\node at ($(b1)+(120:0.125)$) [] {$B\!$};
\node at ($(c1)+(0:0.125)$) [] {$C\!$};
\lsVerts{0}{0}{0}{5}\draw[directedEdge](v1)--(v2);\draw[directedEdge](v2)--(v3);\draw[directedEdge](v3)--(v1);
\node at ($(v1)!0.5!(v2)+(180:0.2)$) [] {{\footnotesize$\ell_b$}};
\node at ($(v2)!0.5!(v3)+(60:0.25)$) [] {{\footnotesize$\ell_c$}};
\node at ($(v3)!0.5!(v1)+(-60:0.25)$) [] {{\footnotesize$\ell_a$}};
\node at (v0) [] {$I$};
}}

\newcommand{\oneMassTriInts}{\tikzBox{\triVerts\triEdges\singleLeg{(v1)}{-120}{a}\singleLeg{(v2)}{120}{b}\massiveLeg{(v3)}{0}{}{}
\node at ($(c1)+(0:0.125)$) [] {$C\!$};
\lsVerts{0}{0}{0}{5}\draw[directedEdge](v1)--(v2);\draw[directedEdge](v2)--(v3);\draw[directedEdge](v3)--(v1);
\node at ($(v1)!0.5!(v2)+(180:0.2)$) [] {{\footnotesize$\ell_b$}};
\node at ($(v2)!0.5!(v3)+(60:0.25)$) [] {{\footnotesize$\ell_c$}};
\node at ($(v3)!0.5!(v1)+(-60:0.25)$) [] {{\footnotesize$\ell_a$}};
\node at (v0) [] {$I$};
}}

\newcommand{\twoMassBubbleInt}{\tikzBoxBub{\bubVerts\bubEdges\massiveLeg{(v1)}{180}{}{}\massiveLeg{(v2)}{0}{}{}
\node at ($(a1)+(-180:0.125)$) [] {$A\!$};
\node at ($(b1)+(0:0.05)$) [] {$B\!$};
\draw[directedEdge] (e1b)--(e1a);\draw[directedEdge] (e2a)--(e2b);
\node at ($(e1a)!0.5!(e1b)+(-90:0.2)$) [] {{\footnotesize$\ell_a\,\,$}};
\node at ($(e2a)!0.5!(e2b)+(90:0.2)$) [] {{\footnotesize$\,\,\ell_b$}};\lsVerts{0}{0}{5}{5}
}}

\newcommand{\twoMassBubbleIntBoth}{\tikzBoxBub{\bubVerts\bubEdges\optLeg{(v1)}{180}{}{}\optLeg{(v2)}{0}{}{}
\node at ($(a1)+(-180:0.125)$) [] {$A\!$};
\node at ($(b1)+(0:0.05)$) [] {$B\!$};
\draw[directedEdge] (e1b)--(e1a);\draw[directedEdge] (e2a)--(e2b);
\node at ($(e1a)!0.5!(e1b)+(-90:0.2)$) [] {{\footnotesize$\ell_a\,\,$}};
\node at ($(e2a)!0.5!(e2b)+(90:0.2)$) [] {{\footnotesize$\,\,\ell_b$}};\lsVerts{0}{0}{5}{5}
}}

\newcommand{\zeroMassBubbleInt}{\tikzBoxBub{\bubVerts\bubEdges\singleLeg{(v1)}{180}{a}\massiveLeg{(v2)}{0}{}{}
\node at ($(b1)+(0:0.05)$) [] {$B\!$};
\draw[directedEdge] (e1b)--(e1a);\draw[directedEdge] (e2a)--(e2b);
\node at ($(e1a)!0.5!(e1b)+(-90:0.2)$) [] {{\footnotesize$\ell_a\,\,$}};
\node at ($(e2a)!0.5!(e2b)+(90:0.2)$) [] {{\footnotesize$\,\,\ell_b$}};\lsVerts{0}{0}{5}{5}
}}

\newcommand{\mhvSingletLSa}{\tikzBox{\boxVerts\boxEdges\singleLegIn{(v1)}{-135}{i}\optLeg{(v2)}{135}{}{}\singleLegIn{(v3)}{45}{j}\optLeg{(v4)}{-45}{}{}
\lsVerts{1}{2}{1}{2}\draw[directedEdge] (v1)--(v2);\draw[directedEdge] (v1)--(v4);\draw[directedEdge] (v3)--(v2);\draw[directedEdge] (v3)--(v4);}}

\newcommand{\mhvSingletLSb}{\tikzBox{\boxVerts\boxEdges\singleLegIn{(v1)}{-135}{i}\optLegs{(v2)}{135}{}{}\singleLegIn{(v2)}{135}{j}\singleLegOut{(v3)}{45}{c}\optLeg{(v4)}{-45}{}{}
\lsVerts{1}{2}{1}{2}\draw[directedEdge] (v1)--(v2);\draw[directedEdge] (v1)--(v4);\draw[directedEdge] (v2)--(v3);\draw[directedEdge] (v3)--(v4);}}

\newcommand{\mhvSingletLSc}{\tikzBox{\boxVerts\boxEdges\singleLegIn{(v1)}{-135}{i}\optLeg{(v2)}{135}{}{}\optLegs{(v4)}{-45}{}{}\singleLegIn{(v4)}{-45}{j}\singleLegOut{(v3)}{45}{c}
\lsVerts{1}{2}{1}{2}\draw[directedEdge] (v1)--(v2);\draw[directedEdge] (v1)--(v4);\draw[directedEdge] (v3)--(v2);\draw[directedEdge] (v4)--(v3);}}

\newcommand{\mhvNonSingletLSbox}{\tikzBox{\boxVerts\boxNonSinglet\boxEdges\singleLegOut{(v1)}{-135}{a}\optLegs{(v2)}{135}{}{}\optLegs{(v4)}{-45}{}{}\singleLegIn{(v2)}{135}{i}\singleLegIn{(v4)}{-45}{j}\singleLegOut{(v3)}{45}{c}
\lsVerts{1}{2}{1}{2}}}

\newcommand{\mhvNonSingletLSboxA}{\tikzBox{\boxVerts\boxEdges\singleLegOut{(v1)}{-135}{a}\optLegs{(v2)}{135}{}{}\optLegs{(v4)}{-45}{}{}\singleLegIn{(v2)}{135}{i}\singleLegIn{(v4)}{-45}{j}\singleLegOut{(v3)}{45}{c}
\lsVerts{1}{2}{1}{2}\draw[directedEdge] (v1)--(v2);\draw[directedEdge] (v2)--(v3);\draw[directedEdge] (v3)--(v4);\draw[directedEdge] (v4)--(v1);
\node at (v0) [] {{\color{nmhvred}`$+$'}};}}

\newcommand{\mhvNonSingletLSboxB}{\tikzBox{\boxVerts\boxEdges\singleLegOut{(v1)}{-135}{a}\optLegs{(v2)}{135}{}{}\optLegs{(v4)}{-45}{}{}\singleLegIn{(v2)}{135}{i}\singleLegIn{(v4)}{-45}{j}\singleLegOut{(v3)}{45}{c}
\lsVerts{1}{2}{1}{2}\draw[directedEdge] (v2)--(v1);\draw[directedEdge] (v3)--(v2);\draw[directedEdge] (v4)--(v3);\draw[directedEdge] (v1)--(v4);
\node at (v0) [] {{\color{nmhvred}`$-$'}};}}

\newcommand{\mhvNonSingletLStriA}{\tikzBox{\triVerts\triNonSinglet\triEdges\singleLegIn{(v1)}{-120}{i}\singleLegOut{(v2)}{120}{b}\optLegsM{(v3)}{0}{c}{}\singleLegIn{(v3)}{0}{j}
\lsVerts{2}{1}{2}{5}}}

\newcommand{\mhvNonSingletLStriB}{\tikzBox{\triVerts\triNonSinglet\triEdges\singleLegIn{(v2)}{120}{i}\singleLegOut{(v1)}{-120}{a}\optLegsM{(v3)}{0}{c}{}\singleLegIn{(v3)}{0}{j}
\lsVerts{1}{2}{2}{5}}}

\newcommand{\mhvNonSingletLStriC}{\tikzBox{\triVerts\triNonSinglet\triEdges\optLegsM{(v2)}{120}{b}{}\optLegsM{(v3)}{0}{c}{}\singleLegIn{(v2)}{120}{i}\singleLegOut{(v1)}{-120}{a}\singleLegIn{(v3)}{0}{j}\lsVerts{1}{2}{2}{5}}}

\newcommand{\mhvSoftCollinearLS}{\tikzBox{\triVerts\draw[dashed](v1)--(v2);\draw[int](v2)--(v3);\draw[int](v3)--(v1);\singleLeg{(v1)}{-120}{a}\singleLeg{(v2)}{120}{a{+}1}\optLegsM{(v3)}{0}{c}{}
\lsVerts{0}{0}{2}{5}}}

\newcommand{\mhvNonSingletLSbubA}{\tikzBoxBub{\bubVerts\bubNonSinglet\bubEdges\optLegsM{(v1)}{180}{a}{}\optLegsM{(v2)}{0}{b}{}\singleLegIn{(v1)}{180}{i}\singleLegIn{(v2)}{0}{j}\lsVerts{2}{2}{5}{5}}}

\newcommand{\genNonSingletMasslessBubbleA}{\tikzBoxBub{\bubVerts\bubNonSinglet\bubEdges\optLegs{(v2)}{0}{}{}\singleLegIn{(v1)}{180}{a}\lsVerts{2}{4}{5}{5}}}

\newcommand{\genSingletMasslessBubbleA}{\tikzBoxBub{\bubVerts\bubEdges\optLegs{(v2)}{0}{}{}\singleLegIn{(v1)}{180}{a}\draw[directedEdge](e1a)--(e1b);\draw[directedEdge](e2a)--(e2b);\lsVerts{1}{4}{5}{5}}}

\newcommand{\genNonSingletMasslessBubbleB}{\tikzBoxBub{\bubVerts\bubNonSinglet\bubEdges\optLegs{(v2)}{0}{}{}\singleLegOut{(v1)}{180}{a}\lsVerts{1}{4}{5}{5}}}

\newcommand{\genSingletMasslessBubbleB}{\tikzBoxBub{\bubVerts\bubEdges\optLegs{(v2)}{0}{}{}\singleLegOut{(v1)}{180}{a}\draw[directedEdge](e1b)--(e1a);\draw[directedEdge](e2b)--(e2a);\lsVerts{2}{4}{5}{5}}}

\newcommand{\generalRuleForMasslessBubbles}{\tikzBoxBub{\bubVerts\bubEdges\optLegsM{(v2)}{0}{b}{}\singleLeg{(v1)}{180}{a}\node at (v1) [compositeDot] {};\lsVerts{0}{4}{5}{5}}\equivR\tikzBoxBub{\bubVerts\bubEdges\coordinate (v3) at ($(v2)+(0:20pt)$);\optLegsM{(v3)}{0}{b}{}\singleLeg{(v1)}{180}{a}\node at (v1) [compositeDot] {};\node at (v2) [compositeDot] {};\draw[int] (v2)--(v3);\lsVerts{0}{0}{4}{5}}\hspace{26pt}\bigger{\Rightarrow}\hspace{-10pt}\tikzBoxBub{\optLegsM{(v0)}{0}{b}{}\singleLeg{(v0)}{180}{a}\node at (v0) [fullamp] {};
}}

\newcommand{\sixPointNonSingletTriA}{\tikzBox{\triVerts\triNonSinglet\triEdges\singleLegOut{(v1)}{-120}{3}\singleLegIn{(v2)}{120}{4}\singleLegIn{(v3)}{0+1.8*\legSpread}{5}\singleLegIn{(v3)}{0+0.6*\legSpread}{6}\singleLegOut{(v3)}{0-0.6*\legSpread}{1}\singleLegOut{(v3)}{0-1.8*\legSpread}{2}
\lsVerts{1}{2}{3}{5}}}

\newcommand{\sixPointNonSingletTriB}{\tikzBox{\triVerts\triNonSinglet\triEdges\singleLegIn{(v1)}{-120}{6}\singleLegOut{(v2)}{120}{1}\singleLegOut{(v3)}{0+1.8*\legSpread}{2}\singleLegOut{(v3)}{0+0.6*\legSpread}{3}\singleLegIn{(v3)}{0-0.6*\legSpread}{4}\singleLegIn{(v3)}{0-1.8*\legSpread}{5}
\lsVerts{2}{1}{3}{5}}}

\newcommand{\sixPointNonSingletTriC}{\tikzBox{\triVerts\triNonSinglet\triEdges\singleLegOut{(v1)}{-120}{2}\singleLegOut{(v2)}{120+0.6*\legSpread}{3}\singleLegIn{(v2)}{120-0.6*\legSpread}{4}\singleLegIn{(v3)}{0+1.2*\legSpread}{5}\singleLegIn{(v3)}{0-0*\legSpread}{6}\singleLegIn{(v3)}{0-1.2*\legSpread}{1}
\lsVerts{1}{2}{3}{5}}}

\newcommand{\sixPointNonSingletTriD}{\tikzBox{\triVerts\triNonSinglet\triEdges\singleLegOut{(v1)}{-120}{2}\singleLegOut{(v2)}{120+1.2*\legSpread}{3}\singleLegIn{(v2)}{120-0*\legSpread}{4}\singleLegIn{(v2)}{120-1.2*\legSpread}{5}\singleLegIn{(v3)}{0+0.6*\legSpread}{6}\singleLegIn{(v3)}{0-0.6*\legSpread}{1}
\lsVerts{1}{3}{2}{5}}}

\newcommand{\sixPointNonSingletTriE}{\tikzBox{\triVerts\triNonSinglet\triEdges\singleLegIn{(v1)}{-120}{5}\singleLegIn{(v2)}{120+0.6*\legSpread}{6}\singleLegOut{(v2)}{120-0.6*\legSpread}{1}\singleLegOut{(v3)}{0+1.2*\legSpread}{2}\singleLegOut{(v3)}{0-0*\legSpread}{3}\singleLegIn{(v3)}{0-1.2*\legSpread}{4}
\lsVerts{2}{2}{2}{5}}}

\newcommand{\sixPointNonSingletTriF}{\tikzBox{\triVerts\triNonSinglet\triEdges\singleLegIn{(v1)}{-120}{5}\singleLegIn{(v2)}{120+1.2*\legSpread}{6}\singleLegOut{(v2)}{120-0*\legSpread}{1}\singleLegOut{(v2)}{120-1.2*\legSpread}{2}\singleLegOut{(v3)}{0+0.6*\legSpread}{3}\singleLegIn{(v3)}{0-0.6*\legSpread}{4}
\lsVerts{2}{2}{2}{5}}}

\newcommand{\sixPointNonSingletBubA}{\tikzBoxBub{\bubVerts\bubNonSinglet\bubEdges\singleLegOut{(v1)}{180+1.2*\legSpread}{2}\singleLegOut{(v1)}{180}{3}\singleLegIn{(v1)}{180-1.2*\legSpread}{4}\singleLegIn{(v2)}{0+1.2*\legSpread}{5}\singleLegIn{(v2)}{0}{6}\singleLegOut{(v2)}{0-1.2*\legSpread}{1}\lsVerts{2}{3}{5}{5}}}

\newcommand{\sixPointNonSingletBubB}{\tikzBoxBub{\bubVerts\bubNonSinglet\bubEdges\singleLegOut{(v1)}{180+1.2*\legSpread}{3}\singleLegIn{(v1)}{180}{4}\singleLegIn{(v1)}{180-1.2*\legSpread}{5}\singleLegIn{(v2)}{0+1.2*\legSpread}{6}\singleLegOut{(v2)}{0}{1}\singleLegOut{(v2)}{0-1.2*\legSpread}{2}\lsVerts{3}{2}{5}{5}}}

\newcommand{\sixPointNonSingletBubC}{\tikzBoxBub{\bubVerts\bubNonSinglet\bubEdges\singleLegOut{(v1)}{180+0.6*\legSpread}{3}\singleLegIn{(v1)}{180-0.6*\legSpread}{4}\singleLegIn{(v2)}{0+1.8*\legSpread}{5}\singleLegIn{(v2)}{0+0.6*\legSpread}{6}\singleLegOut{(v2)}{-0.6*\legSpread}{1}\singleLegOut{(v2)}{0-1.8*\legSpread}{2}\lsVerts{2}{3}{5}{5}}}

\newcommand{\sixPointNonSingletBubD}{\tikzBoxBub{\bubVerts\bubNonSinglet\bubEdges\singleLegOut{(v1)}{180+1.8*\legSpread}{2}\singleLegOut{(v1)}{180+0.6*\legSpread}{3}\singleLegIn{(v1)}{180-0.6*\legSpread}{4}\singleLegIn{(v1)}{180-1.8*\legSpread}{5}\singleLegIn{(v2)}{0.6*\legSpread}{6}\singleLegOut{(v2)}{0-0.6*\legSpread}{1}\lsVerts{3}{2}{5}{5}}}

\newcommand{\fourMassContourA}{\tikzBox{\boxVerts\boxEdges\massiveLeg{(v1)}{-135}{}{}\massiveLeg{(v2)}{135}{}{}\massiveLeg{(v3)}{45}{}{}\massiveLeg{(v4)}{-45}{}{}
\node at ($(a1)+(-135:0.2)$) [] {$A$};
\node at ($(b1)+(135:0.2)$) [] {$B$};
\node at ($(c1)+(45:0.2)$) [] {$C$};
\node at ($(d1)+(-45:0.2)$) [] {$D$};
\node at ($(v1)!0.5!(v2)$) [] {\includegraphics[angle=-90]{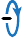}};
\node at ($(v2)!0.5!(v3)$) [] {\includegraphics[angle=180]{cut_1}};
\node at ($(v3)!0.5!(v4)$) [] {\includegraphics[angle=90]{cut_1}};
\node at ($(v4)!0.5!(v1)$) [] {\includegraphics[angle=0]{cut_1}};
\node at ($(v1)!0.5!(v2)+(180:0.2)$) [] {{\footnotesize$$}};
\node at ($(v2)!0.5!(v3)+(90:0.25)$) [] {{\footnotesize$$}};
\node at ($(v3)!0.5!(v4)+(0:0.25)$) [] {{\footnotesize$$}};
\node at ($(v4)!0.5!(v1)+(-90:0.4)$) [] {{\footnotesize$\,\,\ell\!\hspace{-0.5pt}=\!\hspace{-0.5pt}{\color{cut1}\ell^*_1\hspace{0pt}}$}};
\lsVerts{0}{0}{0}{0}
\node at (v0) [] {${\color{cut1}1}$};
}}
\newcommand{\fourMassContourB}{\tikzBox{\boxVerts\boxEdges\massiveLeg{(v1)}{-135}{}{}\massiveLeg{(v2)}{135}{}{}\massiveLeg{(v3)}{45}{}{}\massiveLeg{(v4)}{-45}{}{}
\node at ($(a1)+(-135:0.2)$) [] {$A$};
\node at ($(b1)+(135:0.2)$) [] {$B$};
\node at ($(c1)+(45:0.2)$) [] {$C$};
\node at ($(d1)+(-45:0.2)$) [] {$D$};
\node at ($(v1)!0.5!(v2)$) [] {\includegraphics[angle=90]{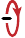}};
\node at ($(v2)!0.5!(v3)$) [] {\includegraphics[angle=0]{cut_2}};
\node at ($(v3)!0.5!(v4)$) [] {\includegraphics[angle=-90]{cut_2}};
\node at ($(v4)!0.5!(v1)$) [] {\includegraphics[angle=180]{cut_2}};
\node at ($(v1)!0.5!(v2)+(180:0.2)$) [] {{\footnotesize$$}};
\node at ($(v2)!0.5!(v3)+(90:0.25)$) [] {{\footnotesize$$}};
\node at ($(v3)!0.5!(v4)+(0:0.25)$) [] {{\footnotesize$$}};
\node at ($(v4)!0.5!(v1)+(-90:0.4)$) [] {{\footnotesize$\,\,\ell\!\hspace{-0.5pt}=\!\hspace{-0.5pt}{\color{cut2}\ell^*_2\hspace{0pt}}$}};
\lsVerts{0}{0}{0}{0}
\node at (v0) [] {${\color{cut2}2}$};
}}

\newcommand{\threeMassContourA}{\tikzBox{\boxVerts\boxEdges\singleLeg{(v1)}{-135}{a}\massiveLeg{(v2)}{135}{}{}\massiveLeg{(v3)}{45}{}{}\massiveLeg{(v4)}{-45}{}{}
\node at ($(b1)+(135:0.2)$) [] {$B$};
\node at ($(c1)+(45:0.2)$) [] {$C$};
\node at ($(d1)+(-45:0.2)$) [] {$D$};
\node at ($(v1)!0.5!(v2)$) [] {\includegraphics[angle=-90]{cut_1}};
\node at ($(v2)!0.5!(v3)$) [] {\includegraphics[angle=180]{cut_1}};
\node at ($(v3)!0.5!(v4)$) [] {\includegraphics[angle=90]{cut_1}};
\node at ($(v4)!0.5!(v1)$) [] {\includegraphics[angle=0]{cut_1}};
\node at ($(v1)!0.5!(v2)+(180:0.2)$) [] {{\footnotesize$$}};
\node at ($(v2)!0.5!(v3)+(90:0.25)$) [] {{\footnotesize$$}};
\node at ($(v3)!0.5!(v4)+(0:0.25)$) [] {{\footnotesize$$}};
\node at ($(v4)!0.5!(v1)+(-90:0.4)$) [] {{\footnotesize$\,\,\ell\!\hspace{-0.5pt}=\!\hspace{-0.5pt}{\color{cut1}\ell^*_1\hspace{0pt}}$}};
\node at (v1) [whiteDot] {};
\node at (v2) [ddot] {};\node at (v3) [ddot] {};\node at (v4) [ddot] {};
\node at (v0) [] {${\color{cut1}1}$};
}}
\newcommand{\threeMassContourB}{\tikzBox{\boxVerts\boxEdges\singleLeg{(v1)}{-135}{a}\massiveLeg{(v2)}{135}{}{}\massiveLeg{(v3)}{45}{}{}\massiveLeg{(v4)}{-45}{}{}
\node at ($(b1)+(135:0.2)$) [] {$B$};
\node at ($(c1)+(45:0.2)$) [] {$C$};
\node at ($(d1)+(-45:0.2)$) [] {$D$};
\node at ($(v1)!0.5!(v2)$) [] {\includegraphics[angle=90]{cut_2}};
\node at ($(v2)!0.5!(v3)$) [] {\includegraphics[angle=0]{cut_2}};
\node at ($(v3)!0.5!(v4)$) [] {\includegraphics[angle=-90]{cut_2}};
\node at ($(v4)!0.5!(v1)$) [] {\includegraphics[angle=180]{cut_2}};
\node at ($(v1)!0.5!(v2)+(180:0.2)$) [] {{\footnotesize$$}};
\node at ($(v2)!0.5!(v3)+(90:0.25)$) [] {{\footnotesize$$}};
\node at ($(v3)!0.5!(v4)+(0:0.25)$) [] {{\footnotesize$$}};
\node at ($(v4)!0.5!(v1)+(-90:0.4)$) [] {{\footnotesize$\,\,\ell\!\hspace{-0.5pt}=\!\hspace{-0.5pt}{\color{cut2}\ell^*_2\hspace{0pt}}$}};
\node at (v1) [blueDot] {};
\node at (v2) [ddot] {};\node at (v3) [ddot] {};\node at (v4) [ddot] {};
\node at (v0) [] {${\color{cut2}2}$};
}}

\newcommand{\twoMassHardContourA}{\tikzBox{\boxVerts\boxEdges\singleLeg{(v1)}{-135}{a}\singleLeg{(v2)}{135}{b}\massiveLeg{(v3)}{45}{}{}\massiveLeg{(v4)}{-45}{}{}
\node at ($(c1)+(45:0.2)$) [] {$C$};
\node at ($(d1)+(-45:0.2)$) [] {$D$};
\node at ($(v1)!0.5!(v2)$) [] {\includegraphics[angle=-90]{cut_1}};
\node at ($(v2)!0.5!(v3)$) [] {\includegraphics[angle=180]{cut_1}};
\node at ($(v3)!0.5!(v4)$) [] {\includegraphics[angle=90]{cut_1}};
\node at ($(v4)!0.5!(v1)$) [] {\includegraphics[angle=0]{cut_1}};
\node at ($(v1)!0.5!(v2)+(180:0.2)$) [] {{\footnotesize$$}};
\node at ($(v2)!0.5!(v3)+(90:0.25)$) [] {{\footnotesize$$}};
\node at ($(v3)!0.5!(v4)+(0:0.25)$) [] {{\footnotesize$$}};
\node at ($(v4)!0.5!(v1)+(-90:0.4)$) [] {{\footnotesize$\,\,\ell\!\hspace{-0.5pt}=\!\hspace{-0.5pt}{\color{cut1}\ell^*_1\hspace{0pt}}$}};
\node at (v1) [whiteDot] {};
\node at (v2) [blueDot] {};\node at (v3) [ddot] {};\node at (v4) [ddot] {};
\node at (v0) [] {${\color{cut1}1}$};
}}
\newcommand{\twoMassHardContourB}{\tikzBox{\boxVerts\boxEdges\singleLeg{(v1)}{-135}{a}\singleLeg{(v2)}{135}{b}\massiveLeg{(v3)}{45}{}{}\massiveLeg{(v4)}{-45}{}{}
\node at ($(c1)+(45:0.2)$) [] {$C$};
\node at ($(d1)+(-45:0.2)$) [] {$D$};
\node at ($(v1)!0.5!(v2)$) [] {\includegraphics[angle=90]{cut_2}};
\node at ($(v2)!0.5!(v3)$) [] {\includegraphics[angle=0]{cut_2}};
\node at ($(v3)!0.5!(v4)$) [] {\includegraphics[angle=-90]{cut_2}};
\node at ($(v4)!0.5!(v1)$) [] {\includegraphics[angle=180]{cut_2}};
\node at ($(v1)!0.5!(v2)+(180:0.2)$) [] {{\footnotesize$$}};
\node at ($(v2)!0.5!(v3)+(90:0.25)$) [] {{\footnotesize$$}};
\node at ($(v3)!0.5!(v4)+(0:0.25)$) [] {{\footnotesize$$}};
\node at ($(v4)!0.5!(v1)+(-90:0.4)$) [] {{\footnotesize$\,\,\ell\!\hspace{-0.5pt}=\!\hspace{-0.5pt}{\color{cut2}\ell^*_2\hspace{0pt}}$}};
\node at (v1) [blueDot] {};
\node at (v2) [whiteDot] {};\node at (v3) [ddot] {};\node at (v4) [ddot] {};
\node at (v0) [] {${\color{cut2}2}$};
}}
\newcommand{\twoMassEasyContourA}{\tikzBox{\boxVerts\boxEdges\singleLeg{(v1)}{-135}{a}\massiveLeg{(v2)}{135}{}{}\singleLeg{(v3)}{45}{c}\optLeg{(v4)}{-45}{}{}
\node at ($(b1)+(135:0.2)$) [] {$B$};
\node at ($(d1)+(-45:0.2)$) [] {$D$};
\node at ($(v1)!0.5!(v2)$) [] {\includegraphics[angle=-90]{cut_1}};
\node at ($(v2)!0.5!(v3)$) [] {\includegraphics[angle=180]{cut_1}};
\node at ($(v3)!0.5!(v4)$) [] {\includegraphics[angle=90]{cut_1}};
\node at ($(v4)!0.5!(v1)$) [] {\includegraphics[angle=0]{cut_1}};
\node at ($(v1)!0.5!(v2)+(180:0.2)$) [] {{\footnotesize$$}};
\node at ($(v2)!0.5!(v3)+(90:0.25)$) [] {{\footnotesize$$}};
\node at ($(v3)!0.5!(v4)+(0:0.25)$) [] {{\footnotesize$$}};
\node at ($(v4)!0.5!(v1)+(-90:0.4)$) [] {{\footnotesize$\,\,\ell\!\hspace{-0.5pt}=\!\hspace{-0.5pt}{\color{cut1}\ell^*_1\hspace{0pt}}$}};
\node at (v1) [whiteDot] {};
\node at (v2) [ddot] {};\node at (v3) [whiteDot] {};\node at (v4) [ddot] {};
\node at (v0) [] {${\color{cut1}1}$};
}}
\newcommand{\twoMassEasyContourB}{\tikzBox{\boxVerts\boxEdges\singleLeg{(v1)}{-135}{a}\massiveLeg{(v2)}{135}{}{}\singleLeg{(v3)}{45}{c}\optLeg{(v4)}{-45}{}{}
\node at ($(b1)+(135:0.2)$) [] {$B$};
\node at ($(d1)+(-45:0.2)$) [] {$D$};
\node at ($(v1)!0.5!(v2)$) [] {\includegraphics[angle=90]{cut_2}};
\node at ($(v2)!0.5!(v3)$) [] {\includegraphics[angle=0]{cut_2}};
\node at ($(v3)!0.5!(v4)$) [] {\includegraphics[angle=-90]{cut_2}};
\node at ($(v4)!0.5!(v1)$) [] {\includegraphics[angle=180]{cut_2}};
\node at ($(v1)!0.5!(v2)+(180:0.2)$) [] {{\footnotesize$$}};
\node at ($(v2)!0.5!(v3)+(90:0.25)$) [] {{\footnotesize$$}};
\node at ($(v3)!0.5!(v4)+(0:0.25)$) [] {{\footnotesize$$}};
\node at ($(v4)!0.5!(v1)+(-90:0.4)$) [] {{\footnotesize$\,\,\ell\!\hspace{-0.5pt}=\!\hspace{-0.5pt}{\color{cut2}\ell^*_2\hspace{0pt}}$}};
\node at (v1) [blueDot] {};
\node at (v2) [ddot] {};\node at (v3) [blueDot] {};\node at (v4) [ddot] {};
\node at (v0) [] {${\color{cut2}2}$};
}}

\newcommand{\triangleCutsA}{\node at ($(v1)!0.5!(v2)$) [] {\includegraphics[scale=1,angle=90]{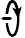}};
\node at ($(v2)!0.5!(v3)$) [] {\includegraphics[scale=1,angle=-210]{cut_3}};
\node at ($(v3)!0.5!(v1)$) [] {\includegraphics[scale=1,angle=210]{cut_3}};}
\newcommand{\triangleCutsB}{\node at ($(v1)!0.5!(v2)$) [] {\includegraphics[scale=1,angle=90]{cut_1}};
\node at ($(v2)!0.5!(v3)$) [] {\includegraphics[scale=1,angle=-210]{cut_1}};
\node at ($(v3)!0.5!(v1)$) [] {\includegraphics[scale=1,angle=210]{cut_1}};}
\newcommand{\triangleCutsC}{\node at ($(v1)!0.5!(v2)$) [] {\includegraphics[scale=1,angle=-90]{cut_2}};
\node at ($(v2)!0.5!(v3)$) [] {\includegraphics[scale=1,angle=-30]{cut_2}};
\node at ($(v3)!0.5!(v1)$) [] {\includegraphics[scale=1,angle=30]{cut_2}};}

\newcommand{\threeMassTriContourA}{\tikzBox{\triVerts\triEdges\triangleCutsA\massiveLeg{(v1)}{-120}{}{}\massiveLeg{(v2)}{120}{}{}\massiveLeg{(v3)}{0}{}{}
\node at ($(a1)+(-120:0.145)$) [] {$A\!$};
\node at ($(b1)+(120:0.125)$) [] {$B\!$};
\node at ($(c1)+(0:0.125)$) [] {$C\!$};
\lsVerts{0}{0}{0}{5}
\node at (v0) [] {{\scriptsize${\color{black}1\,}$}};
\node at ($(v0)-(90:0.65)+(0.45,0)$) [] {{\scriptsize${\color{black}\ell^*\!\!\to\!\infty}$}};
\node at ($(v0)-(90:0.95)+(0.45,0)$) [] {{\scriptsize${\color{black}\text{(odd)}}$}};
}}

\newcommand{\threeMassTriContourB}{\tikzBox{\triVerts\triEdges\triangleCutsA\massiveLeg{(v1)}{-120}{}{}\massiveLeg{(v2)}{120}{}{}\massiveLeg{(v3)}{0}{}{}
\node at ($(a1)+(-120:0.145)$) [] {$A\!$};
\node at ($(b1)+(120:0.125)$) [] {$B\!$};
\node at ($(c1)+(0:0.125)$) [] {$C\!$};
\lsVerts{0}{0}{0}{5}
\node at (v0) [] {{\scriptsize${\color{black}2\,}$}};
\node at ($(v0)-(90:0.65)+(0.65,0)$) [] {{\scriptsize${\color{black}\ell^*_1\!\!\to\!\infty}$}};
\node at ($(v0)-(90:0.95)+(0.65,0)$) [] {{\scriptsize${\color{black}\text{(double-pole)}}$}};
\node at ($(v0)-(90:1.25)+(0.65,0)$) [] {{\scriptsize${\color{black}\text{(odd)}}$}};
}}

\newcommand{\threeMassTriContourC}{\tikzBox{\triVerts\triEdges\triangleCutsA\massiveLeg{(v1)}{-120}{}{}\massiveLeg{(v2)}{120}{}{}\massiveLeg{(v3)}{0}{}{}
\node at ($(a1)+(-120:0.145)$) [] {$A\!$};
\node at ($(b1)+(120:0.125)$) [] {$B\!$};
\node at ($(c1)+(0:0.125)$) [] {$C\!$};
\lsVerts{0}{0}{0}{5}
\node at (v0) [] {{\scriptsize${\color{black}3\,}$}};
\node at ($(v0)-(90:0.65)+(0.65,0)$) [] {{\scriptsize${\color{black}\ell^*_2\!\!\to\!\infty}$}};
\node at ($(v0)-(90:0.95)+(0.65,0)$) [] {{\scriptsize${\color{black}\text{(double-pole)}}$}};
\node at ($(v0)-(90:1.25)+(0.65,0)$) [] {{\scriptsize${\color{black}\text{(even)}}$}};
}}

\newcommand{\twoMassTriContourA}{\tikzBox{\triVerts\triEdges\triangleCutsA\singleLeg{(v1)}{-120}{a}\massiveLeg{(v2)}{120}{}{}\massiveLeg{(v3)}{0}{}{}
\node at ($(b1)+(120:0.125)$) [] {$B\!$};
\node at ($(c1)+(0:0.125)$) [] {$C\!$};
\lsVerts{0}{0}{0}{5}\node at (v1) [compositeDot] {};
\node at (v0) [] {{\scriptsize${\color{black}1\,}$}};\triangleCutsA
}}

\newcommand{\twoMassTriContourB}{\tikzBox{\triVerts\triEdges\triangleCutsB\singleLeg{(v1)}{-120}{a}\massiveLeg{(v2)}{120}{}{}\massiveLeg{(v3)}{0}{}{}
\node at ($(b1)+(120:0.125)$) [] {$B\!$};
\node at ($(c1)+(0:0.125)$) [] {$C\!$};
\node at (v1) [whiteDot] {};\node at (v2) [ddot] {};\node at (v3) [ddot] {};
\node at (v0) [] {{\scriptsize${\color{cut1}2\,}$}};
\node at ($(v0)-(90:0.65)+(0.45,0)$) [] {{\scriptsize${\color{cut1}\ell^*_1\!\!\to\!\infty}$}};
\node at ($(v0)-(90:0.95)+(0.45,0)$) [] {{\scriptsize${\color{cut1}\text{(double-pole)}}$}};
}}

\newcommand{\twoMassTriContourC}{\tikzBox{\triVerts\triEdges\triangleCutsC\singleLeg{(v1)}{-120}{a}\massiveLeg{(v2)}{120}{}{}\massiveLeg{(v3)}{0}{}{}
\node at ($(b1)+(120:0.125)$) [] {$B\!$};
\node at ($(c1)+(0:0.125)$) [] {$C\!$};
\node at (v1) [blueDot] {};\node at (v2) [ddot] {};\node at (v3) [ddot] {};
\node at (v0) [] {{\scriptsize${\color{cut2}3\,}$}};
\node at ($(v0)-(90:0.65)+(0.45,0)$) [] {{\scriptsize${\color{cut2}\ell^*_2\!\!\to\!\infty}$}};
\node at ($(v0)-(90:0.95)+(0.45,0)$) [] {{\scriptsize${\color{cut2}\text{(double-pole)}}$}};
}}

\newcommand{\oneMassTriContourA}{\tikzBox{\triVerts\draw[dashed](v1)--(v2);\draw[int](v2)--(v3);\draw[int](v3)--(v1);\triangleCutsA\singleLeg{(v1)}{-120}{a}\singleLeg{(v2)}{120}{b}\massiveLeg{(v3)}{0}{}{}
\node at ($(c1)+(0:0.125)$) [] {$C\!$};
\lsVerts{0}{0}{0}{5}
\node at (v0) [] {{\scriptsize${\color{black}1\,}$}};
}}
\newcommand{\oneMassTriContourB}{\tikzBox{\triVerts\triEdges\triangleCutsB\singleLeg{(v1)}{-120}{a}\singleLeg{(v2)}{120}{b}\massiveLeg{(v3)}{0}{}{}
\node at ($(c1)+(0:0.125)$) [] {$C\!$};
\node at (v0) [] {{\scriptsize${\color{cut1}2\,}$}};
\node at (v1) [whiteDot] {};\node at (v2) [blueDot] {};\node at (v3) [ddot] {};
\node at ($(v0)-(90:0.65)+(0.45,0)$) [] {{\scriptsize${\color{cut1}\ell^*_1\!\!\to\!\infty}$}};
\node at ($(v0)-(90:0.95)+(0.45,0)$) [] {{\scriptsize${\color{cut1}\text{(double-pole)}}$}};
}}
\newcommand{\oneMassTriContourC}{\tikzBox{\triVerts\triEdges\triangleCutsC\singleLeg{(v1)}{-120}{a}\singleLeg{(v2)}{120}{b}\massiveLeg{(v3)}{0}{}{}
\node at ($(c1)+(0:0.125)$) [] {$C\!$};
\node at (v0) [] {{\scriptsize${\color{cut2}3\,}$}};
\node at (v1) [blueDot] {};\node at (v2) [whiteDot] {};\node at (v3) [ddot] {};
\node at ($(v0)-(90:0.65)+(0.45,0)$) [] {{\scriptsize${\color{cut2}\ell^*_2\!\!\to\!\infty}$}};
\node at ($(v0)-(90:0.95)+(0.45,0)$) [] {{\scriptsize${\color{cut2}\text{(double-pole)}}$}};
}}

\newcommand{\twoMassBubbleContour}{\tikzBoxBub{\bubVerts\bubEdges\massiveLeg{(v1)}{180}{}{}\massiveLeg{(v2)}{0}{}{}
\node at ($(a1)+(-180:0.125)$) [] {$A\!$};
\node at ($(b1)+(0:0.05)$) [] {$B\!$};\lsVerts{0}{0}{5}{5}
\node at ($(e1a)!0.44!(e1b)$) [] {\includegraphics[scale=1,angle=180]{cut_3}};
\node at ($(e2a)!0.56!(e2b)$) [] {\includegraphics[scale=1,angle=0]{cut_3}};
\node at ($(v0)-(90:0.65)+(0,0)$) [] {{\scriptsize${\color{black}\ell^*\!\!\to\!\infty}$}};
\node at ($(v0)-(90:0.95)+(0.0,0)$) [] {{\scriptsize${\color{black}\text{(single-pole$\Rightarrow$double-pole)}}$}};
}}

\newcommand{\zeroMassBubbleContour}{\tikzBoxBub{\bubVerts\bubEdges\singleLeg{(v1)}{180}{a}\massiveLeg{(v2)}{0}{}{}\node at ($(b1)+(0:0.05)$) [] {$B\!$};
\node at ($(b1)+(0:0.05)$) [] {$B\!$};
\node at ($(e1a)!0.44!(e1b)$) [] {\includegraphics[scale=1,angle=180]{cut_3}};
\node at ($(e2a)!0.56!(e2b)$) [] {\includegraphics[scale=1,angle=0]{cut_3}};
\node at (v1) [compositeDot] {};\lsVerts{0}{0}{5}{5}
\node at ($(v0)-(90:0.65)+(0,0)$) [] {{\scriptsize${\color{black}\ell^*\!\!\to\!\infty}$}};
\node at ($(v0)-(90:0.95)+(0.0,0)$) [] {{\scriptsize${\color{black}\text{(double-pole)}}$}};
}}

\newcommand{\fourPointNonSingletEg}{\tikzBox{\boxVerts\boxNonSinglet\boxEdges\singleLegOut{(v1)}{-135}{1}\singleLegIn{(v2)}{135}{2}\singleLegOut{(v3)}{45}{3}\singleLegIn{(v4)}{-45}{4}
\lsVerts{1}{2}{1}{2}}}

\newcommand{\fourPointNonSingletPartA}{\tikzBox{\boxVerts\boxEdges\singleLegOut{(v1)}{-135}{1}\singleLegIn{(v2)}{135}{2}\singleLegOut{(v3)}{45}{3}\singleLegIn{(v4)}{-45}{4}
\lsVerts{1}{2}{1}{2}\draw[directedEdge] (v1)--(v2);\draw[directedEdge] (v2)--(v3);\draw[directedEdge] (v3)--(v4);\draw[directedEdge] (v4)--(v1);}}
\newcommand{\fourPointNonSingletPartB}{\tikzBox{\boxVerts\boxEdges\singleLegOut{(v1)}{-135}{1}\singleLegIn{(v2)}{135}{2}\singleLegOut{(v3)}{45}{3}\singleLegIn{(v4)}{-45}{4}
\lsVerts{1}{2}{1}{2}\draw[directedEdge] (v2)--(v1);\draw[directedEdge] (v3)--(v2);\draw[directedEdge] (v4)--(v3);\draw[directedEdge] (v1)--(v4);}}

\newcommand{\threePointMHVbar}{
\scalebox{1.4}{\tikzset{ext/.style={black,line width=0.7*\lineThickness,line cap=round}}\tikzset{directedEdge/.style={draw=none,decoration={markings,mark connection node=connode,mark=at position 0.5 with {\node[transform shape, scale=0.146*\figScale,shape=dart,aspect=0.5,fill=black,draw] (connode) {};}},postaction={decorate}}}
\fwbox{40pt}{\tikzBoxBub{\singleLegIn{(0,0)}{180}{\text{{\scriptsize$i$}}}\singleLegOut{(0,0)}{60}{\text{{\scriptsize$a$}}}\singleLegOut{(0,0)}{-60}{\text{{\scriptsize$b$}}}\node at (0,0) [fullmhvBar] {};}}\tikzset{ext/.style={black,line width=\lineThickness,line cap=round}\tikzset{directedEdge/.style={draw=none,decoration={markings,mark connection node=connode,mark=at position 0.5 with {\node[transform shape, scale=0.205*\figScale,shape=dart,aspect=0.5,fill=black,draw] (connode) {};}},postaction={decorate}}}}}}
\newcommand{\threePointMHV}{
\scalebox{1.4}{\tikzset{ext/.style={black,line width=0.7*\lineThickness,line cap=round}}\tikzset{directedEdge/.style={draw=none,decoration={markings,mark connection node=connode,mark=at position 0.5 with {\node[transform shape, scale=0.146*\figScale,shape=dart,aspect=0.5,fill=black,draw] (connode) {};}},postaction={decorate}}}
\fwbox{40pt}{\tikzBoxBub{\singleLegOut{(0,0)}{0}{\text{{\scriptsize$a$}}}\singleLegIn{(0,0)}{120}{\text{{\scriptsize$j$}}}\singleLegIn{(0,0)}{-120}{\text{{\scriptsize$i$}}}\node at (0,0) [fullmhv] {};}}\tikzset{ext/.style={black,line width=\lineThickness,line cap=round}\tikzset{directedEdge/.style={draw=none,decoration={markings,mark connection node=connode,mark=at position 0.5 with {\node[transform shape, scale=0.205*\figScale,shape=dart,aspect=0.5,fill=black,draw] (connode) {};}},postaction={decorate}}}}}}

\newcommand{\threePointMHVbarBare}{
\scalebox{1.4}{\tikzset{ext/.style={black,line width=0.7*\lineThickness,line cap=round}}\tikzset{directedEdge/.style={draw=none,decoration={markings,mark connection node=connode,mark=at position 0.5 with {\node[transform shape, scale=0.146*\figScale,shape=dart,aspect=0.5,fill=black,draw] (connode) {};}},postaction={decorate}}}
\fwbox{40pt}{\tikzBoxBub{\singleLegIn{(0,0)}{180}{}\singleLegOut{(0,0)}{60}{\text{{\scriptsize$$}}}\singleLegOut{(0,0)}{-60}{\text{{\scriptsize$$}}}\node at (0,0) [fullmhvBar] {};}}\tikzset{ext/.style={black,line width=\lineThickness,line cap=round}\tikzset{directedEdge/.style={draw=none,decoration={markings,mark connection node=connode,mark=at position 0.5 with {\node[transform shape, scale=0.205*\figScale,shape=dart,aspect=0.5,fill=black,draw] (connode) {};}},postaction={decorate}}}}}}
\newcommand{\threePointMHVBare}{
\scalebox{1.4}{\tikzset{ext/.style={black,line width=0.7*\lineThickness,line cap=round}}\tikzset{directedEdge/.style={draw=none,decoration={markings,mark connection node=connode,mark=at position 0.5 with {\node[transform shape, scale=0.146*\figScale,shape=dart,aspect=0.5,fill=black,draw] (connode) {};}},postaction={decorate}}}
\fwbox{40pt}{\tikzBoxBub{\singleLegOut{(0,0)}{180}{\text{{\scriptsize$$}}}\singleLegIn{(0,0)}{60}{\text{{\scriptsize$$}}}\singleLegIn{(0,0)}{-60}{\text{{\scriptsize$$}}}\node at (0,0) [fullmhv] {};}}\tikzset{ext/.style={black,line width=\lineThickness,line cap=round}\tikzset{directedEdge/.style={draw=none,decoration={markings,mark connection node=connode,mark=at position 0.5 with {\node[transform shape, scale=0.205*\figScale,shape=dart,aspect=0.5,fill=black,draw] (connode) {};}},postaction={decorate}}}}}}

\let\olditemize\itemize\renewcommand{\itemize}{\vspace{-2pt}\olditemize\setlength{\itemsep}{1pt}\setlength{\parskip}{0pt}\setlength{\parsep}{-0pt}}
\let\oldenumerate\enumerate\renewcommand{\enumerate}{\vspace{-4pt}\oldenumerate\setlength{\itemsep}{1pt}\setlength{\parskip}{0pt}\setlength{\parsep}{0pt}}
\makeatletter\renewcommand\section{\addtocontents{toc}{\protect\addvspace{-2.25\p@}}\@startsection {section}{1}{\z@}{-0.0ex \@plus .2ex \@minus 0.2ex}{1ex \@plus.1ex\@minus .5ex}{\normalfont\large\bfseries}}
\renewcommand\subsection{\addtocontents{toc}{\protect\addvspace{-2.5\p@}}\@startsection {subsection}{1}{\z@}{0.5ex \@plus .2ex \@minus 0.2ex}{0.75ex \@plus.1ex\@minus .5ex}{\normalfont\bfseries}}

\definecolor{rindou1}{rgb}{0.4431,0.2862,0.7960}
\definecolor{rindou2}{rgb}{0.0078,0.1215,0.4392}
\definecolor{lapis}{rgb}{0.0.0470,0.2941,0.5568}
\definecolor{emerald}{rgb}{0.31, 0.78, 0.47}
\definecolor{pinegreen}{rgb}{0.0, 0.47, 0.44}
\definecolor{jade}{rgb}{0.0, 0.66, 0.42}
\definecolor{teal}{rgb}{0.0, 0.5, 0.5}
\newcommand{\eq}[1]{\vspace{-0.5pt}\begin{equation}#1\vspace{-0.5pt}\end{equation}}
\newcommand{\eqs}[1]{\vspace{-0.5pt}\begin{equation}\begin{split}#1\end{split}\vspace{-0.5pt}\end{equation}}
\newcommand{\fwbox}[2]{\text{\makebox[#1][c]{$\hspace{-150pt}\displaystyle#2\hspace{-150pt}$}}}
\newcommand{\fwboxL}[2]{\text{\makebox[#1][l]{$#2$}}}
\newcommand{\fwboxR}[2]{\text{\makebox[#1][r]{$#2$}}}
\newcommand{\equivR}{\fwbox{14.5pt}{\hspace{-0pt}\fwboxR{0pt}{\raisebox{0.47pt}{\hspace{1.25pt}:\hspace{-4pt}}}=\fwboxL{0pt}{}}}
\newcommand{\equivL}{\fwbox{14.5pt}{\fwboxR{0pt}{}=\fwboxL{0pt}{\raisebox{0.47pt}{\hspace{-4pt}:\hspace{1.25pt}}}}}
\newcommand{\fig}[3]{\raisebox{#1}{\includegraphics[scale=#2]{#3}}}

\newcommand{\bigger}[1]{\raisebox{-0.95pt}{\scalebox{1.25}{$#1$}}}
\newcommand{\mi}{\raisebox{0.75pt}{\scalebox{0.75}{$\hspace{-0.5pt}\,-\,\hspace{-0.5pt}$}}}

\newcommand{\calA}{\mathcal{A}}
\newcommand{\calN}{\mathcal{N}}

\newcommand{\calI}{\mathcal{I}}

\newcommand{\calO}{\mathcal{O}}
\newcommand{\lam}[1]{\lambda_{#1}}
\newcommand{\lamt}[1]{\widetilde{\lambda}_{#1}}
\renewcommand{\phi}{\varphi}
\renewcommand{\bar}{\overline}
\renewcommand{\hat}{\widehat}
\newcommand{\br}[1]{\left[\!\!\,\left[#1\right]\!\!\,\right]}
\renewcommand{\tilde}{\widetilde}
\newcommand{\ab}[1]{\langle #1\rangle}
\renewcommand{\sb}[1]{[ #1]}
\newcommand{\asb}[1]{\langle #1]}

\newcommand{\dbar}{\fwboxL{7.2pt}{\raisebox{4.5pt}{\fwboxL{0pt}{\scalebox{1.5}[0.75]{\hspace{1.25pt}\text{-}}}}d}}

\newcommand{\Li}[1]{\text{Li}_{2}\!\left(\hspace{-0.5pt}#1\hspace{-0.5pt}\right)}
\newcommand{\symN}{sYM${}_{\calN}$}
\newcommand{\pX}{{\color{hred}p_X}}
\newcommand{\X}{{\color{hred}X}}
\def\checkmark{\tikz\fill[scale=0.4](0,.35) -- (.25,0) -- (1,.7) -- (.25,.15) -- cycle;} 

\definecolor{hblue}{rgb}{0,0,0.575}
\definecolor{hred}{rgb}{0.575,0.0,0.225}
\definecolor{hgreen}{rgb}{0.0,0.4,0.2}
\definecolor{hteal}{rgb}{0.0,0.545,0.7451}
\definecolor{dim}{rgb}{0.6,0.6,0.6}

\title{\texorpdfstring{{\huge \mbox{Integrands of Less-Supersymmetric\hspace{-10pt}}}\\[-6pt]{\huge\mbox{Yang-Mills at One Loop}}\\[-2pt]}{Integrands of Less-Supersymmetric Yang-Mills at One Loop}}
\author[a,b]{\vspace{-18pt}Jacob~L.~Bourjaily,}\emailAdd{bourjaily@psu.edu}
\author[c]{Enrico~Herrmann,}\emailAdd{eh10@g.ucla.edu}
\author[a]{Cameron~Langer,}\emailAdd{ckl5552@psu.edu}
\author[a]{Kokkimidis~Patatoukos,}\emailAdd{kzp326@psu.edu}
\author[d]{Jaroslav~Trnka,}\emailAdd{trnka@ucdavis.edu}
\author[d]{Minshan~Zheng}\emailAdd{mszheng@ucdavis.edu}
\affiliation[a]{Institute for Gravitation and the Cosmos, Department of Physics,\\Pennsylvania State University, University Park, PA 16802, USA}
\affiliation[b]{Niels Bohr International Academy and Discovery Center, Niels Bohr Institute,\\University of Copenhagen, Blegdamsvej 17, DK-2100, Copenhagen \O, Denmark}
\affiliation[c]{Mani L. Bhaumik Institute for Theoretical Physics,
UCLA Department of Physics and Astronomy, Los Angeles, CA 90095, USA}
\affiliation[d]{Center for Quantum Mathematics and Physics (QMAP),\\Department of Physics, University of California, Davis, CA 95616, USA}

\abstract{%
We construct a prescriptive, bubble power-counting basis of one-loop integrands suitable for representing amplitude integrands in less-supersymmetric ($1\!\leq\!\calN\!\leq\!4$) Yang-Mills theory. With the exception of massless bubbles, all integrands have unambiguous, leading singularities as coefficients defined in field theory; for the massless bubbles on external legs, we find two natural choices which lead to different integrands that highlight distinct aspects of field theory. For concreteness, we give the all-multiplicity integrands for MHV amplitudes, and the split-helicity amplitude integrand for six-particle NMHV. The basis we construct is mostly pure and is divided into to separately UV- and IR-finite sectors of fixed transcendental weight, resulting in UV- and IR-finite ratio functions of $n$-particle helicity amplitudes. 
}
\preprint{}

\begin{document}
\maketitle
\pagenumbering{roman}

\pagenumbering{roman}

\pagenumbering{arabic}
\vspace{0pt}%

\newpage
\section{Introduction and Overview}
\label{sec:introduction}\vspace{0pt}

Important recent progress in our understanding of scattering amplitudes in quantum field theory originated from considering the structure of loop amplitudes at the level of the \emph{integrand}---the unintegrated sum of Feynman diagrams, whose analytic structure is determined by unitarity in terms of \emph{on-shell} processes. In particular, these investigations at one loop led directly to the discovery of BCFW tree-level recursion relations \cite{Britto:2004ap,Britto:2005fq}, dual conformal-(and ultimately Yangian-)invariance of planar maximally supersymmetric Yang-Mills theory \cite{Drummond:2006rz,Alday:2007hr,Drummond:2008vq,Drummond:2009fd}, and the correspondence between leading singularities and subspaces of Grassmannian manifolds \cite{Arkani-Hamed:2016byb}. 

The origins of generalized unitarity \cite{Bern:1994zx,Bern:1994cg,Britto:2004nc} are extremely simple to understand: loop integrands, being rational differential forms on the space of loop momenta, can be expanded into a basis of such forms with coefficients that are loop-momentum independent. For any process in any particular quantum field theory and at any fixed loop order and spacetime dimension, the space of \emph{all} scattering amplitude integrands (arbitrary multiplicity and external particle content) spans a finite-dimensional space of `master' integrands. Once these integrands are integrated they can be recycled for arbitrary scattering amplitudes of interest in the theory. 

A familiar illustration of the power of this idea is the `no-triangle property' for amplitudes in maximally supersymmetric Yang-Mills and gravity at one loop \mbox{\cite{Bern:2005bb,Bjerrum-Bohr:2006xbk,Bern:2007xj,Bjerrum-Bohr:2008qoa}}. Specifically, this means that all amplitudes in these theories are expressible in a basis of `scalar box' integrals (those that scale like four propagators at infinite loop momentum). This basis was called $\mathfrak{B}_4^{(4)}$ in ref.~\cite{Bourjaily:2020qca}, and it is a classic result of Passarino and Veltman \cite{Passarino:1978jh} that all one loop \emph{integrals} involving more than four-propagators can be expanded into those with four or fewer. Thus, at one-loop in these theories, the scalar box integrals suffice for representing all scattering amplitudes. 

More generally, the size of the basis required to represent amplitudes in a quantum field theory remains an important and open question. For example, it is known that scattering amplitudes in both the Standard Model and pure Yang-Mills are expressible in terms of the basis of integrands $\mathfrak{B}_0$---integrands that scale like a loop-independent constant at infinite momentum---which is the basis described by OPP in ref.~\cite{Ossola:2006us,Mastrolia:2010nb}; but it is not known whether this is the \emph{smallest} space of loop integrands needed to express amplitudes in these theories. 

In this work, we consider the case of one-loop amplitudes in less-than-maximally supersymmetric ($1\!\leq\!\calN\!<\!4$) Yang-Mills theory (`\symN\!'). We show that these amplitudes can be expressed in the space $\mathfrak{B}_2^{(4)}$---the space of integrands with `bubble' power-counting in four dimensions. We do this by constructing a particular, \emph{prescriptive} \cite{Bourjaily:2017wjl} basis for $\mathfrak{B}_2^{(4)}$ with several special features, and show how amplitudes in \symN~can be represented in this basis. 

More precisely, we focus on scattering amplitudes of pure $\calN\!=\!1,2$ vector multiplets without additional matter. In terms of on-shell multiplets, one can label on-shell scattering states in terms of helicity super-multiplets \cite{Elvang:2011fx}. In the planar limit, we expect a well-defined notion of the \emph{integrand} due to the fact that planarity, or equivalently (leading) color ordering, induces a fixed cyclic ordering of the external momenta of the scattering states, which in turn allows us to define unique labels for the loop-variables to any order in perturbation theory. These variables are given either by choosing an origin of loop-momentum space, going to dual coordinates \cite{Drummond:2008vq}, or (in strictly four spacetime dimensions) introducing momentum twistors \cite{Hodges:2009hk}, all of which have played a major role in recent developments for maximally supersymmetric amplitudes, and beyond. One key advantage of the global labels that originated in $\calN{=}4$ sYM arises from multiple different definitions of the integrand, either in terms of a standard diagrammatic representation or via loop-level on-shell recursion relations \cite{ArkaniHamed:2010kv}. For less-supersymmetric amplitudes in the planar limit, these recursion relations should exist, but are associated with various subtleties \cite{Benincasa:2015zna}.

One goal of this work is to uniquely define the one-loop integrands for less than maximally ($1\!\leq\!\calN{\leq}4$) supersymmetric Yang-Mills theory (the pure Yang-Mills case has new features which we leave for future work). The situation is significantly different from the case of maximal supersymmetry because of the presence of poles at infinity as indicated by having triangles and bubbles in the one-loop expansion. We show that the standard cuts considered in the context of generalized unitarity fix the integrand up to massless bubbles contributions. These terms integrate to zero but are nevertheless important at the integrand-level; and we illustrate two choices of contours which can be used to fix their coefficients.

\subsection*{Organization and Outline}
%
This work is organized as follows. In \mbox{section \ref{sec:bubble_power_counting_basis}}, we review the ingredients of basis-integrand construction and the role of prescriptivity \cite{Bourjaily:2017wjl}. We describe our particular choice of basis for $\mathfrak{B}_2^{(4)}$ in \mbox{section \ref{subsec:bubble_basis}}, and highlight how it is stratified by its UV/IR structure and its transcendental weight in \mbox{section \ref{subsubsec:bubble_basis_stratifications}}.

Because the basis we construct is prescriptive, the coefficient of every integrand is a `leading singularity' in field theory: i.e. the integral of the amplitude along some particular compact contour (at one loop, always a `residue'). In less-than-maximally supersymmetric Yang-Mills theory, leading singularities require more information to specify than in $\calN\!=\!4$ sYM. We review these ingredients in \mbox{section \ref{sec:ls_in_n_leq_4_sym}}. In particular, we find that one loop amplitude integrands in \symN~can be represented as a combination of the corresponding amplitude integrands in $\calN\!=\!4$ sYM (which have better power-counting), plus corrections involving only those diagrams with so-called `non-singlet' helicity flow. In \mbox{section \ref{subsec:massless_bubble_ambiguities}}, we discuss some subtleties that arise in the case of leading singularities associated with massless bubble integrals, and suggest two natural paths to defining a unique integrand.

In \mbox{section \ref{sec:amplitudes}}, we apply our diagonalized bubble power-counting basis of integrands to write down a closed formula for all-multiplicity MHV amplitudes in \mbox{section \ref{subsec:mhv_formulae}}. We further illustrate these ideas with a particular six-point NMHV amplitude integrand in \mbox{section \ref{subsec:nmhv_6point}}, and discuss how this representation of amplitudes \emph{manifests} the finiteness of many observables in these theories before concluding in \mbox{section \ref{sec:discussion}}. 

Finally, in \mbox{appendix \ref{integrand_basis_details}} we provide full details for our integral basis, and each basis element's result from loop integration. These results, as well as the all-multiplicity MHV amplitude integrand, are also provided as ancillary files attached to this work.

\section{A Prescriptive, Bubble Power-Counting Basis at One Loop}\label{sec:bubble_power_counting_basis}\vspace{0pt}

The fundamental principle behind generalized unitarity \cite{Bern:1994zx,Bern:1994cg,Britto:2004nc} is that loop amplitude integrands $\calA$ are elements of a vector space of differential forms on the space of loop momenta; as such, they may be expanded into a \emph{basis} $\mathfrak{B}$ (large enough that $\calA\!\subset\!\mathfrak{B}$) of such forms, 
\eq{\calA=\sum_{\mathfrak{b}_i\in\mathfrak{B}}\,c_i\,\mathfrak{b}_i\,,\label{general_integrand_expansion}}
where the coefficients $c_i$ are loop-momentum-independent `on-shell' functions determined by \emph{generalized unitarity}: i.e.~the left and right-hand sides of eq.~(\ref{general_integrand_expansion}) agree on all contour integrals which `encircle' loop-dependent Feynman propagators. 

In principle, an \emph{arbitrary} spanning set of Feynman integrands (rational differential forms involving some number of Feynman propagators and arbitrary functions of loop momenta in the numerators) can be chosen for a basis in (\ref{general_integrand_expansion}). In this case, the determination of the coefficients $c_i$ amounts to a problem of linear algebra: suppose that one has some spanning set of integration contours $\{\Omega_j\}$ on which the \emph{period matrix}
\eq{\oint\limits_{\Omega_j}\!\mathfrak{b}_i\equivL \mathbf{M}_{i,j}\,\label{period_matrix}}
were known or determined to be full-rank. Then the coefficients of amplitudes $a_i$ would be determined by the system of equations
\eq{\begin{split}\fwbox{0pt}{a_j\equivR\oint\limits_{\Omega_j}\!\calA=\oint\limits_{\Omega_j}\Big(\sum_{\mathfrak{b}_i\in\mathfrak{B}}\,c_i\,\mathfrak{b}_i\Big)=\sum_{\mathfrak{b}_i\in\mathfrak{B}}\,c_i\Big(\oint\limits_{\Omega_j}\!\mathfrak{b}_i\Big)=\sum_ic_i\mathbf{M}_{i,j}}\\
\fwbox{0pt}{\bigger{\,\Rightarrow\,}c_j=\sum_{i}a_i.\left[\mathbf{M}^{-1}\right]_{i,j}\,.}\end{split}\label{coefficient_fixing_equation}}
Typically, the cycles chosen to determine coefficients are those involving as many `residue' contours as possible---those which encircle a number of Feynman propagators, poles at infinity, collinear regions, and so-on. Because these contours enclose physical poles, the periods of \emph{amplitude integrands} $a_j$ defined in (\ref{coefficient_fixing_equation}) are called \emph{leading singularities} \cite{Cachazo:2008vp} and can be determined in terms of on-shell (tree) amplitudes. The story of these coefficients is one with a very rich history. 

Setting aside the potential computational complexity involved in inverting the period matrix $\mathbf{M}_{i,j}$ defined in (\ref{period_matrix}), it is worth emphasizing that most \emph{seemingly natural} choices for bases of master integrands (those involving some Feynman graph's worth of propagators and a spanning-set of `Lorentz-invariant scalar products' in their numerators) lead to very poor integrals---ones that can deeply obscure many interesting and important features of scattering amplitudes. Thus, it is worthwhile to try and find a \emph{good} set of integrands for any basis. 

\subsection{Brief Review of Prescriptive Integrand Bases for Amplitudes}
\label{subsec:prescriptivity_in_brief}\vspace{0pt}

A \emph{prescriptive} integrand basis is one chosen to be cohomologically \emph{dual} to a spanning set of maximal-dimensional compact contours of integration. That is, a basis is prescriptive provided that there exists a set of compact, maximal-dimensional integration contours $\{\Omega_j\}$ such that 
\eq{\oint\limits_{\Omega_j}\!\mathfrak{b}_i=\delta_{i,j}\,.\label{prescriptivity_condition}}
When this is the case, the coefficients $c_i$ of the amplitude integrand (\ref{general_integrand_expansion}) are \emph{leading singularities} of field theory because the inversion of the period matrix (\ref{period_matrix}) is trivial:
\eq{c_i = a_i\equivR \oint\limits_{\Omega_i}\!\!\ \calA\,.}
Prescriptive integrand bases have been shown to possess many desirable properties. In particular, they often evaluate to \emph{pure} functions (those satisfying nilpotent systems of differential equations, see e.g.~\cite{Broedel:2018qkq,Henn:2013pwa}), and hence are comparatively easy to integrate. 

To be clear, prescriptive integrand bases are fairly straightforwardly constructed. Starting from an arbitrary basis of loop integrands $\mathfrak{B}^0$ and an arbitrary spanning-set of contour integrals $\{\Omega_j\}$, a prescriptive basis can be obtained according a simple `rotation' of the basis:
\eq{\mathfrak{b}_i\equivR \sum_{k}\left[\mathbf{M}^{-1}\right]_{i,k}\mathfrak{b}_k^0\quad\text{where}\quad \mathbf{M}_{k,j}\equivR\oint\limits_{\Omega_j}\!\mathfrak{b}^0_k\,.}

It should be clear how important the role of the cycle basis is in the above discussion: different choices of contours $\{\Omega_j\}$ can result in strikingly different bases of integrands. Thus, there is relatively little uniqueness here. For our particular purposes in this work, we chose a maximal subset of contours to expose IR and UV divergences, resulting in a basis stratified by divergences. As stressed previously, this choice is by no means unique and one could think about alternative bases inspired by other physical or mathematical properties. 

In what follows, we review the elements involved in defining a particular set of Feynman integrands for a basis---as defined by (some proxy for) `power-counting'. Then we illustrate the kinds of choices made for a dual set of cycles, and how these choices affect the resulting integrand basis.

\newpage
\subsection{\texorpdfstring{Defining a Bubble Power-Counting Basis $\mathfrak{B}_{2}^{(4)}$}{Defining a Bubble Power-Counting Basis}}
\label{subsec:bubble_basis}\vspace{0pt}

As described in ref.~\cite{Bourjaily:2020qca}, one can construct a basis of `bubble-power-counting' integrands at one loop as follows. Start with any Feynman graph involving some number of $p\!\geq\!2$ propagators and consider the vector space of loop-dependent polynomials in the numerator
\eq{[\ell]^{(p-2)}\quad\text{with}\quad[\ell]^q\equivR\underset{Q_i\in\mathbb{R}^d}{\text{span}}\!\Big\{(\ell-Q_1)^2\cdots (\ell-Q_q)^2\Big\}\,.}
That is, $[\ell]^q$ represents that linear span of all $q$-fold products of inverse propagators. Thus, the space of $\mathfrak{B}_2$ is defined as the linear span of all Feynman integrals with $p$ propagators and a product of $p{-}2$ inverse propagators in the numerator. 

Graphically, if we use 
\vspace{-10pt}\eq{\fwbox{65pt}{\tikzBox{\draw[int](0,0)--(1,0);\draw[markedEdgeR](0,0)--(1,0);\node[anchor=north] at (0.5,0) {$\vec{\ell}$};}}\equivR\frac{[\ell]}{\ell^2}\,,\label{vector_space_of_decorated_edge}\vspace{-18pt}}
to denote the vector space of inverse-propagators times some propagator, then 
\eq{\vspace{-0pt}\fig{-18pt}{1}{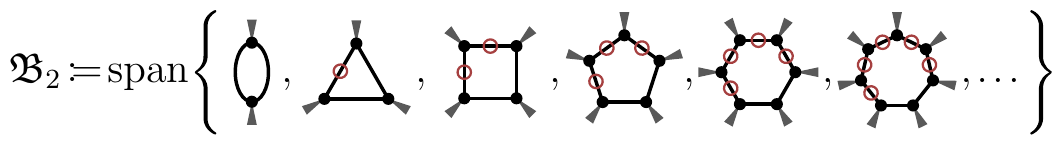}\,.\vspace{-0pt}}
As described in ref.~\cite{Bourjaily:2020qca}, this space is finite dimensional for any fixed spacetime dimension (or multiplicity). In four dimensions, all integrands involving more than four propagators are expressible in terms of those with four or fewer. In particular, the basis $\mathfrak{B}_2^{(4)}$ is spanned by the following vector spaces of loop integrands:
\eq{\begin{split}
\fwboxL{200pt}{\hspace{-50pt}\boxIntegrandsSchematic\equivR\frac{[\ell]^2}{\ell^2(\ell+p_A)^2(\ell+p_{AB})^2(\ell+p_{ABC})^2}\,,}\\
\fwboxL{200pt}{\hspace{-100pt}\triangleIntegrandsSchematic\equivR\frac{[\ell]^1}{\ell^2(\ell+p_A)^2(\ell+p_{AB})^2}\,,\hspace{5pt}\text{and}\hspace{5pt}\bubbleIntegrandsSchematic\,\equivR\frac{1/2}{\ell^2(\ell+p_A)^2}\,.}\end{split}}
Throughout this work, we always leave implicit the factor of `$\dbar^4\ell$' in these integration measures. For each set of leg distributions, these spaces of integrands have rank (in four-dimensions) of $20\!=\!{\color{topCount}2}{+}18$, $6\!=\!{\color{topCount}3}{+}3$, and $1$, respectively. What we mean by this, for example, in is that the 6-dimensional vector space $[\ell]^1$ of loop-dependent numerators for the triangle integrands can be viewed as spanned by $3$ `{\color{topCount}top-level}' degrees of freedom and $3$ contact terms---spanned by the inverse-propagators appearing in the graph. Similarly, of the 20-dimensional vector space $[\ell]^2$ of numerators for the box integrands, all but {\color{topCount}2} can be spanned by contact terms: $\binom{4}{2}\!=\!6$ double-contact terms (with one degree of freedom each), and $\binom{4}{1}\!=\!4$ single contact terms with 3 top-level degrees of freedom each. Labeling only the top-level degrees of freedom for each denominator topology (those numerators not spanned by the contact terms of the integral) , our bubble power-counting basis consists of {\color{topCount}2} numerators per box, {\color{topCount}3} numerators per triangle, and a single numerator per bubble, denoted by $\calI_{A,B,C,D}^i$, $\calI_{A,B,C}^I$, and $\calI_{A,B}$, respectively. We may represent each of these integrands graphically as follows:
\vspace{0.2cm}
\eq{\fwbox{0pt}{\hspace{-30pt}\boxIntegrandSchematic\!\!\!\!\bigger{\Leftrightarrow}\frac{\mathfrak{n}^i_{A,B,C,D}(\ell)}{\ell_a^2\,\ell_b^2\,\ell_c^2\,\ell_d^2}\,,\hspace{-7pt}\triangleIntegrandSchematic\bigger{\Leftrightarrow}\frac{\mathfrak{n}^I_{A,B,C}(\ell)}{\ell_a^2\,\ell_b^2\,\ell_c^2}\,,\;\bubbleIntegrandSchematic\,\bigger{\Leftrightarrow}\frac{1/2}{\ell_a^2\,\ell_b^2}}}
where $i\!\in\!\{1,2\}$ indexes the {\color{topCount}top-level} degrees of freedom of each box, and $I\!\in\!\{1,2,3\}$ indexes the {\color{topCount}top-level} degrees of freedom of each triangle. To be clear, the sets $\{A\},\ldots$ represent arbitrary \emph{non-empty} collections of external momenta flowing into the vertex, with $p_A\equivR\sum_{a\in A}p_a$ and $s_A\equivR p_A^2=(\sum_{a\in A}p_a)^2$.

Later on, we will have reason to distinguish between sets of external momenta that are `massive' (sets consisting of more than one massless leg) from those which are massless. When $\{A\}$ consists of a single element, we will denote it by $a\equivR\{a\}=\{A\}$, and similarly for the other momenta labels. More generally, we refer to `$a$' as the \emph{first} label in the set $\{A\}\equivR\{a,\ldots\}$, and so-on. Due to our focus on planar (color-ordered) amplitudes, the sets are endowed with a natural ordering of external legs. \\

To determine the specific numerators for the basis, we start from a spanning set of contours and fix the precise numerators according to the prescriptivity condition (\ref{prescriptivity_condition}). It is worth emphasizing how the particular numerators are chosen using these conditions. For example, in the case of box-integrands, two particular numerators are chosen not simply by the condition
\eq{\oint\limits_{\Omega_{A,B,C,D}^j}\hspace{-10pt}\calI_{A,B,C,D}^i=\delta_{i,j}\,,\label{eg_top_level_dof_fixing}}
but \emph{also} by the requirement that it vanish on all triangle-topology contours of its contact terms 
\eq{\oint\limits_{\fwboxL{0pt}{\Omega_{(A{+}B),C,D}^J}}\hspace{-4pt}\calI_{A,B,C,D}^i=\oint\limits_{\fwboxL{0pt}{\Omega_{A,(B{+}C),D}^J}}\hspace{-4pt}\calI_{A,B,C,D}^i=\oint\limits_{\fwboxL{0pt}{\Omega_{A,B,(C{+}D)}^J}}\hspace{-4pt}\calI_{A,B,C,D}^i=\oint\limits_{\fwboxL{0pt}{\Omega_{(D{+}A),B,C}^J}}\hspace{-4pt}\calI_{A,B,C,D}^i=0\,;\label{eg_level_one_dof_fixing}}
and similarly for all the contours for its bubble-topology, double-contact terms:
\eq{\oint\limits_{\fwboxL{0pt}{\Omega_{(A{+}B{+}C),D}}}\hspace{-04pt}\calI_{A,B,C,D}^i=\hspace{-0pt}\oint\limits_{\fwboxL{0pt}{\Omega_{A,(B{+}C{+}D)}}}\hspace{-4pt}\calI_{A,B,C,D}^i=\hspace{-0pt}\oint\limits_{\fwboxL{0pt}{\Omega_{B,(C{+}D{+}A)}}}\hspace{-4pt}\calI_{A,B,C,D}^i=\hspace{-0pt}\oint\limits_{\fwboxL{0pt}{\Omega_{(D{+}A{+}B),C}}}\hspace{-4pt}\calI_{A,B,C,D}^i=\hspace{-0pt}\oint\limits_{\fwboxL{0pt}{\Omega_{(A{+}B),(C{+}D)}}}\hspace{-4pt}\calI_{A,B,C,D}^i=\hspace{-0pt}\oint\limits_{\fwboxL{0pt}{\Omega_{(D{+}A),(B{+}C)}}}\hspace{-4pt}\calI_{A,B,C,D}^i=0\,.\label{eg_level_two_dof_fixing}}
Thus, of the $\text{rank}([\ell]^2)\!=\!{\color{topCount}2}{+}{\color{black}18}$ degrees of freedom required to specify the basis numerators $\mathfrak{n}^{i}_{A,B,C,D}$, only {\color{topCount}2} are fixed by (\ref{eg_top_level_dof_fixing}), while $3\!\times\!4$ of the remaining degrees of freedom are determined by (\ref{eg_level_one_dof_fixing}) and $6\!\times\!1$ are fixed by the analogous equations (\ref{eg_level_two_dof_fixing}) for bubble contact-terms. This is what we mean by saying that an integrand basis $\mathfrak{B}_{2}^{(4)}$ is \emph{dual} to a spanning set of particular cycles.  

Of course, in order to construct \emph{specific} integrand numerators, we must specify the contour conditions which define the basis prescriptively as described above. We do this in the following subsection. However, it should be clear that, independent from the precise contour definition, scattering amplitude integrands in this basis will be represented according to 
%
\eq{\begin{split}\ \calA=&\phantom{+}\sum_{A,B,C,D}\sum_{i=1}^2a_{A,B,C,D}^i\calI_{A,B,C,D}^i{+}\sum_{A,B,C}\sum_{I=1}^3a_{A,B,C}^I\calI_{A,B,C}^I{+}\sum_{A,B}a_{A,B}\calI_{A,B}\end{split}
}%
where 
\eq{a_{A,B,C,D}^i\equivR\oint\limits_{\fwboxL{0pt}{\hspace{-10pt}\Omega_{A,B,C,D}^i}}\calA\,,\quad a_{A,B,C}^I\equivR\oint\limits_{\fwboxL{0pt}{\hspace{-10pt}\Omega_{A,B,C}^I}}\calA\,,\quad a_{A,B}\equivR\oint\limits_{\fwboxL{0pt}{\hspace{-10pt}\Omega_{A,B}}}\calA\,.}
To be any more specific, we must specify the contour conditions which define our basis prescriptively.

\subsubsection{A Spanning Set of Maximal-Dimension Contours}
\label{subsubsec:bubble_basis_contours}\vspace{0pt}

It is interesting to note that the basis of bubble power-counting integrands in four dimensions can be viewed as $\mathfrak{B}_2^{(4)}\!\simeq\mathfrak{B}_3^{(4)}\!\oplus\mathfrak{B}_2^{(3)}$. That is, we may consider the new integrands in $\mathfrak{B}_2^{(4)}$ relative to those of $\mathfrak{B}_3^{(4)}$ to be those associated with a bubble power-counting basis in \emph{three} dimensions---merely \emph{reinterpreted} in four dimensions. This is also motivated by the fact that all the new integrals required have less than maximal transcendental weight when integrated in $4D$, but would be of maximal-weight in $3D$; these weight drops are related to the presence of double-poles when the integrands are interpreted in $4D$. Provided the integrands of $\mathfrak{B}_3^{(4)}$ are full-weight, they will automatically be diagonal with respect to the integrands in $\mathfrak{B}_{2}^{(3)}$---that is, they will vanish on all contours involving double-poles. 

The basis elements \emph{without} double-poles---those of $\mathfrak{B}_3^{(4)}\!\subset\!\mathfrak{B}_2^{(4)}$---are easiest to discuss, which is why we start with their defining contours. The basis elements in this category are the chiral boxes $\calI^{i}_{A,B,C,D}$ as well as the scalar triangle integrands $\calI^{I=1}_{A,B,C}$. All other basis elements have double-poles at infinity and will be considered momentarily in section~\ref{subsubsubsec:weight_drop_contours}. A summary of our defining set of contours is also provided in Table \ref{basisContourTable} of appendix \ref{app:cycle_basis}.  

The contours defining the chiral box integrands can be represented graphically according to:
\vspace{-2pt}\eq{
\Omega^i_{A,B,C,D}\equivR \left\{\fourMassContourA\,, \fourMassContourB\right\}\,;
\vspace{-2pt}}
these are simply the contours encircling the two solutions $\{\ell^*_1, \ell^*_2\}$ to the quadruple-cut equations $\ell^2_a{=}\ell^2_b{=}\ell^2_c{=}\ell^2_d{=}0$. Only the box integrals have four-propagators to have a non-vanishing contour integral on such a cut, and each box integrand involves a unique set of such propagators; as such, all other basis elements automatically vanish on these contours. 

The chirality of box-integrand contours can be seen more clearly in cases where massless corners are present, for which we may indicate the parity of the contour using blue or white vertices. For example, we denote the three-mass box contours as
\vspace{-6pt}\eq{
\Omega^i_{a,B,C,D}\equivR \left\{\threeMassContourA\,, \threeMassContourB\right\}\,,
\vspace{-4pt}}
which highlights that these contours involve $\ell^*_1\!=\!\lam{a}\lamt{X}$ and $\ell^*_2\!=\!\lam{X}\lamt{a}$, respectively, and the precise form of $\lam{X}$ and $\lamt{X}$ is irrelevant for the moment.  

Next, consider the contours for the \emph{scalar} triangle integrands. Most interesting are the cases where there is at least one massless leg, since the associated dual basis integrands can have IR singularities. For example, we define the two-mass scalar-triangle integrals' contours by
\vspace{-6pt}\eq{
\Omega^{I=1}_{a,B,C} \equivR \twoMassTriContourA\,,
\vspace{-5pt}}
where the circle is a graphical notation for the collinearity condition $\ell_a\!\sim\!p_a$ imposed in addition to the triple cut $\ell^2_a\!=\!\ell^2_b\!=\!\ell^2_c\!=\!0$. Let us mention that this particular contour is spurious, and thus no scattering amplitude has support here. Furthermore, demanding that the chiral box integrands vanish on $\Omega^{I=1}_{a,B,C}$ guarantees that they are free of this particular collinear singularity associated with IR divergences. 

A similar discussion also applies for the scalar one-mass triangle contour $\Omega^{I=1}_{a,b,C}$ (see subsection \ref{subsubsec:bubble_basis_numerators} for further details). The contour choice in Table \ref{basisContourTable} for the scalar triangles renders all boxes \emph{locally} IR-finite as in \cite{Bourjaily:2021ewt} by demanding that the chiral box integrands vanish in all collinear or soft regions of loop-momentum space. This choice leads to the same numerators that have been described in the context of $\calN{=}4$ sYM in \cite{Bourjaily:2013mma}.

\newpage
\paragraph{Defining Contours for Lower-Weight Integrands}
\label{subsubsubsec:weight_drop_contours}~\\[-12pt]\vspace{0pt}

The second class of basis integrands and their associated contours involves certain double-poles at infinite loop momentum. These are the objects we turn to now.

The key observation to define a bubble power-counting basis in four dimensions is that $\mathfrak{B}_2^{(4)}\!\simeq\mathfrak{B}_3^{(4)}\!\oplus\mathfrak{B}_2^{(3)}$. That is, the additional integrands needed, relative to a triangle power-counting basis in four dimensions, are scalar bubbles and triangle integrals with single-inverse-propagator loop-dependence in their numerators which define $\mathfrak{B}_2^{(3)}$; both of these are \emph{naturally} defined in three dimensions---and for more than merely pragmatic reasons. 

Consider for example the scalar bubble integral. With the appropriate normalization of the numerator in terms of powers of $s_A$, the bubble integrates to a pure weight-one function in either two or three dimensions. Moreover, it is possible to write it as a wedge-product of $d\!\log$-differential forms in either case: (for a more detailed discussion, see e.g.~\cite{Arkani-Hamed:2017ahv,Herrmann:2019upk})
\begin{align}
\begin{split}
I^{(D{=}2)}_{A,B} & = \int\!\!\dbar^2\ell\,\,\frac{1}{2}\frac{s_A}{\ell^2_a\ \ell^2_b}
= \frac14 \int\!\!
   \dbar\!\log \frac{\ell^2_a}{\ell^2_b} \wedge 
   \dbar\!\log \frac{(\ell_a{-}\ell^*_a)^2}{(\ell_a{-}\widetilde{\ell}^*_a)^2} 
   \\
I^{(D{=}3)}_{A,B} & = \int\!\!\dbar^3\ell\,\, \frac{1}{2}\frac{\sqrt{s_A}}{\ell^2_a\ \ell^2_b}
= \frac14 \int\!\!
   \dbar\!\log \ell^2_a \wedge 
   \dbar\!\log \ell^2_b \wedge 
   \dbar\!\log \frac{\ell\!\cdot\!q}{\ell\!\cdot\!\overline{q}}
\end{split}
\end{align}
where, in the two-dimensional bubble, $\ell^*_a$ and $\widetilde{\ell}^*_a$ are the two solutions to the maximal cut equation $\ell^2_a\!=\!\ell^2_b\!=\!0$ and the bubble has no pole at infinity, $\ell\!\to\!\infty$. The three-dimensional bubble is slightly more complicated and has a single pole at $\ell\!\to\!\infty$. This can be thought of as a dual conformal triangle in $D=3$ where one of the dual points is taken to be infinity \cite{Herrmann:2019upk}. In suitable coordinates (embedding space), infinity is treated on the same footing as any other point which makes this analysis very clear. Here, we refrain from introducing embedding coordinates (see \cite{Bourjaily:2019exo}) and work in momentum space directly which leads to the appearance of the two null-vectors $q$ and $\overline{q}$ normalized by $q\cdot\overline{q}=1$ which are defined by the relations $q\cdot p_A\!=\!\overline{q}\cdot p_A\!=\!0$. (Technically, this is easiest to implement by choosing light-cone coordinates transverse to $p_A$. Furthermore, the $d\!\log$ form remains valid for massive internal propagators where $\ell^2_{a,b}\!\to\!\ell^2_{a,b}{-}m^2$ which will become important for our discussion in $D{=}4$.) In $D{=}3$, we consider for example the triple cut of the bubble which encircles the two propagators and furthermore encloses the odd combination (parity-even) of simple poles at $\ell\!\to\!\infty$ which is clear from the $d\log$ form in $3D$ where one cuts the two propagators $\ell^2_a\!=\!\ell^2_b=0$ and then encircles the parity-even combination of $\ell\!\cdot\!q\!=\!0$ and $\ell\!\cdot\!\overline{q}\!=\!0$. In three dimensions, this is a leading singularity of the scalar bubble integrand.

In contrast, in four dimensions, the scalar bubble integrand has a double-pole---signaling a weight-drop in the resulting integral \cite{Arkani-Hamed:2014via,Bern:2014kca}. This is reflected in the fact that the bubble can be written explicitly by decomposing the four-dimensional space of loop-momenta into a three-dimensional subspace and one additional direction, say $\ell^i$ orthogonal to the momentum $p_A$ (as well as $q,\, \overline{q}$) entering the bubble and to the three-dimensional slice. This effectively means that we can think about the $4D$ bubble as a $3D$ bubble where the propagators become massive, with mass $m^2\equivR(\ell^i)^2$. Since our $3D$ $d\!\log$ form was valid for internal massive legs, we find
\eq{I^{(D{=}4)}_{A,B} = 
\int\!\!\dbar^3\ell\,\dbar\ell^i\,\,\frac{1}{2}\frac{\sqrt{s_A}}{\ell^2_a\ \ell^2_b} 
= \frac{1}{4} \int\!\! \dbar\!\log \ell^2_a \wedge \dbar\!\log \ell^2_b \wedge  
   \dbar\!\log \frac{\ell\!\cdot\! q}{\ell\!\cdot\!\overline{q}} \wedge \dbar\ell^i\,,
}
so that the triple-cut residue results in a `double-pole at infinity': an integrand which is \emph{independent} of the remaining loop integration parameter. 
\eq{\oint\limits_{\substack{\ell_a^2=\ell_b^2=0\\\vec{\ell}\to\infty\\\text{(odd)}}}\!\!\!\!\!\dbar^4\vec{\ell}\,\,\frac{1}{\ell_a^2\ell_b^2}\propto \int\!\!\dbar\!\ell^i\,,}
where $\ell^i$ is whatever component of $\vec{\ell}$ \emph{not} eliminated in the three integration cycles. Thus, for this integrand the differential of the form `$\dbar\ell^i$' looks like a \emph{total derivative} on the cut. Moreover, this differential form has a double-pole at infinity. Unlike $\dbar\!\log$, $\dbar\ell^i$ is \emph{not} scale invariant and thus the coefficient of the double-pole is not uniquely defined. As this example should make clear, the particular component for the final integration, say $\ell^i$, is completely arbitrary: any three components of $\vec{\ell}$ can be eliminated in the first integrations, always resulting in an integrand of the form $\dbar\ell^i$ in the remaining variable. Thus, there is no \emph{particular} double-pole: there is a three-dimensional (four-dimensional, modulo rescaling) family of such double-poles. Perhaps a more invariant way of describing a bubble integrand in four dimensions would be to start with the fact that in three dimensions, it is purely polylogarithmic: it is a $d\!\log$-form. Going from three to four dimensions amounts to appending a trivial $d\ell^i$ onto this polylogarithmic three-form. 

To be clear, the bubble integral is polylogarithmic on \emph{any} three-dimensional subspace chosen---which we may denote as $(\ell^i)^{\perp}$ for any component $\ell^i$ of $\vec{\ell}$. Considering that $\ell^i\equivR\ell^\mu\!\cdot\!e^i_{\mu}$ for some basis vector $e^i_{\mu}$, it is clear that we can view the complementary space as the solutions to $\ell\!\cdot\!\pX=0$ for any $\pX$. For reasons of simplicity, it turns out to be beneficial to take $\pX$ to be null. In this construction, we define a three-dimensional subspace of loop momenta according to 
\eq{\dbar^4\ell\mapsto \dbar^3\hat{\ell}\quad\text{where}\quad \hat{\ell}\in(\pX)^{\perp}\,.\label{projection_equation}}
Noting that the null-space $(\pX)^{\perp}$ of $\pX$ is defined by $\ell\!\cdot\!\pX=0$, we see that this can be interpreted more concretely as:
\eq{\dbar^3\hat{\ell}\equivR\dbar^4\ell\,\,\,\delta(\ell\!\cdot\!\pX)\,.}
Although this three-dimensional subspace depends on $\pX$, we will choose the \emph{same} subspace for all integrands with double-poles. Thus, when we say that $\mathfrak{B}_2^{(4)}\!\simeq\mathfrak{B}_3^{(4)}\!\oplus\mathfrak{B}_2^{(3)}$, we consider the basis $\mathfrak{B}_2^{(3)}$ to be defined as in (\ref{projection_equation}) for all integrands and consider contours to be taken over \emph{this} three-dimensional space $\hat{\ell}$.

\subsubsection{Illustrations of the Resulting Numerators in the Basis}
\label{subsubsec:bubble_basis_numerators}\vspace{0pt}

In order to make some of the abstract definitions of the previous subsections more concrete, we consider a few illustrative examples that highlight all relevant features. The complete list of one-loop basis integrands with bubble power-counting is summarized in Table \ref{basisNumerators} of appendix \ref{app:integrand_numerators}. First, we consider the two-mass-easy box integrands
\begin{align}
\label{eq:2mass_box_integrand}
\twoMassEasyInts\rule[-32pt]{0pt}{70pt}\bigger{\!\leftrightarrow\;}\mathcal{I}_{a,B,c,D}^{i}\equivR\frac{\mathfrak{n}_{a,B,c,D}^i}{\ell_a^2\,\ell_b^2\,\ell_c^2\,\ell_d^2}\quad\text{with}\quad
\begin{cases}
 \mathfrak{n}^{i=1}_{a,B,c,D} \equivR \br{p_a,\ell_b,\ell_c,p_c} \\
 \mathfrak{n}^{i=2}_{a,B,c,D} \equivR \br{\ell_b,\ell_c,p_c,p_a}
\end{cases}
\end{align}
where we use the kinematic bracket conventions from \cite{Bourjaily:2019iqr,Bourjaily:2019gqu} to denote contractions of momenta
\eq{\br{a_1,a_2,\cdots,c_1,c_2}\equivR\!\Big[(a_1\!\cdot\!a_2)^{\alpha}_{\phantom{\alpha}\beta}\cdots(c_1\!\cdot\!c_2)^{\gamma}_{\phantom{\gamma}\alpha}\!\Big]\,,\label{definition_of_br}}
where $(a_1\!\cdot\!a_2)^{\alpha}_{\phantom{\alpha}\beta}\equivR a_1^{\alpha\,\dot{\alpha}}\epsilon_{\dot{\alpha}\dot{\gamma}}a_2^{\dot{\gamma}\gamma}\epsilon_{\gamma\beta}$ and $a^{\alpha\dot\alpha}\equivR a^{\mu}\sigma_{\mu}^{\alpha\dot{\alpha}}$ are `$2\!\times\!2$' four-momenta, defined via the Pauli matrices. The `$\br{\cdots}$' object may be more familiar to some readers if written equivalently as `$\mathrm{tr}_+[\cdots]$', are linear in their arguments, and satisfy the following identities
\eq{\br{a_1,a_2,\cdots,c_1,c_2}=\br{c_2,c_1,\cdots,a_2,a_1}=\br{c_1,c_2,a_1,a_2,\cdots}\,.}
Often, they may be simplified using 
\eq{\br{\ldots,x,A,A,y,\ldots}=s_A\br{\ldots,x,y,\ldots}\,,
\quad\text{with}\br{}=2\,.}

The two chiral-box numerators are normalized to unity on the following maximal-dimensional cycles 
\begin{align}
\label{eq:2mass_easy_box_countours}
    \fwboxR{0pt}{\Omega_{a,B,c,D}^{{\color{cut1}1}}\equivR\hspace{-5pt}}\twoMassEasyContourA,\qquad \qquad
    \hspace{10pt}\fwboxR{0pt}{\Omega_{a,B,c,D}^{{\color{cut2}2}}\equivR\hspace{-5pt}}\twoMassEasyContourB\,,
\end{align}
where ${\color{cut1}\ell^*_1}$ and ${\color{cut2}\ell^*_2}$ are the two solutions to the maximal cut equations of the box $\ell^2_a\!=\!\ell^2_b\!=\!\ell^2_c\!=\!\ell^2_d\!=\! 0$ and the white and blue vertices in the contour prescription in $(\ref{eq:2mass_easy_box_countours})$ indicates the chirality of the solution at that vertex. In particular ${\color{cut1}\ell^*_1}\!\sim\!\lam{a} \lamt{X}$ and ${\color{cut2}\ell^*_2}\!\sim\!\lam{X} \lamt{a}$. Due to the chirality of the solution and the order of the momenta in the brackets of $\mathfrak{n}^{i}_{a,B,c,D}$ in (\ref{eq:2mass_box_integrand}), the integrand basis elements are diagonal on the respective contours. In order to claim that our basis is truly prescriptive, it remains to be checked that both integrand basis elements vanish on all other defining contours summarized in Table \ref{basisContourTable} in appendix \ref{integrand_basis_details}. 

First, we should note that the chiral boxes scale at infinity like scalar triangle integrals, i.e.\ they have at most \emph{single} poles at $\ell\!\to\!\infty$. This implies that these integrands trivially vanish on all contours that involve the instruction of taking a double-pole at infinity. This implies that the chiral boxes vanish on all defining contours for chiral triangles (to be discussed in detail shortly) as well as on the bubble-integral contours. The only remaining question is therefore associated with the defining contours for the scalar triangle subtopologies $\Omega^1_{a,B,C}$ in the language of Table \ref{basisContourTable}. For the example considered above, all triangle subtopologies have one massless leg. Our particular choice of the one-mass scalar-triangle contour involves the collinear limit around the massless corner of the triangle. Fortunately, the chiral box numerators in (\ref{eq:2mass_box_integrand}) vanish in the collinear limit where $\ell_a\!\propto\!p_a$ or $\ell_c\!\propto\!p_c$ due to the properties of $\br{\cdots}$. Crucially, the fact that these chiral boxes have only single poles at $\ell\!\to\!\infty$ together with the fact that they vanish in the collinear regions $\ell_a\!\propto\!p_a$ or $\ell_c\!\propto\!p_c$ renders these objects both UV and IR finite. These integrands have been integrated in \cite{Bourjaily:2013mma} and for the convenience of the reader we give their result in terms of polylogarithms in Table \ref{basisIntegralsTable}.

A second illustrative example to consider is the one-mass triangle sector
\begin{align}
\label{eq:1mass_triangle_integrands}
\oneMassTriInts\rule[-32pt]{0pt}{70pt}\bigger{\leftrightarrow}
\begin{cases}
\mathfrak{n}^{I=1}_{a,b,C}\equivR {-}s_C 
\\
\mathfrak{n}^{I=2}_{a,b,C}\equivR 
\frac{1}{2 \br{p_{a}{-}p_b,\pX}} 
\left(2 \br{p_a,\ell_b,p_b,\pX} + \ell^2_b \br{p_a{-}p_b,\pX}\right)
\\
\mathfrak{n}^{I=3}_{a,b,C}\equivR  \frac{1}{2 \br{p_a{-}p_b,\pX}} 
\left(2 \br{\pX,p_a,\ell_b,p_b} + \ell^2_b \br{p_a{-}p_b,\pX}\right)
\end{cases}\!\!.
\end{align}
The chiral numerators $\mathfrak{n}^{I=2,3}_{a,b,C}$ are written in a way to make the collinear and UV properties manifest. In particular, the ordering of momenta in $\br{p_a,\ell_b,p_b,\pX}$ and its conjugated version guarantees that these integrand elements are IR finite in the collinear regions $\ell_b\!\propto\!p_a, p_b$ as well as in the soft region $\ell_b\!\sim\!0$.  

These integrands in (\ref{eq:1mass_triangle_integrands}) are constructed to be dual to the following defining contours
\begin{align}
    \Omega^I_{a,b,C} \equivR
    \left\{
    \oneMassTriContourA,\oneMassTriContourB,\oneMassTriContourC 
    \right\}
\end{align}
where the first contour $\Omega^{I=1}_{a,b,C}$ represents the soft-collinear leading singularity that sets $\ell_b\!=\!0$ and uniquely selects the scalar one-mass triangle. (All box integrands are chiral and their numerators guarantee the vanishing in the soft-collinear configuration.)

The chiral contours $\Omega^{I=2,3}_{a,b,C}$ warrant some further explanation. This is the first time in our discussion where we have to deal with the double-poles at infinity that are naturally associated with a weight drop at the integrated level. These were discussed abstractly in section \ref{subsubsubsec:weight_drop_contours} and we would like to concretely give our definition for the chiral one-mass triangles here. The way to think about the chiral contours such as $\Omega^{I=2}_{a,b,C}$ that involve the double-pole at infinity is as follows. First, one projects $\ell_a$ into a particular direction
\begin{align}
\label{eq:eg_projection_condition}
 \frac{\br{\ell_a,\pX}}{\br{p_a{-}p_b,\pX}}=0,
\end{align}
which leaves us with a three-dimensional surface for $\ell_a$ perpendicular to the above projection constraint. The particular normalization of the projection (\ref{eq:eg_projection_condition}) is related to our choice of projection and enters in the overall normalization of our integrand. The remaining three parameters of $(\ell_a)^\perp$ are then fixed on the triple-cut \mbox{$\ell^2_a\!=\!\ell^2_b\!=\!\ell^2_c\!=\!0$}. Together with the projection condition (\ref{eq:eg_projection_condition}), the three on-shell constraints therefore localize all four degrees of freedom of $\ell_a$. There are two different solutions to the constraints which we denote by $\ell^*_{1,\infty}$ and $\ell^*_{2,\infty}$, where the additional subscript signals that we are interested in the leading behavior of $\ell\!\to\!\infty$. Taking into account the proper Jacobian factor $\mathcal{J}$, our numerators evaluated on the leading singularity solutions are unit 
\begin{align}
    \left. \frac{\mathfrak{n}^{I=2}_{a,b,C}}{\mathcal{J}}\right|_{\ell_a =\ell^*_{1,\infty} } \hspace{-.8cm} = 1 \,,
    \qquad 
     \left. \frac{\mathfrak{n}^{I=3}_{a,b,C}}{\mathcal{J}}\right|_{\ell_a =\ell^*_{1,\infty} }\hspace{-.8cm} = 0 \,,
     \qquad
     \left. \frac{\mathfrak{n}^{I=2}_{a,b,C}}{\mathcal{J}}\right|_{\ell_a =\ell^*_{2,\infty} }\hspace{-.8cm} = 0 \,,
    \qquad 
     \left. \frac{\mathfrak{n}^{I=3}_{a,b,C}}{\mathcal{J}}\right|_{\ell_a =\ell^*_{2,\infty} } \hspace{-.8cm}= 1 \,.
\end{align}

One additional point worth discussing is the explicit presence of the bubble-contact term $+\ell^2_b$ in the definition of our one-mass chiral triangle numerators $\mathfrak{n}^{I=2,3}_{a,b,C}$. This term is there in order to have the chiral triangles vanish on the massive bubble contour $\Omega_{a{+}b,C}$, which we define presently. 

Consider the generic massive bubble topology $\Omega_{A,B}$. Just as in the chiral triangle sector, we begin by projecting $\ell_a$ onto the three-dimensional subspace
\eq{\frac{\br{\ell_a,\pX}}{\br{p_A,\pX}}=0.
}
Next, two additional degrees of freedom are fixed by localizing to the bubble cut $\ell_a^2\!=\!\ell_b^2\!=\!0$. In fact, there are two solutions to these three combined conditions, which we may denote as $\ell^*_{1},\ell^*_{2}$. (If one parametrizes $\ell_a$ in a basis of spinors involving the null momentum $\pX$, these two solutions are \emph{chiral} and involve loop momenta proportional to either $\lambda_{\X}$ or $\tilde\lambda_{\X}$, as we illustrate more explicitly in section~\ref{subsec:mhv_formulae}.) The final degree of freedom is fixed on the double-pole at infinity, with the bubble integral normalized on the parity-even combination of these two evaluations. 

The downside of the explicit presence of this bubble contact term is that the chiral triangles are rendered UV divergent. In principle, one could avoid this feature by explicitly removing the bubble contribution. However, this would come at the cost that the resulting basis would no longer be dual to a particular choice of contours, rendering the basis \emph{non-}prescriptive. This would have the effect that in the representation of an amplitude, the coefficients of bubbles, say, would need to be the difference of bubble-cut leading singularities and whatever \emph{pollution} arises from the triangle integrals. 

Of course, one could start from a prescriptive basis to determine coefficients simply and then rotate into a non-prescriptive one in order to highlight other aspects of interest---such as a better separation between UV and IR divergent integrands. We should note in passing that we have in fact constructed such a possibly preferential basis---one in which the \emph{only} UV-divergent integrals are the bubbles, and for which all integrals are pure. Such choices, however, are far from unique, and leave open the generally broad questions of aesthetic and technical preferences, and so we leave such potentially illuminating rotations to future work.

\paragraph{Bubble Integrands and Integrals Involving Massless Legs}
\label{subsubsubsec:massless_bub_integrands}~\\[-12pt]\vspace{0pt}

Most of the integrand and contour definitions are conceptually very simple, although the exact details and choices made required a nontrivial amount of work. There is however, one additional cases that is often neglected:  bubbles involving massless external legs:
\begin{align}
\label{eq:zeroMassBubInt}
    \zeroMassBubbleInt\,.
\end{align}
In dimensional regularization, this integral is somewhat special in the sense that it is scaleless and integrates to zero in a nontrivial way. UV and IR divergences cancel one another in the form $0\!=\!\frac{1}{\epsilon_{{\rm UV}}}{-}\frac{1}{\epsilon_{{\rm IR}}}$.  In traditional generalized unitarity constructions, these terms are neglected at first and a tentative amplitude is computed. Once one separates UV from IR divergences, e.g.\ by introducing a small mass regulator, one can compare the resulting IR or UV divergences of the tentative amplitude to general expectations and ultimately adjust the coefficients of these massless bubbles to match the expected results. We will come back to this point in more detail in subsection \ref{subsec:massless_bubble_ambiguities}.

From the viewpoint of prescriptivity and basis-building, however, these integrals pose no subtlety whatsoever: they are defined in exactly the same ways as the massive bubble integrands---except that the collinear condition is imposed on the three-point vertex instead of taking a residue at infinity---thereby highlighting the region of loop-momentum space that is responsible for these integrals' IR divergences.

What is genuinely subtle, however, is the meaning of leading singularities defined on such a contour---which affects the coefficients of these integrands in the representation of amplitudes. We review this issue in some detail in \mbox{section \ref{subsec:massless_bubble_ambiguities}}, and pose two possible definitions one may take for these coefficients. 

\newpage\subsubsection{Stratification of UV/IR Structure and Transcendental Weight}
\label{subsubsec:bubble_basis_stratifications}\vspace{0pt}

The discussion of the previous subsection \ref{subsubsec:bubble_basis_numerators} should have made clear that our prescriptive basis integrands have certain desirable features both from an UV and IR point of view, related to the presence or absence of double-poles at infinity, or (soft-)collinear poles, respectively. Of course, being one-loop integrals, by now it is straight forward to explicitly check these integrand-level statements by simply integrating all basis elements. The results loop integration (in dimensional regularization) for every integrand in our basis is given in Table \ref{basisIntegralsTable}. 

It is worth highlighting several interesting features. In the decomposition of the basis $\mathfrak{B}_2^{(4)}\!\simeq\!\mathfrak{B}_3^{(4)}\!\!\oplus\!\mathfrak{B}_{2}^{(3)}$, the integrands in the $\mathfrak{B}_3^{(4)}$ subspace are all pure, weight-2 functions that are free of any regions of UV-divergence; moreover, only the scalar triangle integrals involving massless legs are IR divergent---all others are \emph{locally} finite. These general features are summarized in \mbox{Table \ref{triangle_subbasis}}. 
\begin{table}[t]
\caption{Properties of basis integrands defined in $\mathfrak{B}_3^{(4)}\!\subset\!\mathfrak{B}_{2}^{(4)}$. All these integrals are are \emph{pure} and weight-2 in transcendentality when integrated in (4{-}$2\epsilon$) dimensions.\label{triangle_subbasis}}\vspace{-10pt}
$$\fwbox{400pt}{\begin{array}{|l@{$\,$}|@{$$}c@{$\,$}|@{$\,$}c@{$\,$}|@{$\,$}c@{$\,$}|@{$\,$}c@{$\,$}|@{$\,$}c@{$\,$}|@{$\,$}c@{$\,$}|@{$\,$}c@{$\,$}|}\cline{2-8}\multicolumn{1}{l@{$$}|@{$$}}{}&\,\mathcal{I}_{A,B,C,D}^{i}&\mathcal{I}_{a,B,C,D}^{i}&\mathcal{I}_{a,b,C,D}^{i}&\mathcal{I}_{a,B,c,D}^{i}&\mathcal{I}_{A,B,C}^{1}&\mathcal{I}_{a,B,C}^{1}&\mathcal{I}_{a,b,C}^{1}\\\hline
\text{locally-finite}&\checkmark&\checkmark&\checkmark&\checkmark&\checkmark&{\color{dim}\text{---}}&{\color{dim}\text{---}}\\\hline
\text{IR-divergent (UV-finite)}&{\color{dim}\text{---}}&{\color{dim}\text{---}}&{\color{dim}\text{---}}&{\color{dim}\text{---}}&{\color{dim}\text{---}}&\checkmark&\checkmark\\\hline\end{array}}$$
\vspace{-10pt}
\end{table}

In contrast, all those integrands in the $\mathfrak{B}_2^{(3)}$ subspace are weight-one functions when evaluated in $4{-}2\epsilon$ dimensions (as is natural for having maximal weight in $3D$). All but one class evaluates to a pure function. These general features of these integrals are summarized in \mbox{Table \ref{bubble_subbasis}}. 
\begin{table}[b]
$$\fwbox{400pt}{\begin{array}{|l@{$\,$}|@{$$}c@{$\,$}|@{$\,$}c@{$\,$}|@{$\,$}c@{$\,$}|@{$\,$}c@{$\,$}|@{$\,$}c@{$\,$}|@{$\,$}c@{$\,$}|}\cline{2-7}\multicolumn{1}{l@{$$}|@{$$}}{}&\,\mathcal{I}_{A,B,C}^{2}&\mathcal{I}_{A,B,C}^{3}&\mathcal{I}_{a,B,C}^{2,3}&\mathcal{I}_{a,b,C}^{2,3}&\mathcal{I}_{A,B}^1&\mathcal{I}_{a,B}^1\\\hline
\text{locally-finite}&\checkmark&{\color{dim}\text{---}}&{\color{dim}\text{---}}&{\color{dim}\text{---}}&{\color{dim}\text{---}}&{\color{dim}\text{---}}\\\hline
\text{UV-divergent (IR-finite)}&{\color{dim}\text{---}}&\checkmark&\checkmark&\checkmark&\checkmark&{\color{dim}\text{---}}\\\hline
\text{UV- \emph{and} IR-divergent}&{\color{dim}\text{---}}&{\color{dim}\text{---}}&{\color{dim}\text{---}}&{\color{dim}\text{---}}&{\color{dim}\text{---}}&\checkmark\\\hline
\end{array}}$$\vspace{-20pt}
\caption{Properties of basis integrands defined in $\mathfrak{B}_2^{(3)}\!\subset\!\mathfrak{B}_{2}^{(4)}$. All these integrals are weight-1 in transcendentality when integrated in (4{-}$2\epsilon$) dimensions; only $\mathcal{I}_{a,B,C}^{2,3}$ are impure.\label{bubble_subbasis}}\vspace{-10pt}
\end{table}

As discussed above, it is possible to alter the basis of integrands to improve the IR/UV properties of the basis. For example, it is easy to render the integrands $\calI^{2,3}_{a,B,C}$ pure or to make all triangle integrands UV-finite. However, this rotation of the basis would seem to be in conflict with prescriptivity, and make it harder to directly determine the coefficients of an amplitudes in the new basis.  

\newpage
\section{\texorpdfstring{Leading Singularities in $\calN\!\leq\!4$ Super Yang-Mills Theory}{Leading Singularities in N<4 sYM}}
\label{sec:ls_in_n_leq_4_sym}

Having discussed the integrand basis construction at length, we are now in the position to comment on the second key building block in the generalized unitarity expansion of the amplitude which are the coefficient functions. As discussed in section \ref{subsubsec:bubble_basis_contours}, all defining contours are of maximal dimension so that the coefficients of our basis integrands are simply  leading singularities. In this section, we give details on how to compute these leading singularities in less (than maximally) supersymmetric theories.\\[-8pt]

The description of on-shell (super-)states for amplitudes in $\calN\!<\!4$ super Yang-Mills theory are best implemented by considering the states to be truncations of those in $\calN\!=\!4$. This was described in detail in ref.~\cite{Elvang:2011fx}, but is worth reviewing. We denote the fields related by supersymmetry to the $({+})$-helicity gluon by an ordered list (of length 0 to length $\calN$) of indices $I\!\in\!\{1,\ldots,\calN\}$; similarly, we can label the fields related by supersymmetry to the $({-})$-helicity gluon by the complements of the previous labels within the set $\{1,\ldots,4\}$. Thus, a $({+})$-helicity gluon always has a label of $\{\}$ and $({-})$-helicity gluon always has a label of $\{1,2,3,4\}\equivL\bar{\{\}}$; similarly, the (${+}$\textonehalf)-helicity fermions are labeled by sets $\{I\}$ with $I\!\in\!\{1,\ldots,\calN\}$ while the (${-}$\textonehalf)-helicity fermions are labelled by $\bar{\{I\}}\equivR\{1,2,3,4\}\backslash\{I\}$; and similarly for the rest of the states in the theory. Labeling the states in this way, every (non-vanishing) amplitude involves the same number $k$ of each of the indices $\{1,2,3,4\}$ corresponding to an N${}^{k{-}2}$MHV super-amplitude. 

This scheme makes it obvious that for any amount of supersymmetry, the external states can be labelled as particular instances of those of $\calN\!=\!4$---the only difference being in the selection rule for which $R$-charge labels are allowed among the external states. These selection rules have the effect of \emph{requiring} that the indices $\{\calN{+}1,\ldots,4\}$ \emph{all} appear in the labels of some subset of $k$ external states for an N${}^{k{-}2}$MHV amplitude. This amounts to a truncation of some $\calN\!=\!4$ super-function. 

Thus, all processes in an N${}^{k{-}2}$MHV amplitude (or on-shell function) must specify precisely $k$ states related by supersymmetry to the $({-})$-helicity gluon. These are simply `helicity' amplitudes in the case of `pure' ($\calN\!=\!0$) Yang-Mills theory; but the same $\binom{n}{k}$ distinguished labels are required for all component amplitudes for any degree of supersymmetry (other than maximal). 

This can be encoded graphically in an on-shell diagram by \emph{orienting} all its edges. We choose to use an incoming arrow to denote the $\binom{n}{k}$ states related to the $({-})$-helicity gluons (incoming at the vertex), and outgoing arrows to denote those related to the $({+})$-helicity gluons (incoming at a vertex). 

For example the three-point super-amplitudes in \symN\,would require \emph{orientations} as in
\eq{\threePointMHVbarBare\;\threePointMHVBare\,.}
To be clear, these amplitudes may be defined in terms of coherent states as follows:
\eq{\begin{split}\fwboxR{40pt}{\calA_3^{(i)}\equivR}\raisebox{-2pt}{\threePointMHVbar}\,\,\equivR&\frac{\sb{a\,b}^{\fwboxL{0pt}{4{-}\calN}}}{\sb{i\,a}\sb{a\,b}\sb{b\,i}}\hspace{3pt}\delta^{1\times\calN}\!\big(\sb{ab}\tilde{\eta}_i^I{+}\sb{b\,i}\tilde{\eta}_a^I{+}\sb{i\,a}\tilde{\eta}_b^I\big)\delta^{2\times2}\big(\lambda\!\cdot\!\tilde\lambda\big)\\
\fwboxR{40pt}{\calA_{3}^{(i,j)}\equivR}\raisebox{-2pt}{\threePointMHV}\,\,\equivR&\frac{\ab{i\,j}^{\fwboxL{0pt}{4{-}\calN}}}{\ab{i\,j}\ab{j\,a}\ab{a\,i}}\delta^{2\times\calN}\!\big(\lambda_i^{\alpha}\tilde{\eta}_i^I{+}\lambda_j^{\alpha}\tilde{\eta}_j^I{+}\lambda_a^{\alpha}\tilde{\eta}_a^I\big)\hspace{6pt}\delta^{2\times2}\big(\lambda\!\cdot\!\tilde\lambda\big)\end{split}}
in terms of Grassmann variables $\tilde{\eta}_i^I$ for $I\!\in\!\{1,\ldots,\mathcal{N}\}$. The generalization to MHV amplitudes is extremely natural: 
\eq{\calA_{n,0}^{(i,j)}\equivR\frac{\ab{i\,j}^{4{-}\calN}}{\ab{1\,2}\ab{2\,3}\cdots\ab{n\,1}}\delta^{2\times\calN}\!\big(\lambda\!\cdot\!\tilde\eta\big)\,\delta^{2\times2}\!\big(\lambda\!\cdot\!\tilde\lambda\big)\,}
where 
\eq{\delta^{2\times\calN}\!\big(\lambda\!\cdot\!\tilde\eta\big)\equivR\delta^{2\times\calN}\!\Big(\sum_{a}\lambda_a^{\alpha}\tilde\eta_a^{I}\Big)}
is the super-momentum-conserving $\delta$-function and $\delta^{2\times2}\!\big(\lambda{\cdot}\tilde\lambda\big)\equivR\delta^{2\times2}\!\big( \sum^n_{a=1}\lambda^\alpha_a \tilde{\lambda}^{\dot \alpha}_a\big)$ encodes overall momentum conservation. More generally, an N${}^{k{-}2}$MHV superfunction (such as a leading singularity) in $\calN\!=\!4$ super Yang-Mills is related to $\binom{n}{k}$ \emph{oriented} superfunctions in \symN\,according to 
\eq{f(\lambda,\tilde\lambda)\delta^{k\times4}\!\big(C\!\cdot\!\tilde\eta\big)\bigger{\;\Rightarrow\;}
f^{(i_1,\ldots,i_k)}\equivR f(\lambda,\tilde\lambda)\det(c_{i_1},\ldots,c_{i_k})^{4{-}\calN}\delta^{k\times\calN}\!\big(C\!\cdot\!\tilde\eta\big)\,,}
where $C$ represents the $k\!\times\!n$ `boundary-measurement' matrix \cite{Arkani-Hamed:2016byb} and $\{i_r\}$ label the negative helicity super-multiplets. Just as in the three-point amplitudes given above, any decorated on-shell diagram must be oriented such that each N${}^{k-2}$MHV tree-amplitude appearing at a vertex has $k$ `sources'---i.e., incoming arrows.

\subsection{Decorated On-Shell Diagrams: Singlet vs.\ Non-Singlet}
\label{subsec:decorated_on_shell_diagrams}\vspace{0pt}
%
There is a marked difference between on-shell functions in maximally supersymmetric Yang-Mills and its less supersymmetric cousins. This is primarily a result of the distinction between so-called `singlet' and `non-singlet' helicity configurations. In the former case, the $R$-charges of the external states uniquely determine those of the internal states running through the loop, regardless of the amount of supersymmetry. All such singlet on-shell diagrams are therefore $\mathcal{N}$-independent and therefore equal to (truncations of) $\mathcal{N}\!=\!4$ super-functions and may be immediately recycled. In contrast, when there are oriented loops of `helicity' in an on-shell diagram, we must sum over all the states in the supermultiplet which clearly depends on $\mathcal{N}$. 

A prototypical example of a non-singlet decorated on-shell function is the following four-point box diagram with external states $\{2,4\}$ are taken as incoming:
\eq{\label{eq:dec_on_shell_eg}\fourPointNonSingletEg\equivR\fourPointNonSingletPartA+\fourPointNonSingletPartB.}
For each of the two possible `helicity' flows through the graph (each involving a sum over states), it is not difficult to determine the corresponding on-shell function by direct computation. In particular, we find:
\vspace{10pt}\eq{\raisebox{0pt}{$\fwbox{0pt}{\fourPointNonSingletPartA\hspace{-10pt}\!=\,\calA_{4,0}^{(2,4)}\phi^{4-\calN}\,,\;\hspace{10pt}\fourPointNonSingletPartB\hspace{-10pt}\!=\,\calA_{4,0}^{(2,4)}(1{-}\phi)^{4-\calN}}\vspace{-0pt}\,,$}}
where we have defined the cross-ratio
\eq{\phi\equivR\frac{\ab{1\,4}\ab{2\,3}}{\ab{1\,3}\ab{2\,4}}\,.}
Thus, the decorated on-shell diagram (\ref{eq:dec_on_shell_eg}) is, for $1\leq\mathcal{N}<4$,
\eq{
\label{eq:non_singlet_os_func_eg}
\fourPointNonSingletEg\hspace{-10pt}=\calA_{4,0}^{(2,4)}\Big[\phi^{4{-}\calN}+(1\mi\phi)^{4{-}\calN}\Big]\,.}
When $\calN\!=\!4$, the equation above over-counts the sum over states by 2 as both directions of helicity flow are included in the same coherent state. Furthermore, eq.~(\ref{eq:non_singlet_os_func_eg}) is valid for \emph{entire} super-amplitudes: replacing the pre-factor (the gluonic component of the MHV tree amplitude) by the superamplitude gives the correct answer for all components such that the $R$-charges of particles $\{1,3\}$ are in the `$+$' multiplet (related to $g^+$ by some number of supersymmetry generators $\widetilde{Q}_I$'s).

Another example which is directly relevant for the all-multiplicity MHV amplitude presented in section~\ref{sec:amplitudes} is the generic two-mass easy box cut where the states related to the negative helicity gluons have particle label $i,j$. It is easy to verify that the only non-singlet configuration in this case is when both $i$ and $j$ are each in a distinct massive corner:
\eq{\mhvNonSingletLSbox\equivR\mhvNonSingletLSboxA+\mhvNonSingletLSboxB.}
A straightforward calculation yields the result
\eq{\mhvNonSingletLSbox\hspace{-10pt}=\calA_{n,0}^{(i,j)}\Bigg[\left(\frac{\ab{a\,i}\ab{c\,j}}{\ab{a\,c}\ab{i\,j}}\right)^{4-\calN}+\left(1-\frac{\ab{a\,i}\ab{c\,j}}{\ab{a\,c}\ab{i\,j}}\right)^{4-\calN}\Bigg]\,.}
On-shell diagrams of either the triangle or bubble type may be computed in an analogous fashion; the structure of the result is depends on whether the cut is singlet or non-singlet. As mentioned in section~\ref{sec:bubble_power_counting_basis}, the evaluation of field theory on triangle and bubble contours involves double-poles at infinity and requires a projection of the loop momentum onto a particular direction. Pragmatically, one can always derive such contour integrals from the double and triple-cuts of standard unitarity. We illustrate this feature for the massive MHV bubble coefficients in section~\ref{subsec:mhv_formulae}.

\subsection{Generalized Unitarity for Massless Bubble Coefficients}
\label{subsec:massless_bubble_ambiguities}\vspace{0pt}

For gauge theories with $\calN\!<3$ supersymmetry, there is an important subtlety associated with loop integrand and cut topologies which define the massless bubble integrals. If only interested in the integrated amplitudes, these coefficients may be ignored as all such integrands integrate to zero (in dimensional regularization). However, if one were interested in disentangling the UV and IR structure of an amplitude, they play an important role. As such, their coefficients can be determined post-integration by the requirement that this behavior is correct (see e.g.~\cite{Bern:1995db,Arkani-Hamed:2008owk,Elvang:2011fx}). 

To see this subtlety, consider the two-particle, massless cuts of an amplitude. For any N${}^k$MHV degree (and any assignments of external helicities), there always exists one singlet and one non-singlet configuration depending on the parity of the three-particle vertex:
\eq{\Big\{\genSingletMasslessBubbleA\hspace{-5pt},\genNonSingletMasslessBubbleA\hspace{-5pt}\Big\}\;\text{or}\;\Big\{\genNonSingletMasslessBubbleB\hspace{-5pt},\genSingletMasslessBubbleB\hspace{-5pt}\Big\}\,.}
While the singlet cuts are always unambiguous and finite (and in fact always equal to truncated superfunctions of $\mathcal{N}\!=\!4$), the \emph{non-singlet} cuts are unfortunately always \emph{ill-defined}---as they generally diverge. Thus, there is no obvious meaning to these cuts in field theory, making it difficult to compute the leading singularities corresponding to the massless bubble contours: there always exists \emph{some} branch of the bubble-cut on which the amplitude diverges. 

Of course, the massless bubble integrals in our basis have been defined by contours not merely taking the co-dimension 2 residue of the bubble cut, but a contour accessing the double-pole at infinity which starts from the collinear triple-cut in loop-momentum space---the region in which
\eq{\ell^*_a = \alpha\, p_a\,, \qquad \ell^*_b = (1+\alpha)\, p_a\,.}
If this contour were viewed as arising as a co-dimension one residue taken along the well-defined (singlet) triple-cut in every case, then because all such cuts are equal to (truncations of) their $\mathcal{N}\!=\!4$ equivalents, no amplitudes would have support on these double poles. This would suggest that every massless bubble coefficient should be identically zero. This is the first option we consider.

While this choice for interpretation is ensured to match field theory functionally on all of the well-defined (singlet) massless bubble-cuts, it turns out that it fails to match the conventional UV-structure of amplitudes (as deduced using the logic of e.g.~\cite{Bern:1995db,Arkani-Hamed:2008owk,Elvang:2011fx}). In particular, it leads to representations of one-loop amplitudes that exactly misses the standard answer by a multiple of the tree amplitude times the sum of massless bubble integrals. 

Perhaps this missing contribution could be attributed to some (however unconventional) renormalization `scheme'. And it may prove that ignoring all massless bubble contributions turns out to lead to better (more elegant in some way, perhaps) strategies at higher loops. But we must leave such speculation to future work. 

However, there is another way to interpret the leading singularities corresponding to these collinear cuts. Namely, it seems natural to associate the collinear configuration as equivalent to a massless bubble on an external leg, as in:
\eq{
\label{eq:massless_bub_def_collinear}
\generalRuleForMasslessBubbles\,.}
This interpretation naturally suggests that we interpret theory theory for these contours as being proportional directly to the tree amplitude as in \cite{Bern:2004cz}. This reproduces the standard result for one-loop amplitudes' UV and IR structure, and certainly seems like an appropriate `convention' for defining these bubble coefficients. This is the prescription used in the expressions generated for our concrete examples given in the ancillary files for this work.

These kinds of subtleties are much more abundant in pure ($\mathcal{N}\!=\!0)$ Yang-Mills theory, the amplitudes of which are known to require worse power-counting in their bases. While we can certainly define a prescriptive basis $\mathfrak{B}_0$ to express these amplitudes, the coefficients of tadpoles and constants seem intrinsically ambiguous and for similar reasons. There have been some notable recent proposals for how to deal with tadpoles  \cite{Britto:2010um} (see also  \cite{Feng:2021enk}); however, all these proposals begin from some prior knowledge of the loop integrand---i.e.\ start from the (literal) sum of Feynman diagrams in some gauge and using some regularization scheme. This does lead to specific coefficients for any integrand in a basis even as ugly as $\mathfrak{B}_0$, but it does not provide a gauge-invariant, cut-level definition of the coefficients in terms of on-shell, tree-level scattering data. (But see e.g.\ \cite{Baumeister:2019rmh} for some interesting ideas in that direction at higher loops that relies on a particular on-shell renormalization scheme.) Naturally, we must leave such questions---important though they are---to future work.

\newpage
\section{Amplitude Integrands for $\calN\!\leq\!4$ Super Yang-Mills Theory}
\label{sec:amplitudes}
The derivation of a diagonalized basis of integrands in section~\ref{sec:bubble_power_counting_basis} has an immediate application: namely, the construction of prescriptive representations of $1\!\leq\!\mathcal{N}\!\leq\!4$ sYM amplitudes. Achieving this amounts to the computation of the coefficient of each basis element---that is, field theory evaluated on the contours defining the basis. 

As discussed in section~\ref{subsec:decorated_on_shell_diagrams}, there are essentially two cases to consider for each coefficient, depending on the helicity configuration of interest. For a given on-shell diagram, if there is only a single allowed internal helicity flow---i.e., a `singlet' configuration where the external helicities uniquely specify the internal helicity states---then the on-shell function is identical for sYM for \emph{any} $\mathcal{N}$. By virtue of the fact that the $\mathcal{N}\!=\!4$ integrand is free of all poles at infinity, this implies that for all `singlet' cuts, the coefficient of every basis element defined on contours involving infinite loop momentum necessarily vanishes. 

For the `non-singlet' configurations where there are multiple allowed helicity configurations, sYM for $\mathcal{N}\!<\!4$ can have support on single (and double) poles at infinity and the associated coefficients are generically non-vanishing (and non-trivial). 

In this section, we illustrate the procedure outlined above with two concrete examples: the all-multiplicity MHV ($\mathcal{A}_{n,\text{1-loop}}^{(i,j)}$) and the six-point split-helicity NMHV ($\mathcal{A}_{6,\text{1-loop}}^{(4,5,6)}$) one-loop integrands.

\subsection{General Structure of Amplitude Integrands}
\label{subsec:generalities}\vspace{0pt}
The general form of a one-loop amplitude integrand expressed in the bubble-power-counting basis defined in section~\ref{sec:bubble_power_counting_basis} is,
\eq{\label{eq:schematic_integrand}\begin{split}\calA=&\phantom{+}\sum_{A,B,C,D}\sum_{i=1}^2a_{A,B,C,D}^i\calI_{A,B,C,D}^i{+}\sum_{A,B,C}\sum_{I=1}^3a_{A,B,C}^I\calI_{A,B,C}^I{+}\sum_{A,B}a_{A,B}\calI_{A,B}\end{split}
}
where the coefficients of each basis element are defined as
\eq{a_{A,B,C,D}^i\equivR\oint\limits_{\fwboxL{0pt}{\hspace{-10pt}\Omega_{A,B,C,D}^i}}\calA\,,\quad a_{A,B,C}^I\equivR\oint\limits_{\fwboxL{0pt}{\hspace{-10pt}\Omega_{A,B,C}^I}}\calA\,,\quad a_{A,B}\equivR\oint\limits_{\fwboxL{0pt}{\hspace{-10pt}\Omega_{A,B}}}\calA\,.}
The box coefficients $a_{A,B,C,D}^i$ are defined on the two quad-cut leading singularities i.e., field theory evaluated on the two solutions to $\ell_a^2=\ell_b^2=\ell_c^2=\ell_d^2=0$. For any singlet configuration, these leading singularities are simply truncations of those defined in $\mathcal{N}\!=\!4$; for the non-singlet configurations, there is a modification resulting from the helicity flow as described above.

Regardless of supersymmetry, all one-mass triangle integrands with scalar numerators have coefficients $a_{a,b,C}^1$ because are defined on the composite `soft-collinear' residue where one internal leg is set to zero on which amplitudes always have support. Moreover, and just as in maximal sYM, the residue of field theory is always equal to the tree amplitude (as this reflects the only universal IR divergence at one loop); that is, $a_{a,b,C}^1\!=\!\mathcal{A}_{n,0}$. For similar reasons, the coefficients of all two-mass scalar triangles are always zero: $a_{a,B,C}^1\!=\!0$. 

The non-singlet cuts of amplitudes \emph{can} generally lead to support on double-poles at infinity, resulting in non-trivial coefficients for triangles with loop-dependent numerators. For any singlet cuts, these coefficients are all zero. The same is true for the all bubble contours defined on double-poles at infinity. Thus, these coefficients depend strongly on how the helicity-flow at each vertex amplitude of the cut flows into the graph, and varies depending on which of the $\binom{n}{k}$ external legs are taken to have `incoming' helicity.

\subsection{\emph{Exempli Gratia}: MHV Amplitude Integrands}
\label{subsec:mhv_formulae}\vspace{0pt}
We can illustrate how these considerations work in the concrete case of MHV amplitudes ($k$\!=\!2) in $\mathcal{N}\!=\!1,2$ super Yang-Mills theory. As with maximal supersymmetry, the only box cuts which have non-vanishing support for these amplitudes are $\Omega_{a,B,c,D}^1$---the (chiral) two-mass-easy contours (and their one-mass degenerations). Of these, most contours admit only a singlet configuration of internal helicity---namely, 
\eq{\label{eq:box_coeffs_mhv2}
\hspace{-.3cm}
\mhvSingletLSa \hspace{-10pt}\equivL a_{i,i{+}1,j,j{+}1}^1=\hspace{-10pt}\mhvSingletLSb\hspace{-10pt}\equivL a_{i,i{+}1,c,c{+}1}^1=\hspace{-10pt}\mhvSingletLSc\hspace{-10pt}=\mathcal{A}_{n,0}^{(i,j)}.}
All of these leading singularities are equal to the tree-level MHV amplitude $\mathcal{A}_{n,0}^{(i,j)}$. 

Among the two-mass-easy boxes, there is only one case which admits a non-singlet configuration:
\eq{\mhvNonSingletLSbox\equivR\mhvNonSingletLSboxA+\mhvNonSingletLSbox\,.}
This example was already encountered in section~\ref{sec:ls_in_n_leq_4_sym}, and leads to the coefficient
\eq{\label{eq:box_coeffs_mhv1}\begin{split}\mhvNonSingletLSbox\hspace{-10pt}=&\,a_{a,a{+}1,c,c{+}1}=\calA_{n,0}^{(i,j)}\Bigg[\left(\frac{\ab{a\,i}\ab{c\,j}}{\ab{a\,c}\ab{i\,j}}\right)^{4{-}\calN}+\left(1-\frac{\ab{a\,i}\ab{c\,j}}{\ab{a\,c}\ab{i\,j}}\right)^{4{-}\calN}\Bigg]\\[-10pt]
=&\,\calA_{n,0}^{(i,j)}\Big[1+(4{-}\calN)\phi(\phi{-}1)\Big]\quad\text{where}\quad\phi\equivR\frac{\ab{a\,i}\ab{c\,j}}{\ab{a\,c}\ab{i\,j}},
\end{split}}
where the final equality follows from the binomial expansion of the exponents in the first line and is valid only for $\mathcal{N}\!=\!1,2$.

Turning now to the triangle configurations, we may start with the scalar one-mass contours, on which all amplitudes have support equal to the tree:
\eq{\label{eq:tri_coeffs_mhv1}\mhvSoftCollinearLS\equivL a_{a,a{+}1,c}^1=\mathcal{A}_{n,0}^{(i,j)}.}
(We have neglected to indicate any helicity information from the left-hand-side for the simple reason that \emph{every} one-mass scalar triangle has the same coefficient, regardless of the helicity configuration under consideration.) 

For the two triangle integrals elements normalized on double-poles, there are just three classes of leg distributions with non-singlet helicity configurations leading to non-zero coefficients. By directly evaluating field theory on the corresponding contours, we find that these non-vanishing coefficients are:
\eq{\label{eq:tri_coeffs_mhv2}\begin{split}\hspace{-40pt}&\mhvNonSingletLStriB\bigger{\Rightarrow}\;\;a_{a,i,C}^2\,\equivR\calA_{n,0}^{(i,j)}(4{-}\calN)\frac{\ab{i\,c}\ab{j\,a}}{\ab{i\,j}\ab{c\,a}}\frac{\ab{i\X}\ab{j\,a}}{\ab{i\,j}\ab{\X a}}\left(\!1{-}\frac{\br{p_a,\pX}}{\br{p_i,\pX}}\right)\,,\\[-10pt]
\hspace{-40pt}&\mhvNonSingletLStriA\bigger{\Rightarrow}\;\;a_{i,b,C}^3\,\equivR\calA_{n,0}^{(i,j)}(4{-}\calN)\frac{\ab{b\,j}\ab{i\X}}{\ab{i\,j}\ab{b\X}}\frac{\ab{b\,j}\ab{i\,i{-}1}}{\ab{i\,j}\ab{b\,i{-}1}}\left(\!1{-}\frac{\br{p_b,\pX}}{\br{p_i,\pX}}\right)\,,\\[-10pt]
\hspace{-40pt}&\fwboxL{320pt}{\mhvNonSingletLStriC\bigger{\Rightarrow}\;\;a_{a,B,C}^2\,\equivR\calA_{n,0}^{(i,j)}(4{-}\calN)\frac{1}{2}\frac{\sb{\X a}\ab{c{-}1\,c}}{\asb{a|p^{}_B|\X}\ab{c{-}1\,a}\ab{c\,a}}\left(\hspace{-2pt}\frac{\ab{i\,a}\ab{j\,a}}{\ab{i\,j}}\hspace{-2pt}\right)^2\,.}
\end{split}
}
Finally, among the massive bubble contours, the only ones with non-singlet helicity flow are those for which $\{i,j\}$ are on opposite sides of the bubble. These coefficients turn out to be 
\eq{\label{eq:bubble_coeffs_mhv}
\hspace{-.6cm}
\begin{split}\hspace{-30pt}\mhvNonSingletLSbubA\bigger{\Rightarrow}\;\;a_{A,B}\,\equivR&\calA_{n,0}^{(i,j)}(4{-}\mathcal{N})\frac{\ab{a{-}1\,a}\ab{b{-}1\,b}}{\ab{i\,j}^2}\Bigg[\!
\frac{\ab{i\X}^2\ab{j\X}^2}{\ab{a{-}1\X}\ab{a\X}\ab{b{-}1\X}\ab{b\X}}\hspace{-25pt} \\
&\hspace{42pt}{+}\frac{\asb{i|p^{}_A|\X}^2\asb{j|p^{}_A|\X}^2}{\asb{a{-}1|p^{}_A|\X}\asb{a|p^{}_A|\X}\asb{b{-}1|p^{}_A|\X}\asb{b|p^{}_A|\X}}\Bigg].\hspace{-80pt}\end{split}\hspace{-.8cm}}
The massless bubble coefficients $a_{a,B}$ were discussed at length in section~\ref{subsec:massless_bubble_ambiguities} and---as emphasized there---we have two options: either $a_{a,B}\!=\!2(\mathcal{N}{-}4)\mathcal{A}_{n,0}^{(i,j)}$ or $a_{a,B}\!=\!0$.

It is worth clarifying how the massive bubble coefficients $a_{A,B}$ in eq.~(\ref{eq:bubble_coeffs_mhv}) may be obtained by a straightforward computation. It is convenient to evaluate field theory on the bubble contour by first computing the two-parameter non-singlet bubble cut, which was in fact given already in \cite{Elvang:2011fx} and may be written as,
\eq{\label{eq:bubble_cut_in_field_thy}\mhvNonSingletLSbubA=\mathcal{A}_{n,0}^{(i,j)}\frac{\ab{a{-}1\,a}\ab{b{-}1\,b}}{J\ab{i\,j}^2}\frac{\left(\ab{i\,\ell_a}\ab{j\,\ell_b}\right)^{4-\mathcal{N}}+\left(\ab{i\,\ell_b}\ab{\ell_a\,j}\right)^{4-\mathcal{N}}}{\ab{a{-}1\,\ell_a}\ab{a\,\ell_a}\ab{b{-}1\,\ell_b}\ab{b\,\ell_b}}\,.
}
A parametrization of $\ell_a,\ell_b$ which is particularly convenient for the projection onto $\br{\ell_a,\pX}\!=\!0$ is given by
\eqs{\label{eq:bubble_param}\ell_a&=\left[s_A\left(\frac{1}{\br{\pX,p_A}}{-}\alpha\right)\lambda_{\X}{+}\beta\left(p_A{\cdot}\tilde\lambda_{\X}\right)\right]\left[\tilde\lambda_{\X}{+}\frac{\alpha}{\beta}\left(p_A{\cdot}\lambda_{\X}\right)\right],\\
\ell_b&=\left[{-}\alpha\,s_A\,\lambda_{\X}{+}\beta\left(p_A{\cdot}\tilde\lambda_{\X}\right)\right]\left[\tilde\lambda_{\X}{-}\frac{1}{\beta}\left(\frac{1}{\br{\pX,p_A}}{-}\alpha\right)\left(p_A{\cdot}\lambda_{\X}\right)\right].
}
The Jacobian of the bubble cut in this parametrization is simply $J\!=\!\beta$, while the projection condition $\br{\ell_a,\pX}\!=\!0$ has two solutions, $\lambda_{\ell_a}\!\sim\!\lambda_{\X}$ and $\tilde\lambda_{\ell_a}\!\sim\!\tilde\lambda_{\X}$, which correspond to $\beta\!\rightarrow\!0$ and $\beta\!\rightarrow\!\infty$, respectively. Our bubble contour prescription amounts to evaluating (\ref{eq:bubble_cut_in_field_thy}) on (\ref{eq:bubble_param}), taking the residue at either $\beta\!\rightarrow\!0$ or $\beta\!\rightarrow\!\infty$, and extracting the coefficient of the double-pole at $\alpha\!\rightarrow\!\infty$. We define the bubble leading singularity to be the even combination of these two field-theory evaluations, which are precisely the two terms appearing in (\ref{eq:bubble_coeffs_mhv}). 

The basis of integrands and the collection of non-vanishing coefficients in (\ref{eq:box_coeffs_mhv1}), (\ref{eq:box_coeffs_mhv2}), (\ref{eq:tri_coeffs_mhv2}), (\ref{eq:tri_coeffs_mhv2}) and (\ref{eq:bubble_coeffs_mhv}), together with a prescription for the massless bubble coefficients, constitutes the MHV one-loop amplitude integrand in the form of (\ref{eq:schematic_integrand}). Combining all terms, one can (numerically) check that the $\pX$ dependence drops out of the integrand via a nontrivial cancellation between all terms.

Using the tabulated integration rules found in appendix~\ref{app:integral_results_details}, we find the $n$-point MHV integral to be of the form,
\eq{\begin{split}\int\!\!\!\dbar^{4{-}2\epsilon}\ell\,\,\calA_{n,1\text{(-loop)}}^{(i,j)}\equivL{-}\calA_{n,0}^{(i,j)}&\Bigg[n\left(\frac{1}{\epsilon^2}{+}\frac{1}{\epsilon}\log(\mu^2){+}\frac{1}{2}\log(\mu^2)^2\right)\\
&\,\,{+}\left(\frac{1}{\epsilon}{+}\log(\mu^2){+}2\right)\left((4{-}\calN){-}\sum_{a=1}^n\log(s_{a,a+1})\right)\\
&\,\, {+}\hat{\calA}_{n,1}^{(i,j)}\Bigg]{+}\mathcal{O}(\epsilon)\,.\end{split}\label{one_loop_mhv_amps}
}
Here, the expression $\hat{\mathcal{A}}_{n,1}^{(i,j)}$ is implicitly defined to be the UV- and IR-finite part of the one-loop amplitude divided by the tree amplitude. 

It is worth remarking that the while the expression in (\ref{one_loop_mhv_amps}) is \emph{correct}, it is not entirely manifest in our representation. In particular, the expression on the second line does not follow \emph{manifestly} from the basis we have constructed. Nevertheless, we have explicitly checked its correctness.

\subsection{\emph{Exempli Gratia}: a Six-Point NMHV Amplitude Integrand}
\label{subsec:nmhv_6point}\vspace{0pt}
As another example of prescriptive unitarity with bubble power-counting, we consider the six-particle split-helicity NMHV amplitude integrand with particles $\{4,5,6\}$ to be those related by supersymmetry generators to negative helicity states. 

First, it is easy to see that for the particular helicity configuration we've considered, \emph{every} non-vanishing box diagram is of the singlet type. This implies that the box coefficients are given by extracting the $(\tilde{\eta}_4)^4(\tilde{\eta}_5)^4(\tilde{\eta}_6)^4$ component of the $R$-invariants appearing in the $\calN{=}4$ superamplitude. 

Just as in the MHV example discussed above, the coefficients of the one-mass scalar triangles is always the tree amplitude, $\mathcal{A}_{6,\text{1-loop}}^{(4,5,6)}$. It turns out that the non-vanishing chiral triangle and bubble coefficients, can all be expressed compactly in terms of the following two superfunctions ($R$-invariants)
\eq{\begin{split}f_1^{}
\equivR&\frac{\asb{6|p^{}_{45}|3}^{\fwboxL{0pt}{4{-}\calN}}\fwboxL{0pt}{\hspace{20pt}}}{\ab{12}\sb{45}\asb{2|p^{}_{34}|5}s_{345}\asb{6|p^{}_{45}|3}\ab{61}\sb{34}}\,\,\delta^{3\times\calN}\!\big(\!C_1\!\cdot\!\tilde{\eta}\big)\,\,\delta^{2\times2}\!\big(\lambda\!\cdot\!\tilde\lambda\big)\,,\\
f_3^{}
\equivR&\frac{\asb{4|p^{}_{56}|1}^{\fwboxL{0pt}{4{-}\calN}}\fwboxL{0pt}{\hspace{20pt}}}{\ab{34}\sb{61}\asb{4|p^{}_{56}|1}s_{561}\asb{2|p^{}_{16}|5}\ab{23}\sb{56}}\,\,\delta^{3\times\calN}\!\big(\!C_3\!\cdot\!\tilde{\eta}\big)\,\,\delta^{2\times2}\!\big(\lambda\!\cdot\!\tilde\lambda\big)\,,
\end{split}}
\eq{
\hspace{-.7cm}
\text{where }C_1\equivR\left(\begin{array}{@{}c@{}c@{}c@{}c@{}c@{}c@{}}\lambda_1^1&\lambda_2^1&\lambda_3^1&\lambda_4^1&\lambda_5^1&\lambda_6^1\\
\lambda_1^2&\lambda_2^2&\lambda_3^2&\lambda_4^2&\lambda_5^2&\lambda_6^2\\
\fwbox{20pt}{0}&\fwbox{20pt}{0}&\fwbox{20pt}{\sb{45}}&\fwbox{20pt}{\sb{53}}&\fwbox{20pt}{\sb{34}}&\fwbox{20pt}{0}\end{array}\right), \quad \!\! C_3\equivR\left(\begin{array}{@{}c@{}c@{}c@{}c@{}c@{}c@{}}\lambda_1^1&\lambda_2^1&\lambda_3^1&\lambda_4^1&\lambda_5^1&\lambda_6^1\\
\lambda_1^2&\lambda_2^2&\lambda_3^2&\lambda_4^2&\lambda_5^2&\lambda_6^2\\
\fwbox{20pt}{\sb{56}}&\fwbox{20pt}{0}&\fwbox{20pt}{0}&\fwbox{20pt}{0}&\fwbox{20pt}{\sb{61}}&\fwbox{20pt}{\sb{15}}\end{array}\right).
\hspace{-.6cm}
}
In terms of these two superfunctions, we find that the non-vanishing non-singlet cuts for this amplitude give rise to the following non-vanishing coefficients:
\eq{\begin{split}
\fwboxL{300pt}{\hspace{-60pt}\sixPointNonSingletTriA\bigger{\Rightarrow}\;\;a_{3,4,\{5,6,1,2\}}^{2}\equivR(4{-}\calN)\frac{\br{p_3{-}p_4,\pX}}{\ab{\X3}\sb{4\X}}\!\!\left(\!\!f_1^{}\frac{\asb{6|p_{12}|4}}{\asb{6|p_{12}|3}}{-}f_3^{}\frac{\asb{3|p^{}_{56}|1}}{\asb{4|p^{}_{56}|1}}\right)} 
\\[-10pt]
\fwboxL{300pt}{\hspace{-60pt}\sixPointNonSingletTriB\bigger{\Rightarrow}\;\;a_{6,1,\{2,3,4,5\}}^{3}\equivR(4{-}\calN)\frac{\br{p_6{-}p_1,\pX}}{\ab{\X1}\sb{6\X}}\!\!\left(\!\!f_1^{}\frac{\asb{1|p^{}_{45}|3}}{\asb{6|p^{}_{45}|3}}{-}f_3^{}\frac{\asb{4|p_{23}|6}}{\asb{4|p_{23}|1}}\right)}
\end{split}}
\eq{\begin{split}
\fwboxL{300pt}{\hspace{-60pt}\sixPointNonSingletTriC\bigger{\Rightarrow}\;\;a_{2,\{3,4\},\{5,6,1\}}^{2}\equivR(4{-}\calN)f_3\frac{1}{2}\frac{\ab{2\,4}\asb{2|p^{}_{34}|1}\sb{\X2}}{\asb{2|p^{}_{34}|\X}\asb{4|p^{}_{56}|1}}}\\[-10pt]
\fwboxL{300pt}{\hspace{-60pt}\sixPointNonSingletTriD\bigger{\Rightarrow}\;\;a_{2,\{3,4,5\},\{6,1\}}^{2}\equivR(4{-}\calN)f_1\frac{1}{2}\frac{\ab{2\,6}\asb{2|p^{}_{16}|3}\sb{\X2}}{\asb{2|p^{}_{16}|\X}\asb{6|p^{}_{45}|3}}}
\end{split}}
\eq{\begin{split}
\fwboxL{300pt}{\hspace{-60pt}\sixPointNonSingletTriE\bigger{\Rightarrow}\;\;a_{5,\{6,1\},\{2,3,4\}}^{3}\equivR(4{-}\calN)f_3\frac{1}{2}\frac{\sb{1\,5}\asb{4|p^{}_{16}|5}\ab{\X5}}{\asb{4|p^{}_{56}|1}\asb{\X|p^{}_{16}|5}}}\\
\fwboxL{300pt}{\hspace{-60pt}\sixPointNonSingletTriF\bigger{\Rightarrow}\;\;a_{5,\{6,1,2\},\{3,4\}}^{3}\equivR(4{-}\calN)f_1\frac{1}{2}\frac{\sb{3\,5}\asb{6|p^{}_{34}|5}\ab{\X5}}{\asb{6|p^{}_{45}|3}\asb{\X|p^{}_{34}|5}}}
\end{split}}
%
%
\eq{\begin{split} 
\fwboxL{300pt}{\hspace{-60pt}\sixPointNonSingletBubA\bigger{\Rightarrow}\;\;a_{\{2,3,4\},\{5,6,1\}}\equivR \frac{(4{-}\calN)f_3}{\asb{4|p_{561}|1}}\Big[
\frac{\ab{24}\langle 2 | [p_{561},\pX] p_{561}|1]}{\langle 2|p_{561}\pX|2\rangle}} \\
\fwboxR{300pt}{+\frac{\asb{4|p_{561}|5}\, [1|[\pX,p_{561}]|5]}{[5|\pX p_{561}|5]}\Big] }\\[-10pt]
\fwboxL{300pt}{\hspace{-60pt}\sixPointNonSingletBubB\bigger{\Rightarrow}\;\;a_{\{3,4,5\},\{6,1,2\}}\equivR \frac{(4{-}\calN)f_1}{\asb{6|p_{345}|3}}\Big[
\frac{\ab{26}\langle 2 | [p_{345},\pX] p_{345}|3]}{\langle 2|p_{345} \pX|2\rangle}}\\ 
\fwboxR{300pt}{+\frac{\asb{6|p_{345}|5}\, [5|[\pX,p_{345}]|3]}{[5|\pX p_{345}|5]}}
\Big]
\end{split}}
where we introduced a `commutator' $\langle a|[p_A,p_B]p_C|b]\equivR\langle a|p_A p_B p_C|b]{-}\langle a|p_B p_A p_C|b]$ to write more compact expressions for the bubble-integrand coefficients. The final two non-zero bubble massive bubble coefficients are:
\eq{\begin{split}
\hspace{-15pt}\sixPointNonSingletBubC & \bigger{\Rightarrow}\;\;a_{\{3,4\},\{5,6,1,2\}}\equivR \\ (4{-}\calN)\Bigg\{ &\left[\frac{\asb{6|p_{12}|4}\br{p_3{-}p_4,\X}}{\ab{\X3}\sb{4\X}}{-}\frac{\asb{6|p_{34}|5}\, [3|[\pX,p_{34}]|5]}{[5|\pX p_{34}|5]}\right] \frac{f_1}{\asb{6|p_{45}|3}} \\
+& \left[\frac{\asb{3|p_{56}|1}\br{p_3{-}p_4,\X}}{\ab{\X3}\sb{4\X}}
{-}\frac{\asb{2|p_{34}|1}\,\langle 2|[\pX,p_{34}]|4\rangle}{\langle 2|p_{34}\pX|2\rangle}\right] \frac{f_3}{\asb{4|p_{56}|1}}
\Bigg\}
\end{split}}
\eq{\begin{split}
\hspace{-9pt}\sixPointNonSingletBubD &\bigger{\Rightarrow}\;\;a_{\{6,1\},\{2,3,4,5\}}\equivR \\ (4{-}\calN)\Bigg\{&\left[\frac{\asb{1|p_{45}|3}\br{p_6{-}p_1,\X}}{\ab{\X6}\sb{1\X}}{+}\frac{\asb{2|p_{16}|3}\, \langle 2| [\pX,p_{16}]|6\rangle}{\langle 2|p_{16} \pX|2\rangle}
\right]\frac{f_1}{\asb{6|p_{12}|3}} \\
&+ \left[\frac{\asb{4|p_{23}|6}\br{p_6{-}p_1,\X}}{\sb{\X6}\ab{1\X}}
{+}\frac{\asb{4|p_{16}|5}\, [1|[\pX,p_{16}]|5]}{[5|\pX p_{16}|5]}
\right]
\frac{f_3}{\asb{4|p_{23}|1}}
\Bigg\}
\end{split}}
Plugging these coefficients into the expansion of the amplitude in eq.~(\ref{eq:schematic_integrand}), we obtain the integrand for the six-point split-helicity NMHV amplitude integrand with particles $4,5$ and $6$ being related by supersymmetry to negative helicity gluons.

\subsection{Finite Observables at One Loop}
\label{subsec:finiteness_of_ratio_functions}\vspace{0pt}

It is widely appreciated that four-dimensional scattering amplitudes for massless particles are problematic due to the presence of long-distance (infrared) divergences associated to low energy (soft) or unresolved collinear radiation, see e.g.~\cite{Sterman:1993hfp}. For inclusive enough physical observables such as cross-sections, all such divergences cancel when real radiation effects are taken into account consistently as a consequence of the KLN theorem \cite{Kinoshita:1962ur,Lee:1964is} in QED and its generalizations. Another example of an IR-finite observable is the energy-energy correlation function, see e.g.~\cite{Basham:1978bw,Basham:1978zq}. The IR structure of general gauge theories is still an important subject of current study; both formally (see e.g.~\cite{Strominger:2013lka,Strominger:2017zoo}) as well as phenomenologically in the form of efficient IR subtraction schemes for high-precision predictions for collider observables \mbox{\cite{Giele:1993dj,Frixione:1995ms,Catani:1996vz,Gehrmann-DeRidder:2005btv,Somogyi:2006da}}. 

From an amplitudes perspective, it is possible to determine which diagrams can contribute to IR divergences and which ones remain finite. This analysis amounts to investigating all soft and collinear regions of a given diagram, taking into account potential numerator factors that can dampen IR singularities. It turns out that the situation is especially simple for one-loop integrals where one can easily account for all possible singular regions which suffices for the present discussion. For general gauge theories, the infrared structure has been completely understood up to two-loop order by Catani \cite{Catani:1998bh} with numerous subsequent progress, see e.g.~\cite{Kidonakis:1998nf,Aybat:2006mz,Dixon:2008gr,Gardi:2009qi,Becher:2009cu,Almelid:2015jia}.

The universality of IR divergences of gauge theory scattering amplitudes at one loop implies that all divergences should be proportional to the tree amplitude. Together with the requirement that UV divergences in a renormalizable gauge theory should be canceled by appropriate counter terms also implies that the one-loop UV divergences is also proportional to the tree-level amplitude. Motivated by this discussion, we can organize the $n$-particle one-loop amplitude in the following form:
\eq{
\label{eq:1loop_div_fin_split}
\begin{split}
\calA_{n,1}&\equivL
\calA_{n,1}^{\text{fin}}
{+}\calA_{n,0}\big(
        \calI_{\text{div}}^{\text{UV}}
     {+}\calI_{\text{div}}^{\text{IR}}
    \big)
\\
&\equivL\calA_{n,0}\Big(
    \hat{\calA}^{\text{fin}}_{n,1}
    {+}\calI_{\text{div}}^{\text{UV}}
    {+}\calI_{\text{div}}^{\text{IR}}\Big)\,
\quad\text{with}\quad
\hat{\calA}_{n,1}\equivR\calA_{n,1}/\calA_{n,0}\,,
\end{split}}
where we suppress the explicit helicity-labels of the (super-)amplitudes as well as the MHV-degree $k$. The universality of IR divergences is more general than the specific one-loop example discussed above and is encoded in the following factorization formula (see e.g.~\cite{Sterman:1995fz,Almelid:2015jia}) for massless parton scattering amplitudes
\begin{align}
    \calA_n(\{p_i\},\alpha_s) = Z_n(\{p_i\},\mu,\alpha_s) \calA^{\text{fin}}_n(\{p_i\},\mu,\alpha_s)
\end{align}
where all IR singularities are factorized in $Z_n$ in the form of poles in dimensional regularization $\epsilon\!=\!(D-4)/2$. The above equation depends on a factorization scale $\mu$ and the running coupling constant $\alpha_s\equivR\alpha_s(\mu^2)$. 

Motivated by this decomposition, it is natural to introduce the IR-finite ratio function where the universal IR singularities cancel. A priori, we can write the ratio of two $n$-point amplitudes $\calA^{(a)}_n$, and $\calA^{(b)}_n$ to all orders in perturbation theory:
\begin{align}
    \mathcal{P}^{(a,b)}_n = \frac{\calA^{(a)}_n}{\calA^{(b)}_n}.
\end{align}

In maximally supersymmetric theories, there is only a single independent super amplitude for a given N$^{(k{-}2)}$MHV sector and one takes IR finite ratios between amplitudes of different $k$ charge. In this case, the labels `$a$' and `$b$' denote the respective $k$-charge of the amplitudes and it is common to always divide by the $k\!=\!2$ MHV amplitude and denote the resulting ratio function by $\mathcal{P}^{(k)}_n$. The IR-finiteness of $\mathcal{P}^{(k)}_n$ underlies several important features of the integrated results for the maximally supersymmetric theory, including dual conformal invariance \cite{Drummond:2006rz,Bern:2006ew,Bern:2007ct,Alday:2007hr,Drummond:2008vq}. These simplifications, together with a number of conceptual and technological advances enabled Dixon and collaborators to obtain function level results to very high loop order, see e.g.~\cite{Dixon:2011pw,Caron-Huot:2019vjl,Dixon:2020cnr}.

For the $\calN\!=\!1,2$ supersymmetric amplitudes under consideration, we can furthermore take nontrivial ratios of (super-) amplitudes within the same N$^{(k-2)}$MHV $k$ sector due to the distinction between the positive and negative helicity gluon supermultiplet and write e.g. 
\eq{
\mathcal{P}^{(2)}_4 =\frac{\calA^{(2)}_4(1^-,2^-,3^+,4^+)}
                          {\calA^{(2)}_4(1^-,2^+,3^-,4^+)}
                    \equivL \frac{\calA^{(1,2)}_4}{\calA^{(1,3)}_4}      \,,
}
where the $\pm$ labels the relevant supermultiplet of particle $i$. We omit labeling the ratios by the individual helicities of the contributing amplitudes to avoid cluttering the equations and introduced the shorthand notation $\calA^{(i,j)}_n$ for MHV amplitudes to indicate the position of the negative helicity supermultiplets. 

All ratios can be expanded perturbatively in the coupling constant $g$ and yield IR-finite quantities at each order in perturbation theory, e.g. up to two-loop order we find 
\begin{align}
\label{eq:ratio_fct_pert_exp}
\hspace{-.5cm}
    \mathcal{P}^{(a,b)}_n 
    &\equivL \frac{\calA^{(a)}_{n,0} {+} \alpha_s \calA^{(a)}_{n,1}
               {+}\alpha^2_s\calA^{(a)}_{n,2} + \calO(\alpha^3_s)}
        {\calA^{(b)}_{n,0} {+} \alpha_s \calA^{(b)}_{n,1}
               {+} \alpha^2_s \calA^{(b)}_{n,2} + \calO(\alpha^3_s)}
     = \mathcal{P}^{(a,b)}_{n,0} {+} \alpha_s \, \mathcal{P}^{(a,b)}_{n,1} 
      {+} \alpha^2_s \, \mathcal{P}^{(a,b)}_{n,2} {+} \calO(\alpha^3_s) 
    \nonumber \\
    & = \frac{\calA^{(a)}_{n,0}}{\calA^{(b)}_{n,0}} 
      + \alpha_s \frac{\left(
                    \calA^{(b)}_{n,0}\,\calA^{(a)}_{n,1}
                 {-}\calA^{(a)}_{n,0}\,\calA^{(b)}_{n,1}
                   \right)}
                 {\left[\calA^{(b)}_{n,0}\right]^2} +
    \\             
    & + \alpha^2_s 
    \frac{\left( \left[\calA^{(b)}_{n,0}\right]^2    \calA^{(a)}_{n,2}
                {-} \calA^{(b)}_{n,0}\, \calA^{(a)}_{n,0}\, \calA^{(b)}_{n,2}
                {+} \calA^{(a)}_{n,0} \left[\calA^{(b)}_{n,1}\right]^2
                {-} \calA^{(b)}_{n,0}\, \calA^{(b)}_{n,1}\, \calA^{(a)}_{n,1}
         \right)}{\left[\calA^{(b)}_{n,0}\right]^3}
    {+} \calO(\alpha^3_s)\,, \nonumber
\hspace{-.5cm}
\end{align}
where we indicate the loop order of various quantities by an additional subscript. The formulae for the ratio of amplitudes in the same MHV sector follow trivially from the above results. In the presentation above, the various factors of the tree-level amplitudes $\calA^{(b)}_{n,0}$ and $\calA^{(a)}_{n,0}$ ensure a uniform helicity weight of all terms in the perturbatively expanded form version of the ratio function. It is often convenient to divide out certain helicity-dependence by removing the tree-level amplitude and work instead with $\hat{\calA}_{n}$ to define
\begin{align}
\label{eq:ratio_fct_pert_exp_hat}
\hspace{-.5cm}
\begin{split}
    \hat{\mathcal{P}}^{(a,b)}_n 
    & = \frac{1 
    {+}\alpha_s \hat{\calA}^{(a)}_{n,1}
    {+}\alpha^2_s\hat{\calA}^{(a)}_{n,2} 
    {+}\calO(\alpha^3_s)}
    {1 
    {+}\alpha_s \hat{\calA}^{(b)}_{n,1}
    {+} \alpha^2_s \hat{\calA}^{(b)}_{n,2} 
    {+}\calO(\alpha^3_s)}
     = \hat{\mathcal{P}}^{(a,b)}_{n,0} 
      {+} \alpha_s  \, \hat{\mathcal{P}}^{(a,b)}_{n,1} 
      {+} \alpha^2_s\, \hat{\mathcal{P}}^{(a,b)}_{n,2} {+} \calO(\alpha^3_s) 
    \\
    & = 1
      {+}\alpha_s  \left(\hat{\calA}^{(a)}_{n,1}{-}\hat{\calA}^{(b)}_{n,1}\right) 
      {+}\alpha^2_s\left(\hat{\calA}^{(a)}_{n,2}{-}\hat{\calA}^{(b)}_{n,2} 
                         {+}\left[\hat{\calA}^{(b)}_{n,1}\right]^2\!\! 
                         {-}\hat{\calA}^{(a)}_{n,1}\hat{\calA}^{(b)}_{n,1}
                    \right)
    {+} \calO(\alpha^3_s)\,.
\end{split}
\hspace{-.5cm}
\end{align}

At one-loop, the IR and UV finiteness of the ratio function is easy to see. From general expectations (and confirmed by our explicit calculation below), both the UV- and IR-divergent parts of the one-loop amplitudes must be proportional to the tree-level amplitudes as in (\ref{eq:1loop_div_fin_split}). Working with the rescaled quantities, we see that the universal factor $\calI^{\text{IR/UV}}_{\text{div}}$ cancels in the difference (\ref{eq:ratio_fct_pert_exp_hat}). Similar arguments also lead to the finiteness of the higher-loop ratio functions.

We may illustrate how this works for the simplest example involving four particles. Before taking the ratios, we give the integrated results for the individual amplitudes ($\calN=1,2$)
\begin{align}
  \hat{\calA}_{4,1}^{(1,2)} & = 
    \frac{4}{\epsilon^2} 
    -\frac{1}{\epsilon}\left[(4-\calN) + 2 \log \frac{s}{\mu^2} + 2 \log \frac{t}{\mu^2}\right] 
    \\& 
    - 2(4-\calN) + \pi^2 + \log^2\frac{s}{t} + (4-\calN) \log\frac{t}{\mu^2} + \log^2\frac{s}{\mu^2}+\log^2\frac{t}{\mu^2}
    \nonumber\\
    \hat{\calA}_{4,1}^{(1,3)} & = 
    \frac{4}{\epsilon^2} 
    -\frac{1}{\epsilon}\left[(4-\calN) + 2 \log \frac{s}{\mu^2} + 2 \log \frac{t}{\mu^2}\right] 
    \\ &
    -\frac{(4-\calN)s}{u}\left[-2+\log \frac{t}{\mu^2}\right] 
    -\frac{(4-\calN)t}{u}\left[-2+\log \frac{s}{\mu^2}\right] 
    \nonumber \\&
    + \frac{s^{4-\calN}+t^{4-\calN}+(-u)^{4-\calN}}{2(-u)^{4-\calN}}\left[\pi^2 + \log^2\frac{s}{t}\right]
    + \log^2\frac{s}{\mu^2} + \log^2\frac{t}{\mu^2}\nonumber
\end{align}
in terms of the usual Mandelstam variables $s\equivR(p_1{+}p_2)^2$,$t\equivR(p_2{+}p_3)^2$,$u\equivR(p_1{+}p_3)^2$. In the one-loop ratio function of (\ref{eq:ratio_fct_pert_exp_hat}), we are supposed to take the difference of the two amplitudes. Both the UV and IR divergences cancel in this difference and we find for $\calN=1,2$ that
\begin{align}
  \hat{\calA}_{4,1}^{(1,2)} - \hat{\calA}_{4,1}^{(1,3)} & =
  \frac{(4-\calN)\, t}{2 u^2} \left[
    s \left(\pi^2 + \log^2\frac{s}{t}\right)
   + 2 u \log \frac{s}{t}
   \right]\,,
\end{align}
where standard Mandelstam invariants $s,t,u$ satisfy $s+t+u=0$. As advertised, this result is IR and UV finite, but of mixed transcendental weight. Compared to the individual amplitudes, the ratio is considerably simpler and does not depend on the dimensional-regularization scale $\mu^2$ anymore.  

Going to higher point is also feasible by inserting the integral values for each of our basis integrands that are summarized in Table \ref{basisIntegralsTable} of appendix \ref{app:integral_results_details}. At five points, the results depend on five independent Mandelstam invariants which leads to more complicated looking results. Since all ingredients are provided with this work, we refrain from writing explicit results here. In general, however, the fact that these ratio functions are UV- and IR-finite follows directly from the general form (\ref{one_loop_mhv_amps}).

\section{General Discussion \& Future Directions}
\label{sec:discussion}

In this paper, we computed one-loop amplitude integrands in color-ordered less-than-maximally supersymmetric ($1\!\leq\!\calN\!<\!4$) Yang-Mills theory (`\symN\!') in the context of generalized unitarity. We constructed a \emph{prescriptive} bubble power-counting integrand basis, and showed how the coefficients of MHV and NMHV amplitudes can be calculated using contour integrals that are dual to that basis. 

While the box, triangle, and massive bubble integral coefficients can be extracted in a standard manner, there is an important subtlety in the case of massless bubbles. This topology is traditionally ignored in unitarity-based approaches due to the fact that scalar massless bubble integrals evaluate to zero in dimensional regularization. In contrast, in this work, it was our primary objective to construct a well-defined \emph{integrand}. This forces  us to specify a prescription for the massless bubble coefficients as well. Here, we have presented two distinct possibilities that appear well motivated from field theory and on-shell function considerations: (a) choose collinear cuts or (b) choose singlet double cuts which are the same for any amount of supersymmetry, including $\calN{=}4$ where these cuts are unambiguously defined. In the first scenario, the massless bubble coefficients are fixed to be tree-level amplitudes. The resulting integrand correctly reproduces both the expected IR and UV divergences upon integration. In the second scenario, we get zero coefficients for the massless bubbles and the integrand has improved behavior at infinity on singlet cuts. While both approaches are justified, each exhibits a different structure for the resulting integrands for amplitudes. We leave it to future work to investigate which of the two directions is preferred from the point of view of defining the \emph{unique} $\calN{<}4$ sYM integrand beyond one loop. 

Having a unique integrand is essential for the formulation of loop-level recursion relations (see e.g.~\cite{Benincasa:2016awv}), or attempts to reproduce it as a certain differential form on a positive geometry. Therefore, our work is a crucial first stepping stone for a possible extension of amplituhedron-like geometric objects \cite{Arkani-Hamed:2013jha} beyond planar maximally supersymmetric Yang-Mills theory.  

As a key extension of our work, it remains to construct a bubble power-counting basis at two-loops, and expand the $n$-point two-loop integrands in planar $\calN{<}4$ sYM theory in this basis. This would push the computational frontier for integrands in less-supersymmetric Yang-Mills theory, and it would provide valuable theoretical data for further investigation. 

Finally, the most difficult and important question is the extension of our work to pure Yang-Mills theory. This requires addressing the problem of tadpole integrals and rational terms.  

\subsubsection*{Acknowledgements}%
\vspace{-4pt}
\noindent This project has been supported by an ERC Starting Grant \mbox{(757978)}, a grant from the Villum Fonden \mbox{(15369)}, and by a grant from the US Department of Energy \mbox{(DE-SC00019066)} (JLB; CL; KP). JT and MZ are supported by the Department of Energy grant \mbox{(DE-SC0009999)}. EH is supported by the U.S. Department of Energy (DOE) under Award Number \mbox{DE-SC0009937}.

\newpage

\appendix
\section{
\texorpdfstring{Complete Bubble Power-Counting Integrand Basis $\mathfrak{B}_2^{(4)}$}
{Complete Bubble Power-Counting Integrand Basis}}
\label{integrand_basis_details}
Following the general strategy of prescriptive unitarity, constructing a bubble power-counting basis of integrands requires the specification of a spanning set of contours $\{\Omega_j\}$. Once this is done, diagonalization results in a basis such that $\oint_{\Omega_j}\mathcal{I}_i=\delta_{i,j}$. In this appendix, we give complete details regarding our choice of integration cycles $\{\Omega_j\}$, the integrands $\{\mathcal{I}_i\}$ to which they are dual, and the integrals that result.\\

\subsection{\texorpdfstring{Spanning-Set of Integration Contours Defining the Basis}{Spanning-Set of Integration Contours Defining the Basis}}
\label{app:cycle_basis}\vspace{-20pt}

\begin{table}[h!]\vspace{-0pt}$$\begin{array}{|@{}c@{}|}\hline
\rule[-40pt]{0.pt}{85pt}\fwbox{430pt}{\hspace{50pt}\fwboxR{0pt}{\Omega^i_{A,B,C,D}\equivR}\hspace{-5pt}\left\{\hspace{-5pt}\fourMassContourA\hspace{-10pt},\hspace{-0pt}\fourMassContourB\hspace{-5pt}\right\}\hspace{-2pt},\hspace{55pt}\fwboxR{0pt}{\Omega^i_{a,B,C,D}\equivR\hspace{-5pt}}\left\{\hspace{-5pt}\threeMassContourA\hspace{-10pt},\threeMassContourB\hspace{-5pt}\right\}\hspace{-5pt}}\\\hline
\rule[-40pt]{0.pt}{85pt}\fwbox{0pt}{\hspace{50pt}\fwboxR{0pt}{\Omega^i_{a,b,C,D}\equivR}\hspace{-5pt}\left\{\hspace{-5pt}\twoMassHardContourA\hspace{-10pt},\hspace{-0pt}\twoMassHardContourB\hspace{-5pt}\right\}\hspace{-2pt},\hspace{55pt}\fwboxR{0pt}{\Omega^i_{a,B,c,D}\equivR\hspace{-5pt}}\left\{\hspace{-5pt}\twoMassEasyContourA\hspace{-10pt},\twoMassEasyContourB\hspace{-5pt}\right\}\hspace{-5pt}}\\\hline\hline
\rule[-40pt]{0.pt}{85pt}\fwboxR{0pt}{\Omega^I_{A,B,C}\equivR}\left\{\threeMassTriContourA,\threeMassTriContourB,\threeMassTriContourC\right\}\\\hline
\rule[-40pt]{0.pt}{85pt}\fwboxR{0pt}{\Omega^I_{a,B,C}\equivR}\left\{\twoMassTriContourA,\twoMassTriContourB,\twoMassTriContourC\right\}\\\hline
\rule[-40pt]{0.pt}{85pt}\fwboxR{0pt}{\Omega^I_{a,b,C}\equivR}\left\{\oneMassTriContourA,\oneMassTriContourB,\oneMassTriContourC\right\}\\[5pt]\hline\hline
\rule[-30pt]{0.pt}{55pt}\fwboxR{0pt}{\Omega_{A,B}\equivR}\left\{\twoMassBubbleContour\right\},\hspace{57pt}\fwboxR{0pt}{\Omega_{a,B}\equivR}\left\{\zeroMassBubbleContour\right\}\\[10pt]\hline\multicolumn{1}{c}{}\\[-38pt]\end{array}$$\caption{\label{basisContourTable}A complete specification of contours to which the integrand basis $\mathfrak{B}^{(4)}_2$ is dual.}\vspace{-10pt}\end{table}
\vspace{\fill}\newpage

\vspace{-10pt}
\subsection{\texorpdfstring{Explicit Numerators for Basis Integrands in $\mathfrak{B}_{2}^{(4)}$}{Explicit Numerators for All Basis Integrands}}
\label{app:integrand_numerators}\vspace{-10pt}
%
\begin{table}[h!]\vspace{-30pt}$$\begin{array}{|@{}l@{}|@{}l@{}|}\multicolumn{1}{@{}l@{}}{\rule{75pt}{0pt}}&\multicolumn{1}{@{}c@{}}{\rule{352pt}{0pt}}\\\hline\fwbox{75pt}{\fourMassInts\rule[-32pt]{0pt}{70pt}}&\begin{array}{@{$\;$}l@{}l@{}l}\\[-15pt]\rule[-0pt]{0pt}{0pt}\mathfrak{n}^{i=1}_{A,B,C,D}&\equivR&\fwboxL{294pt}{\phantom{{-}}\br{p_A,\ell_b,\ell_c,p_C}{-}\frac{1}{2}s_{A\,B}s_{B\,C}\big(1{-}u{-}v{-}\Delta\big)}\\
&&{+}\frac{1}{2}\Big[\br{p_B,p_C}\ell_a^2{-}\br{p_{A\,B},p_C}\ell_b^2{-}\br{p_{B\,C},p_A}\ell_c^2{+}\br{p_B,p_A}\ell_d^2\Big]\\[9pt]
\hline\rule[-5pt]{0pt}{18pt}\mathfrak{n}^{i=2}_{A,B,C,D}&\equivR&{}\br{\ell_b,\ell_c,p_C,p_A}{-}\frac{1}{2}s_{A\,B}s_{B\,C}\big(1{-}u{-}v{-}\Delta\big)\\
&&{+}\frac{1}{2}\Big(\br{p_B,p_C}\ell_a^2{-}\br{p_{A\,B},p_C}\ell_b^2{-}\br{p_{B\,C},p_A}\ell_c^2{+}\br{p_B,p_A}\ell_d^2\Big)\end{array}\\\hline
\multicolumn{1}{@{}|@{}}{}&\text{{\footnotesize\,\,where $\Delta\equivR\!\sqrt{(1{-}u{-}v)^2{-}4\,u\,v}$, $u\equivR\! s_A\,s_C/(s_{A\,B}\,s_{B\,C})$, $v\equivR\! s_B\,s_D/(s_{A\,B}\,s_{B\,C})$}}\\\hline
\fwbox{75pt}{\threeMassInts\rule[-32pt]{0pt}{70pt}}&\begin{array}{@{$\;$}l@{}l@{}l}\\[-17pt]\rule[-16pt]{0.pt}{36pt}
\fwboxL{40pt}{\mathfrak{n}^{i=1}_{a,B,C,D}}&\equivR&\fwboxL{294pt}{\phantom{{-}}\br{p_a,\ell_b,\ell_c,p_C}{+}\frac{1}{2}\Big(\br{p_B,p_C}\ell_a^2{-}\br{p_{a,B},p_C}\ell_b^2\Big)}\\[10pt]
\hline\rule[-16pt]{0.0pt}{36pt}
\mathfrak{n}^{i=2}_{a,B,C,D}&\equivR&\phantom{{-}}\br{\ell_b,\ell_c,p_C,p_a}{+}\frac{1}{2}\Big(\br{p_B,p_C}\ell_a^2{-}\br{p_{a,B},p_C}\ell_b^2\Big)\end{array}\\\hline
\fwbox{75pt}{\twoMassHardInts\rule[-32pt]{0pt}{70pt}}&\begin{array}{@{$\;$}l@{}l@{}l}\\[-17pt]\rule[-16pt]{0.pt}{36pt}
\fwboxL{40pt}{\mathfrak{n}^{i=1}_{a,b,C,D}}&\equivR&\fwboxL{294pt}{\phantom{{-}}\br{p_a,\ell_b,\ell_c,p_C}{-}\frac{1}{2}\br{p_{a,b},p_C}\ell_b^2}\\[10pt]
\hline\rule[-16pt]{0.0pt}{36pt}
\mathfrak{n}^{i=2}_{a,b,C,D}&\equivR&\phantom{{-}}\br{\ell_b,\ell_c,p_C,p_a}{+}\frac{1}{2}\br{p_{a,b},p_C}\ell_b^2\end{array}\\\hline
\fwbox{75pt}{\twoMassEasyInts\rule[-32pt]{0pt}{70pt}}&\begin{array}{@{$\;$}l@{}l@{}l}\\[-17pt]\rule[-16pt]{0.pt}{36pt}
\fwboxL{40pt}{\mathfrak{n}^{i=1}_{a,B,c,D}}&\equivR&\fwboxL{294pt}{\phantom{{-}}\br{p_a,\ell_b,\ell_c,p_c}}\\[10pt]
\hline\rule[-16pt]{0.0pt}{36pt}
\mathfrak{n}^{i=2}_{a,B,c,D}&\equivR&\phantom{{-}}\br{\ell_b,\ell_c,p_c,p_a}\end{array}\\\hline\fwbox{75pt}{\threeMassTriInts\rule[-32pt]{0pt}{70pt}}&\begin{array}{@{$\;$}l@{}l@{}l}\\[-15pt]\rule[-10pt]{0.0pt}{18pt}\fwbox{40pt}{\mathfrak{n}^{I=1}_{A,B,C}}&\equivR&\fwboxL{294pt}{{-}\frac{1}{2}s_{C}\sqrt{(1{-}u{-}v)^2{-}4\,u\,v},\quad\quad\text{{\footnotesize where $u\equivR\! s_A/s_C,\,v\equivR\! s_B/s_C$}}}\\[5pt]
\hline\rule[-5pt]{0.0pt}{19pt}\fwbox{40pt}{\mathfrak{n}^{I=2}_{A,B,C}}&\equivR&{\color{black}\phantom{{-}}\frac{1}{2}\big(\br{p_A,\ell_a,p_C,\pX}{-}\br{\pX,p_A,\ell_a,p_C}\big)/\br{p_B,\pX}}\\[5pt]
\hline\rule[-5pt]{0pt}{19pt}\fwbox{40pt}{\mathfrak{n}^{I=3}_{A,B,C}}&\equivR&{\color{black}\phantom{{-}}\frac{1}{2}\big(\br{p_A,\ell_a,p_C,\pX}{+}\br{\pX,p_A,\ell_a,p_C}{+}s_A\br{p_C,\pX}
}\\
&&{\color{black}\hspace{20pt}{-}\ell_a^2\br{p_A{-}p_C,\pX}{-}\ell_b^2\br{p_C,\pX}{-}\ell_c^2\br{p_A,\pX}\big)/\br{p_A,\pX}}\end{array}\\\hline
\fwbox{75pt}{\twoMassTriInts\rule[-32pt]{0pt}{70pt}}&\begin{array}{@{$\;$}l@{}l@{}l}\\[-15pt]\rule[-10pt]{0.0pt}{18pt}\fwbox{40pt}{\mathfrak{n}^{I=1}_{a,B,C}}&\equivR&\fwboxL{294pt}{\phantom{{-}}s_B{-}s_C}\\[5pt]
\hline\rule[-5pt]{0.0pt}{19pt}\fwbox{40pt}{\mathfrak{n}^{I=2}_{a,B,C}}&\equivR&{\color{black}{-}\big(\br{\ell_a,\ell_b,p_B,\pX}{+}\br{\ell_b,\ell_a,p_C,\pX}\big)/\br{p_a,\pX}}\\[5pt]
\hline\rule[-5pt]{0pt}{19pt}\fwbox{40pt}{\mathfrak{n}^{I=3}_{a,B,C}}&\equivR&{\color{black}{-}\big(\br{\pX,\ell_a,\ell_b,p_B}{+}\br{\pX,\ell_b,\ell_a,p_C}\big)/\br{p_a,\pX}}\end{array}\\\hline
\fwbox{75pt}{\oneMassTriInts\rule[-32pt]{0pt}{70pt}}&\begin{array}{@{$\;$}l@{}l@{}l}\\[-15pt]\rule[-10pt]{0.0pt}{18pt}\fwbox{40pt}{\mathfrak{n}^{I=1}_{a,b,C}}&\equivR&\fwboxL{294pt}{{-}s_C}\\[5pt]
\hline\rule[-5pt]{0.0pt}{19pt}\fwbox{40pt}{\mathfrak{n}^{I=2}_{a,b,C}}&\equivR&{\color{black}{-}\frac{1}{2}\big(\br{\ell_a,\ell_a{+}\ell_b,p_b,\pX}{+}\br{\ell_b,\ell_a,p_C,\pX}\big)/\br{p_{A}{-}p_B,\pX}}\\[5pt]
\hline\rule[-5pt]{0pt}{19pt}\fwbox{40pt}{\mathfrak{n}^{I=3}_{a,b,C}}&\equivR&{\color{black}{-}\frac{1}{2}\big(\br{\pX,\ell_a,\ell_a{+}\ell_b,p_b}{+}\br{\pX,\ell_b,\ell_a,p_C}\big)/\br{p_{A}{-}p_B,\pX}}\end{array}\\\hline
\fwbox{75pt}{\twoMassBubbleIntBoth\rule[-20pt]{0pt}{45pt}}&\;\fwbox{40pt}{\mathfrak{n}^{}_{A,B}}\equivR \mathfrak{n}_{a,B}\;\equivR1/2\\\hline\multicolumn{2}{c}{}\\[-38pt]
\end{array}$$\caption{\label{basisNumerators}Basis integrand numerators for all box, triangle, and bubble integrands in $\mathfrak{B}^{(4)}_2$.}\vspace{-20pt}\end{table}
\vspace{\fill}

\newpage
\subsection{\texorpdfstring{Integrals of Basis Integrands in Dimensional Regularization}{Integrals of Basis Integrands in Dimensional Regularization}}
\label{app:integral_results_details}
\vspace{-5pt}

\begin{table}[h!]\vspace{-34pt}$$\begin{array}{|@{}l@{}|@{}l@{}|}\multicolumn{1}{@{}l@{}}{\rule{75pt}{0pt}}&\multicolumn{1}{@{}c@{}}{\rule{352pt}{0pt}}\\\hline\fwbox{75pt}{\fourMassInts\rule[-32pt]{0pt}{70pt}}&\begin{array}{@{$\;$}l@{}l@{}l}\\[-15pt]\rule[-10pt]{0.0pt}{28pt}\fwboxL{50pt}{\displaystyle\int\!\!\calI^{i=1,2}_{A,B,C,D}}&=&\fwboxL{289pt}{\text{{\normalsize$\displaystyle \phantom{{{-}}}\Li{1{-}\bar{u}}{+}\Li{1{-}\bar{v}}{-}\Li{1}{-}\log\!\left(1{-}\bar{u}\right)\log\!\left(1{-}\bar{v}\right)$}}}\\[-5pt]
&&{+}\frac{1}{2}\log(u)\log(v)\\[5pt]
&&\text{{\footnotesize $\bar{u}\equivR\!\frac{1}{2}(1{+}u{-}v{-}\Delta)$, $\bar{v}s\equivR\!\frac{1}{2}(1{-}u{+}v{-}\Delta)$, $\Delta\equivR\!\sqrt{(1{-}u{-}v)^2{-}4\,u\,v}$}}\\[-2pt]
&&\text{{\footnotesize with $u\equivR\! s_A\,s_C/(s_{A\,B}\,s_{B\,C})$ and $v\equivR\! s_B\,s_D/(s_{A\,B}\,s_{B\,C})$}}\end{array}\\\hline
\fwbox{75pt}{\threeMassInts\rule[-32pt]{0pt}{70pt}}&\begin{array}{@{$\;$}l@{}l@{}l}\\[-13pt]\rule[-20pt]{0.0pt}{28pt}\fwboxL{50pt}{\displaystyle\int\!\!\calI^{I=1,2}_{a,B,C,D}}&=&\fwboxL{289pt}{\text{{\normalsize$\displaystyle\phantom{{-}}\Li{1{-}\frac{s_B\,s_D}{s_{a\,B}\,s_{B\,C}}}{-}\Li{1{-}\frac{s_B}{s_{a\,B}}}{-}\Li{1{-}\frac{s_D}{s_{B\,C}}}$}}}\\
&&\displaystyle{+}\frac{1}{2}\left[\log\!\left(\frac{s_B}{s_{a\,B}}\right)\log\!\left(\frac{s_C}{s_{B\,C}}\right){+}\log\!\left(\frac{s_C}{s_{a\,B}}\right)\log\!\left(\frac{s_D}{s_{B\,C}}\right)\right]\end{array}\\\hline
\fwbox{75pt}{\twoMassHardInts\rule[-32pt]{0pt}{70pt}}&\begin{array}{@{$\;$}l@{}l@{}l}\\[-13pt]\rule[-20pt]{0.0pt}{28pt}\fwboxL{50pt}{\displaystyle\int\!\!\calI^{I=1,2}_{a,b,C,D}}&=&\fwboxL{289pt}{\text{{\normalsize$\displaystyle{-}\Li{1{-}\frac{s_C}{s_{b\,C}}}{-}\Li{1{-}\frac{s_D}{s_{b\,C}}}$}}}\\
&&\displaystyle{-}\frac{1}{2}\left[\log\!\left(\frac{s_D}{s_{b\,C}}\right)\log\!\left(\frac{s_{a\,b}}{s_{b\,C}}\right){-}\frac{1}{2}\log\!\left(\frac{s_D}{s_{a\,b}}\right)\log\!\left(\frac{s_D}{s_{b\,C}}\right)\right]\end{array}\\\hline
\fwbox{75pt}{\twoMassEasyInts\rule[-32pt]{0pt}{70pt}}&\begin{array}{@{$\;$}l@{}l@{}l}\\[-13pt]\rule[-20pt]{0.0pt}{28pt}\fwboxL{50pt}{\displaystyle\int\!\!\calI^{I=1,2}_{a,b,C,D}}&=&\fwboxL{289pt}{\text{{\normalsize$\displaystyle\phantom{{-}}\Li{1{-}\frac{s_{B}\,s_D}{s_{a\,B}\,s_{B\,c}}}{-}\Li{1{-}\frac{s_B}{s_{B\,c}}}{-}\Li{1{-}\frac{s_B}{s_{a\,B}}}$}}}\\
&&\displaystyle{-}\Li{1{-}\frac{s_{D}}{s_{B\,c}}}{-}\Li{1{-}\frac{s_{D}}{s_{a\,B}}}{-}\frac{1}{2}\log\!\left(\frac{s_{a\,B}}{s_{B\,c}}\right)^2\end{array}
\\\hline\fwbox{75pt}{\threeMassTriInts\rule[-32pt]{0pt}{70pt}}&\begin{array}{@{$\;$}l@{}l@{}l}\\[-15pt]\rule[-10pt]{0.0pt}{28pt}\fwboxL{45pt}{\displaystyle\int\!\!\calI^{I=1}_{A,B,C}}&=&\fwboxL{294pt}{\text{{\footnotesize$ \phantom{{\,}}\Li{1{-}\bar{u}}{+}\Li{1{-}\bar{v}}{-}\Li{1}{-}\log\!\left(1{-}\bar{u}\right)\log\!\left(1{-}\bar{v}\right){+}\frac{1}{2}\log(u)\log(v)$}}}\\[-8pt]
&&\text{{\footnotesize $\bar{u}\equivR\!\frac{1}{2}(1{+}u{-}v{-}\Delta)$, $\bar{v}\equivR\!\frac{1}{2}(1{-}u{+}v{-}\Delta)$, $\Delta\equivR\!\sqrt{(1{-}u{-}v)^2{-}4\,u\,v}$}}\\[-2pt]
&&\text{{\footnotesize with $u\equivR\!s_A/s_C$ and $v\equivR\!s_B/s_C$}}\\[0pt]
\hline\rule[-12pt]{0pt}{30pt}\fwboxL{45pt}{\displaystyle\int\!\!\calI^{I=2}_{A,B,C}}&=&\phantom{{\,}}\mathcal{O}(\epsilon),\quad\displaystyle {\color{black}\int\!\!\calI^{I=3}_{A,B,C}={-}\frac{1}{\epsilon}{+}\frac{1}{2}\log\!\left(\frac{s_A\,s_B}{\mu^4}\right){-}2{+}\mathcal{O}(\epsilon)}\end{array}\\\hline
\fwbox{75pt}{\twoMassTriInts\rule[-32pt]{0pt}{50pt}}&\begin{array}{@{$\;$}l@{}l@{}l}\\[-15pt]\rule[-20pt]{0.0pt}{28pt}\fwboxL{45pt}{\displaystyle\int\!\!\calI^{I=1}_{a,B,C}}&=&\fwboxL{294pt}{\text{{\normalsize$\displaystyle\phantom{{\,}}\frac{1}{\epsilon}\log\!\left(\!\frac{s_C}{s_B}\!\right)\!{-}\frac{1}{2}\!\left[\log(s_C)^2\!{-}\!\log(s_B)^2\right]\!{-}\!\log(\mu^2)\!\log\!\left(\!\frac{s_B}{s_C}\!\right)\!\!{+}\mathcal{O}(\epsilon)$}}}\\[-7pt]
\hline\rule[-12pt]{0pt}{20pt}\fwboxL{45pt}{{\color{black}\displaystyle\int\!\!\calI^{I=2,3}_{a,B,C}}}&=&{\color{black}\displaystyle\phantom{{\,}}\frac{1}{\epsilon}{-}\frac{1}{2}\log\!\left(\frac{s_B\,s_C}{\mu^2}\right){+}2{+}\log\!\left(\hspace{-2pt}\frac{s_B}{s_C}\hspace{-2pt}\right)\frac{1}{2\br{p_A,\pX}}\Big[}\\&&{\color{black}\hspace{28pt}\br{p_B{-}p_C,\pX}{-}4\br{p_B,p_A,p_C,\pX}/(s_B{-}s_C)\Big]{+}\mathcal{O}(\epsilon)}\end{array}\\\hline
\fwbox{75pt}{\oneMassTriInts\rule[-32pt]{0pt}{70pt}}&\begin{array}{@{$\;$}l@{}l@{}l}\\[-13pt]\rule[-20pt]{0.0pt}{28pt}\fwboxL{45pt}{\displaystyle\int\!\!\calI^{I=1}_{a,b,C}}&=&\fwboxL{294pt}{\text{{\normalsize$\displaystyle{-}\frac{1}{\epsilon^2}{+}\frac{1}{\epsilon}\log\!\left(\frac{s_C}{\mu^2}\right){-}\frac{1}{2}\log\!\left(\frac{s_C}{\mu^2}\right)^2{+}\mathcal{O}(\epsilon)$}}}\\[-5pt]
\hline\rule[-12pt]{0pt}{34pt}\fwboxL{45pt}{\displaystyle{\color{black}\int\!\!\calI^{I=2,3}_{a,b,C}}}&=&\displaystyle{\color{black}\phantom{{\,}}\frac{1}{2}\left(\frac{1}{\epsilon}{-}\log\!\left(\frac{s_C}{\mu^2}\right){+}2\right){+}\mathcal{O}(\epsilon)}\end{array}
\\\hline\fwbox{75pt}{\twoMassBubbleInt\rule[-23pt]{0pt}{50pt}}&\begin{array}{@{$\;$}l@{}l@{}l}\\[-16pt]\rule[-10pt]{0.0pt}{10pt}\fwboxL{40pt}{\displaystyle\int\!\!\calI^{}_{A,B}}&=&\fwboxL{289pt}{\text{{\normalsize$\displaystyle \phantom{{{-}}}\frac{1}{2}\left(\frac{1}{\epsilon}{-}\log\!\left(\frac{s_A}{\mu^2}\right){+}2\right){+}\mathcal{O}(\epsilon)$}}}\end{array}\\\hline
\fwbox{75pt}{\zeroMassBubbleInt\rule[-20pt]{0pt}{45pt}}&\begin{array}{@{$\;$}l@{}l@{}l}\\[-10pt]\rule[-20pt]{0.0pt}{10pt}\fwboxL{40pt}{\displaystyle\int\!\!\calI^{}_{a,B}}&=&\fwboxL{289pt}{\text{{\normalsize$\displaystyle\phantom{{-}}\mathcal{O}(\epsilon)$}}}\end{array}\\\hline\multicolumn{2}{c}{}\\[-38pt]
\end{array}$$\caption{\label{basisIntegralsTable}Integration results for all box, triangle, and bubble integrands in $\mathfrak{B}_2^{(4)}$.}\vspace{-26pt}\end{table}

\newpage
\section{\texorpdfstring{Summary of Results Provided as Ancillary Files}{Summary of Results Provided as Ancillary Files}}
\label{app:ancillary_files}\vspace{-5pt}
%
For the interested reader, the results described in this work are available as ancillary files which may be downloaded from the abstract page on the \texttt{arXiv}. Three files are provided:
\begin{description}
\item[$\bullet$] \textbf{one\_loop\_bubble\_basis\_data.m}: a plaintext file consisting of the complete bubble power-counting basis of integrands constructed in this work.
\item[$\bullet$] \textbf{one\_loop\_bubble\_basis\_tools.m}: a \textsc{Mathematica} package file consisting of code useful to analyze and evaluate the data of the preceding file.
\item[$\bullet$] \textbf{one\_loop\_n\_leq\_4\_MHV\_amplitudes\_walkthrough.nb}: a \textsc{Mathematica} notebook which illustrates our results and the functionality of the codebase.
\end{description}
The ancillary file \textbf{one\_loop\_bubble\_basis\_data.m} provides analytic expressions for the integrand basis as well as their integrated expressions, packaged as follows:
\begin{description}
\item[$\bullet$] \texttt{{\color{hblue}integrands}}[${\color{nmhvred}legList\_\_}$]: a function which takes as argument a ${\color{nmhvred}legList}$ of external legs---ordered according to the conventions of this work---and returns the list of diagonalized integrands with bubble power-counting written in terms of dual loop momenta. Alternatively, the more abstract expressions found in appendix~\ref{app:integrand_numerators} can be generated by using as argument anywhere between two and four numbers (depending on the topology of interest), each of which is either $1$ or $2$ (indicating either a massless or massive vertex, respectively). In this case, the edges for external momenta are labelled by $p[A],p[B],\ldots$ when massive, and $p[a],p[b],\ldots$ when massless. Moreover, when necessary the first label appearing in a massive vertex is labelled $p[a]$. Internal edges are labelled $\{a,b,c,d\}$. 
\item[$\bullet$] \texttt{{\color{hblue}integrals}}[${\color{nmhvred}legList\_\_}$]: a function which returns the integrated expressions for every element of our bubble power-counting basis of integrands. Those basis elements which are UV/IR divergent are expressed in dimensional regularization with $\epsilon=(4-d)/2$.
\end{description}
The ancillary file \textbf{one\_loop\_bubble\_basis\_tools.m} contains the all-multiplicity MHV amplitude integrand as well as a variety of useful tools for the analysis and numerical evaluation of our results. We provide a brief summary of some key functionality contained in this package: 
\begin{description}
\item[$\bullet$] \texttt{{\color{hblue}completeIntegrandBasis}}[${\color{nmhvred}n\_,p\_:2}$]: returns a symbolic representation of the complete prescriptive basis of integrands for ${\color{nmhvred}n}$-points and ${\color{nmhvred}p}$-gon power-counting, for either ${\color{nmhvred}p}=2,3$. 
\item[$\bullet$] \texttt{\color{hblue}symMHVAmplitudeTerms}[${\color{nmhvred}n\_}$]: returns the ${\color{nmhvred}n}$-point chiral box expansion for the MHV amplitude in maximally supersymmetric Yang-Mills theory \cite{Bourjaily:2013mma}.
\item[$\bullet$] \texttt{{\color{hblue}ymMHVAmplitudeTerms}}[${\color{nmhvred}\mathcal{N}\_:1}$][${\color{nmhvred}n\_,\{i\_,j\_\}:\{1,2\}}$]: returns in symbolic form the ${\color{nmhvred}n}$-point MHV amplitude integrand in $\text{sYM}_{\color{nmhvred}\mathcal{N}}$ with particles ${\color{nmhvred}\{i,j\}}$ related to the negative helicity gluons. 
\item[$\bullet$] \texttt{{\color{hblue}symMHVAmplitudeIntegral}}[${\color{nmhvred}\mathcal{N}\_:1}$][${\color{nmhvred}n\_,\{i\_,j\_\}:\{1,2\}}$]: returns the integrated expression for the ${\color{nmhvred}n}$-point MHV amplitude with particles ${\color{nmhvred}\{i,j\}}$ related to the negative helicity gluons. 
\end{description}

\subsection*{Kinematics and Numerical Evaluation}

\begin{description}
\item[$\bullet$] \texttt{{\color{hblue}randomKinematics}}[${\color{nmhvred}n\_:6}$]: defines the global variable \texttt{Zs} to be a randomly chosen set of momentum twistors, the corresponding list of four momenta, stored in \texttt{pList}, as well as the two-component spinors, stored in \texttt{Ls} and \texttt{Lbs}.
\item[$\bullet$] \texttt{{\color{hblue}evaluate}}[{\color{nmhvred}expression\_\_}]: uses the kinematical data generated by e.g., \texttt{{\color{hblue}randomKinematics}} to evaluate all expressions involving brackets (angle, square, $\br{}$, etc.) and Mandelstam invariants. 
\end{description}

\newpage 
\providecommand{\href}[2]{#2}\begingroup\raggedright\endgroup

\end{document}